\documentclass[11pt,preprint2]{aastex}
\usepackage{amsmath}
\usepackage{longtable}
\usepackage{lscape}

\newcommand{\fermi}{{\it Fermi}}
\newcommand{\swift}{{\it Swift}}
\newcommand{\rxte}{{\it RXTE}}
\newcommand{\integralsc}{ {\it INTEGRAL}}
\newcommand{\maxi}{{\it MAXI}}
\shorttitle{GBM Earth occultation catalog}
\shortauthors{Wilson-Hodge et al.}
\begin{document}
\title{Three years of \fermi\ GBM Earth Occultation Monitoring: Observations of Hard X-ray/Soft Gamma-Ray Sources}
\author{Colleen A. Wilson-Hodge\altaffilmark{1}, Gary L. Case\altaffilmark{2}, Michael L. Cherry\altaffilmark{2}, James
Rodi\altaffilmark{2}, Ascension Camero-Arranz\altaffilmark{3}, Peter Jenke\altaffilmark{1,4}, Vandiver Chaplin\altaffilmark{5},
Elif Beklen\altaffilmark{6}, Mark Finger\altaffilmark{7}, Narayan Bhat\altaffilmark{5}, Michael S. Briggs\altaffilmark{5}, Valerie Connaughton\altaffilmark{5}, Jochen Greiner\altaffilmark{8}, R. Marc Kippen\altaffilmark{9}, Charles A.
Meegan\altaffilmark{8}, William S. Paciesas\altaffilmark{7}, Robert Preece\altaffilmark{5}, Andreas von Kienlin\altaffilmark{8}}
\altaffiltext{1}{ZP 12 Astrophysics Office, NASA Marshall Space 
 Flight Center, Huntsville, AL 35812}
\altaffiltext{2}{Department of Physics and Astronomy, Louisiana 
State University, Baton Rouge, LA, 70803, USA}
\altaffiltext{3}{Instituto de Ciencias del  Espacio (IEEC-CSIC),  
Campus UAB, Torre C5, 2a planta, 08193 Barcelona, Spain}
\altaffiltext{4}{NASA Postdoctoral Program Fellow}
\altaffiltext{5}{University of Alabama in Huntsville, Huntsville, AL  
35899, USA}
\altaffiltext{6}{Physics Department, Suleyman Demirel University, 
32260 Isparta, Turkey}
\altaffiltext{7}{Universities Space Research Association, Huntsville, 
AL 35805, USA}
\altaffiltext{8}{Max-Planck Institut f\"ur Extraterrestische Physik, 
85748, Garching, Germany}
\altaffiltext{9}{Los Alamos National Laboratory, Los Alamos, NM 
87545}

\begin{abstract}
The Gamma ray Burst Monitor (GBM) on board \fermi\ has been providing continuous data to the astronomical community since 2008 August 12. In this paper we present the results of the analysis of the first three years of these continuous data using the Earth occultation technique to monitor a catalog of 209 sources. From this catalog, we detect 99 sources, including 40 low-mass X-ray binary/neutron star systems, 31 high-mass X-ray binary neutron star systems, 12 black hole binaries, 12 active galaxies, 2 other sources, plus the Crab Nebula, and the Sun. Nine of these sources are detected in the 100-300 keV band, including seven black-hole binaries, the active galaxy Cen A, and the Crab. The Crab and Cyg X-1 are also detected in the 300-500 keV band. GBM provides complementary data to other sky-monitors below 100 keV and is the only all-sky monitor above 100 keV. Up-to-date light curves for all of the catalog sources can be found at \url{http://heastro.phys.lsu.edu/gbm/}.
\end{abstract}


\section{Introduction}
The low energy gamma-ray/hard X-ray sky is populated largely by active X-ray binaries, active galactic nuclei, soft gamma ray repeaters, the Crab, and the Sun. \swift/BAT is currently monitoring the sky in the region 14 - 195 keV \citep{Baumgartner2010}. There are multiple all-sky telescopes monitoring the sky at lower energies, for example \rxte/ASM at 2 - 10 keV \citep{Levine1996}, and \maxi/GSC at 1.5 - 20 keV \citep{Matsuoka2009}. At higher energies, where instruments such as \integralsc/IBIS \citep[15-1000 keV]{Krivonos2010} provides frequent but non-continuous observations, the most recent near-continuous all-sky catalog is that of {\it CGRO}/BATSE over the energy range 20 - 1800 keV \citep{Ling2000, Harmon2004}. The BATSE survey was conducted with a set of non-imaging detectors using the Earth occultation technique \citep[EOT,][]{Harmon2002}. The flux from an individual source is measured by detecting the step-like feature in the counting rate as the source passes into or moves out of eclipse by the Earth. Here we present a new catalog of sources over the energy range 8 - 1000 keV utilizing Earth occultation with the set of sodium iodide detectors aboard the Gamma-ray Burst Monitor (GBM) instrument \citep{Meegan2009} on \fermi.

The occultation technique has been used at X-ray energies with the Moon as an occulter to detect the Crab \citep{Bowyer1964, Fukada1975} and GX 9+1 \citep{Davidson1977}, and BATSE demonstrated the use of Earth occultation to provide a wide-field survey of the hard X-ray/low energy gamma-ray sky over the period 1991 -- 2000. Since the launch of {\it Fermi} on June 11, 2008 and the start of science operations on August 12, 2008, GBM has used Earth occultations to observe a pre-defined catalog of sources.  \citet{Case2011a} have presented initial results, with positive identification above 100 keV of six persistent sources (the Crab, Cyg X-1, SWIFT J1753.5-0127, 1E 1740-29, Cen A, and GRS 1915+105) and two transient sources (GX 339-4 and XTE J1752-223). \citet{Wilson2011} used the GBM occultation measurements to demonstrate that the 15 - 100 keV flux from the Crab had decreased by approximately 7\% over the time from launch through August 2010, and \citet{Camero2012} have presented the GBM observations of the transient Be/X-ray binary system A0535+26. The present paper describes the GBM approach in detail and presents the results of the first three years of GBM Earth occultation measurements.

Sections 2 - 3 describe the GBM occultation technique. Since the method requires a comparison of the measured flux from a particular source just before occultation to the flux just after, the times of occultations must be known. In other words, a predetermined catalog of potential sources must be used. The selection of candidate sources is described in Section 4. Systematic effects and the instrument sensitivity are discussed in Sections 5 and 6, and the results of the 3-year GBM occultation catalog are presented in Section 7.

\section{GBM \label{sec:gbm}}

The \fermi\ Gamma-ray Burst Monitor (GBM) consists of 14 detectors: 12 NaI detectors, each 12.7 cm in diameter and 1.27 cm thick (each with effective area $\sim 123$ cm$^2$ at 100 keV); and two BGO detectors, 12.7 cm in diameter and 12.7 cm thick (each with effective area $\sim 120$ cm$^2$ in the 0.15--2 MeV range).  The NaI detectors are located on the corners of the spacecraft, with six detectors oriented such that the normals to their faces are perpendicular to the z-axis of the spacecraft (the LAT is pointed in the $+z$-direction), four detectors pointed at $45^{\circ}$ from the z-axis, and 2 detectors pointed $20^{\circ}$ off the z-axis.  Together, these 12 detectors provide nearly uniform coverage of the unocculted sky in the energy range from 8 keV to 1 MeV. Typically 3-4 NaI detectors view an Earth occultation within 60 degrees of the detector normal vector.  The two BGO detectors are located on opposite sides of the spacecraft and view a large part of the sky in the energy range $\sim150$ keV to $\sim40$ MeV. It should be noted that none of the GBM detectors have direct imaging capability.

GBM has two continuous data types: CTIME data with nominal 0.256-second time resolution and 8-channel spectral resolution and CSPEC data with nominal 4.096-second time resolution and 128-channel spectral resolution. The catalog results presented in this paper use the lower-spectral resolution CTIME data for the NaI detectors. Analyses using the higher resolution CSPEC data, beyond the Crab spectrum presented in Section~\ref{sec:crabspec}, and using the BGO detectors are reserved for future work.

\begin{figure}[t]
\includegraphics[width=3in]{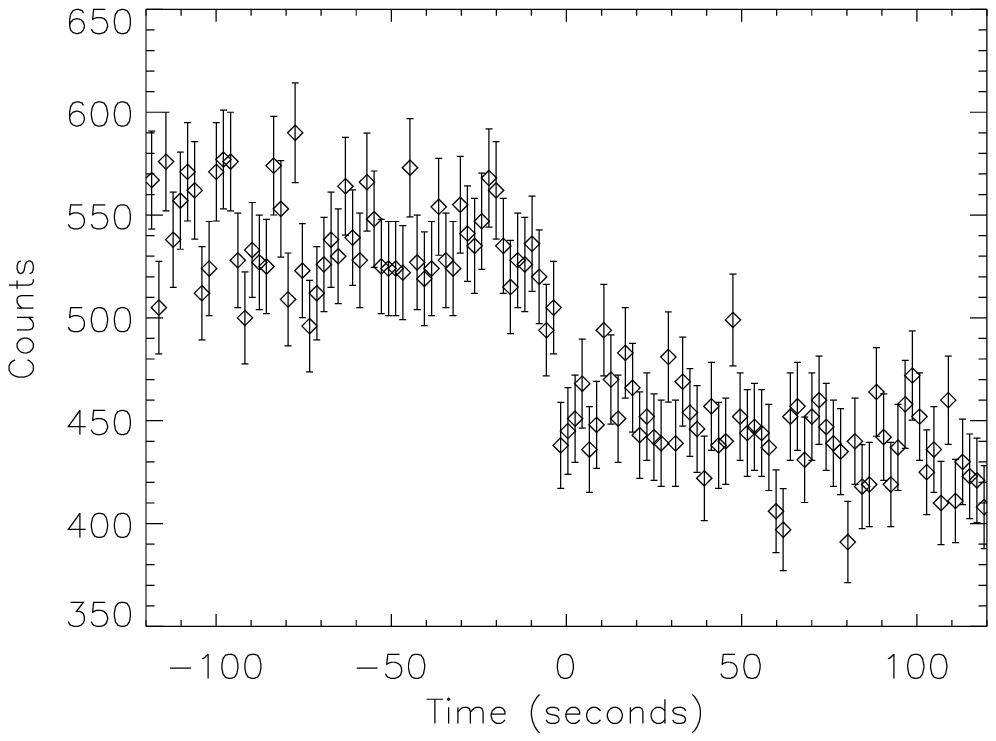}
\caption{\label{crabstep}Single Crab occultation step seen in the CTIME raw counts data of a single GBM NaI detector (NaI 2) in the 12--25 keV band with 2.048-second time bins. The Crab was $4.5^{\circ}$ from the normal to the detector. The time window is centered on the calculated occultation time for 100 keV.}
\end{figure}

\fermi\ was launched into a $i = 25.6^\circ$ inclination orbit at an altitude of 555 km. The diameter of the Earth as seen from \fermi\ is $\sim 135^\circ$, so roughly 30\% of the sky is occulted by the Earth at any one time.  One complete orbit of the spacecraft allows over 85\% of the sky to be observed.  The precession of the orbital plane allows the entire sky to be occulted every $\sim26$ days (half the precession period for the \fermi\ orbit), though the exposure is not uniform.

\section{Step Fitting Technique\label{sec:fitting}}

We have adapted the technique of \citet{Harmon2002} for GBM. This technique involves fitting a model consisting of a quadratic background plus source terms to a short ($\sim 4$ min) window of data centered on the occultation time of the source of interest. For GBM we have incorporated the changing detector response across the fit window into our source terms. The \citet{Ling2000} approach, also used with BATSE, which involved simultaneous fits to an empirical background and numerous source terms over typically an entire day during which each source and the corresponding detector response was assumed constant, is not practical for GBM data due to the rapidly changing detector response. In addition, the \citet{Ling2000} method resulted in apparent hard tails for several weak sources that were not confirmed with other instruments \citep{Harmon2004}.

The primary difference in the implementation of the occultation technique
between GBM and BATSE arises from the different pointing schemes of the
respective missions. {\it CGRO} was three-axis stabilized for each viewing
period, which typically lasted for two weeks. This meant that a source remained
at a fixed orientation with respect to the detectors through an entire viewing
period. In contrast, \fermi\ scans the sky by pointing in a direction $35^\circ$
(August 2008--September 2009) or $50^\circ$ (October 2009--present) north of the
zenith for one orbit; it then rocks to $35^\circ$ or $50^\circ$ south for the
next orbit, continuing to alternate every orbit unless the spacecraft goes into
a pointed mode (which occurs rarely). In addition, the spacecraft performs a
roll about the z-axis as it orbits so the solar panels face the Sun.  Because
the orientation of a source with respect to the GBM detectors varies as a
function of time, the detector response also varies with time. A detailed
instrument modeling and measurement program has been used to develop the GBM
instrument response as a function of direction \citep{Hoover2008,Bissaldi2009,
Meegan2009} and has produced a database of detector responses for each detector
at 272 spacecraft positions. For the occultation analysis we interpolated this
database to produce matrices for the same 272 positions using 137 photon input
energies (from 5 keV to 50 MeV) and using 3-year average CTIME channel energy
edges for each detector and for each of the 8 CTIME channel energy edges
(approximately $<12$, 12-25, 25-50, 50-100, 100-300, 300-500, 500-1000, $>1000$
keV). Custom energy matrices can be constructed using CSPEC data at full
resolution, or in combinations which provide fine resolution in the energy band
of interest while maintaining a small number of output channels in the response
matrix. In the individual step fitting, the spacecraft position of each source in
the fit is computed as a function of time. A new response matrix is interpolated from our database using its three nearest neighbors every time the spacecraft moves by $\sim 2^{\circ}$.

Before any sources are fitted, good time intervals (GTIs) of GBM data are defined. The GTI intervals from the CTIME data files are shortened by 10 s to remove transient events due to GBM high voltage turn-on and turn-off. A spline model is fitted to the 12-25 keV CTIME data to eliminate large background deviations, typically on $\sim 100$s time scales, due to South Atlantic Anomaly entries and exits, bright solar flares, gamma-ray bursts, and other brief bright events from the GTIs.

Mission averaged energy channel edges are used in the detector response matrices. Gain variations in individual detectors due to temperatures or other effects are controlled using an on-board automatic gain control to keep the 511 keV line in a particular on-board channel (one of 4096). The maximum temperature variation observed to date was $\sim 4.35$ C in December 2008, which corresponds to a gain change of 1.36\%, which is negligible. Because scintillators are massive compared to semiconductor detectors, temperature changes have much less impact for GBM than for other X-ray missions using semiconductors or CCDs.

For each day, the occultation times for each source in the catalog are calculated using the known spacecraft positions. The time of each occultation step, $t_0$, is taken to be the time for which the transmission of a 100 keV gamma ray through the atmospheric column is 50\%. A fit window of data, lasting 240 seconds and centered on $t_0$ is used for each step. The time at which the atmospheric transmission reaches 50\% is energy dependent, with lower energies absorbed at lower atmospheric densities so that a setting step will occur earlier than at higher energies (see Fig.~\ref{crabstep}). This energy dependence is accounted for in the calculation of the atmospheric transmission function
$T(E_{ph},t)$, given by 
\begin{equation}
T(E_{ph},t) = \exp[-\mu(E_{ph}) A(h(t))]
\end{equation}
where $\mu(E_{ph})$ is the mass attenuation coefficient of gamma rays at photon energy $E_{ph}$ in air\footnote{\tiny{\url{http://physics.nist.gov/PhysRefData/XrayMassCoef/ComTab/air.html}}} and $A(h(t))$ is the air mass along the line of sight at a given altitude $h(t)$ based on the \citet{atm1976}. This requires instantaneous knowledge of the spacecraft position, the direction to the source of interest as seen from the spacecraft, and a model of the Earth that includes its oblateness. Changes in chemical content with altitude and changes in the atmospheric height with solar activity are not included in the current atmospheric model. Atmospheric density measurements with sounding rockets \citep{Quiroz61} show that at mid-latitudes the standard deviation of the density varies from 4\% at 30 km to about 20\% at 60km and above. To estimate the effect of these variations on our Earth occultation flux measurements, we varied the air mass $A(h(t))$ by 10\% and 50\% in fits to the Crab Nebula. This resulted in changes to the daily Crab flux measurements of $<2$\% in the four bands spanning 12-300 keV. Therefore we ignore the effects of atmospheric variations in our model.

Measuring the flux from the source of interest requires fitting a model to the count rate data. For each detector viewing the source of interest within 60$^{\circ}$ from the detector normal, within the fit time window, the observed count rate $r(t,E_{ch})$ at time $t$ in each energy channel $E_{ch}$ is modeled as
\begin{eqnarray}
r(t,E_{ch}) = b_0(E_{ch})+b_1(E_{ch})*(t-t_0)+ \nonumber\\
  b_2(E_{ch})*(t-t_0)^2 + \sum_{i=1}^{n} a_i(E_{ch})*S_i(t,E_{ch}) 
\end{eqnarray}
where $b_0(E_{ch})$, $b_1(E_{ch})$,and $b_2(E_{ch})$ are quadratic background coefficients, $S_i(t,E_{ch})$ are source models summed for the source of interest and all interfering sources included in the fit, and $a_i(E_{ch})$ are the fitted scale factors for each source model. The background is typically smooth and adequately fitted by the second-order polynomial within the 240 second fit window.
The source count rate models are given by

\begin{equation}
 S(t,E_{ch})  =  R(E_{ph},E_{ch},t)T(E_{ph},t)\int_{E_{ph}} f(E_{ph})dE_{ph} 
\end{equation}
where $E_{ph}$ are photon energies, matching the input side of the time-dependent detector response matrix, $R(E_{ph},E_{ch},t)$.  The assumed energy spectrum for each source integrated over each photon energy bin,  $\int_{E_{ph}} f(E_{ph})dE_{ph}$,  is combined with the atmospheric transmission $T(E_{ph},t))$ and convolved with the detector response $R(E_{ph},E_{ch},t)$ to compute the predicted count rate in each energy channel $E_{ch}$ at time $t$. 

Each energy channel and each detector is fitted independently. For each source term a scaling factor $a_i(E_{ch},t)$ is fitted, along with the quadratic background coefficients. When multiple detectors are included in the fit, the weighted mean for the scaling factor of the source of interest is computed for each energy channel. The mean scaling factor is then multiplied by the predicted flux in each energy channel to obtain flux measurements for the source of interest.

Ideally, each occultation measurement would include the effects of every other source in the sky, but this is not practical to implement. We have adopted a top-down iterative approach for treating interfering sources. Following \citet{Harmon2004}, we have implemented a flare database consisting of times when sources are active and broad levels of activity. For our first iteration, we have used public  \swift/BAT transient monitor data to populate our flare database. Later iterations will incorporate results from GBM and from the \swift/BAT survey over a wider energy range. Currently any source that has a 15--50 keV flux of 50 mCrab or larger in \swift/BAT that persists for at least a few days is included in the database. On any particular day in the flare database, sources are grouped into three classes: (1) $> 500$ mCrab, (2) 150--500 mCrab, or (3) 50--150 mCrab. A source's class changes in the database whenever the flux drops or rises into the next class. If a source changes flux quickly within a single day, the brightest class for that day is retained. Identical source classes are used in the source catalog for persistent sources. Sources in these classes are included as interfering sources if they undergo Earth occultation in the fit window and if they are within $90^{\circ}$, $60^{\circ}$, or $40^{\circ}$ of the detector normal for classes 1, 2, and 3, respectively. Fainter sources are not currently considered in occultation fits except when they are the source of interest. This paper reports results from our first iteration, in which we treat the brightest sources first  to optimize interfering source inclusion.  Our code is flexible, so additional source classes can be added if needed in the future to support analysis of fainter sources. In addition to the flare database, we maintain an eclipse database, containing ephemerides for ten eclipsing sources. When any of these ten sources are in eclipse, they are not included as interfering sources in step fitting, regardless of their levels set in the catalog or flare database. 

At lower energies, especially below about 50 keV, the Sun is a special case of a bright interfering source. We maintain a solar flare database, built from Solar event reports of Geostationary Operational Environment Satellites (GOES) data obtained from the National Oceanic and Atmospheric Administration (NOAA) Space Weather Prediction Center\footnote{\url{http://www.swpc.noaa.gov/}}. In our solar flare database we use classes for the flares, based upon the peak GOES flux in the flare: (1) Class M or X flares, included as an interfering source if the Sun is within $90^{\circ}$ of the detector normal, (2) Class C flares, included as an interfering source if the Sun is within $60^{\circ}$ of the detector normal, and (3) Class B flares, included as an interfering source if the Sun is within $40^{\circ}$ of the detector normal. Since the Sun moves with respect to other sources in the sky, the Sun's position is initially calculated for the center of the day. This position is used to find fit windows that include the Sun. At the time the data are fitted, the Sun position at the time of the fit window is calculated to correctly compute the atmospheric transmission. Small errors in the timing of Sun steps still exist, because the Sun is not a point source and the position of the center of the Solar disk is used. Data containing X or M class flares are usually filtered out from the GTIs before steps are fitted, so only C class and fainter flares are typically included as interfering sources.

Once the steps are fitted, source count rates and errors for the source of interest are written to a Flexible Image Transport System (FITS) file for each detector along with other information about the fit including the flux and error measurement in each energy channel, the step times, sources included in the fit, detectors, and angles to the source of interest at the time of occultation. Light curves and energy spectra are extracted from these files. Light curves used in this catalog paper have been post-filtered using the following criteria: (1) occultation steps are excluded if the source of interest occults within 8 seconds of a bright source, if the occultation transition lasts longer than 20 seconds, or if the spacecraft is rapidly slewing with a spin rate $>0.004$ rad/s. (2) individual steps that are $>10 \sigma$ or $>3.5 \sigma$ from the mean are filtered out for sources with intensities reaching 150-500 mCrab or $<150$ mCrab, respectively. Outliers for sources that reach fluxes $>500$ mCrab are not filtered out to avoid discarding real flares. (3) occultation steps during solar flares are also discarded. 

\section{Source Catalog Selection}

\begin{figure}[!h]
\includegraphics[width=2.65in,angle=90]{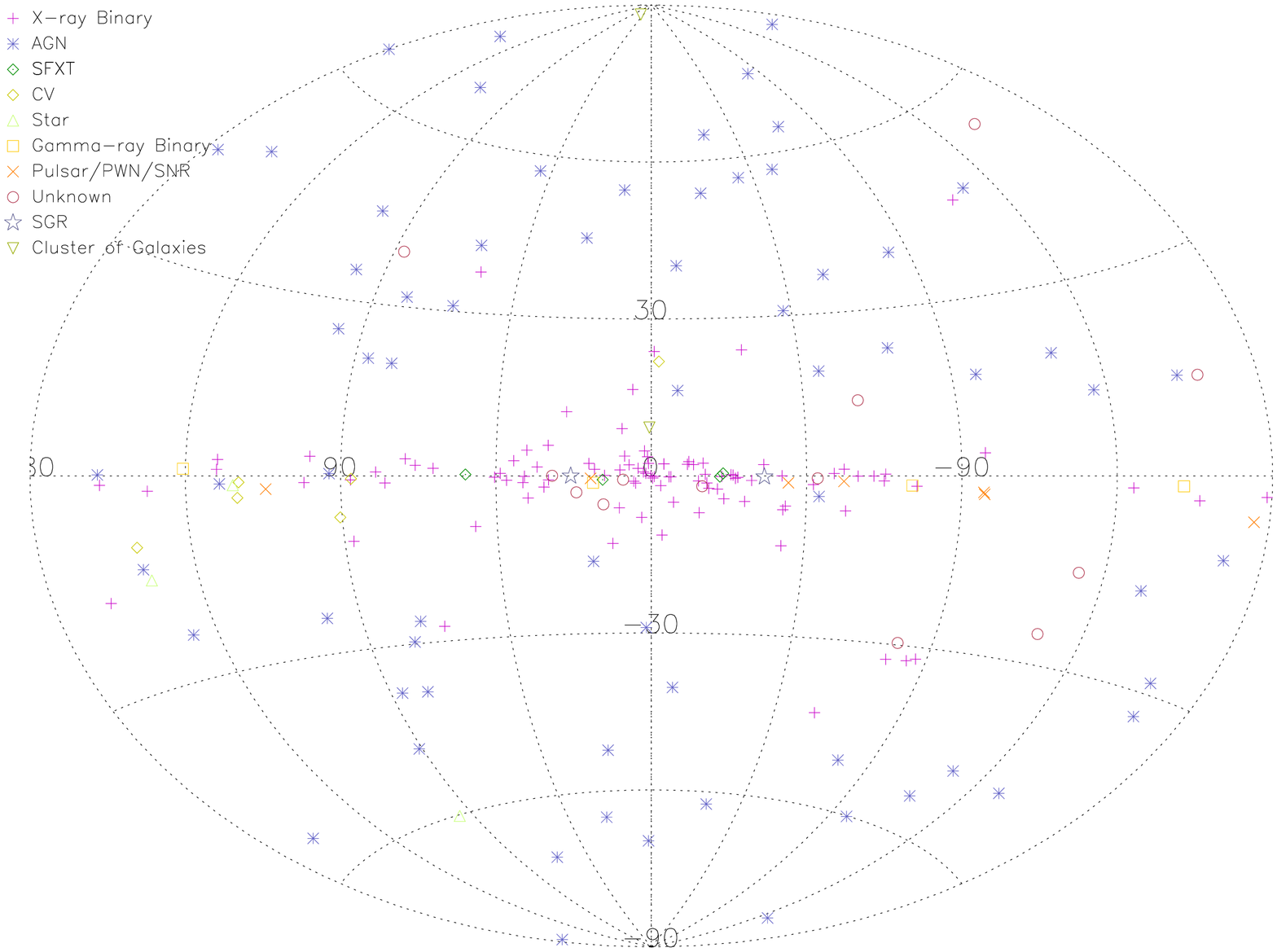}
\caption{Sky map in equatorial coordinates showing the sources included in the GBM occultation catalog. {\label{fig:capmap}}}
\end{figure}

As described in the previous section, the Earth occultation technique being used with GBM is catalog-driven meaning that only those sources included in the catalog are measured. Because GBM is sensitive to lower energies than BATSE, the occultation catalog was expanded to include sources and even source classes that were undetected with BATSE but
seen to be bright in pointed X-ray instruments such as \integralsc, \maxi, \swift/BAT, or \rxte/ASM.   In addition,
extragalactic sources with two spectral components that are shown to be bright in the
\fermi/LAT and to exhibit bright flares on time-scales of days or weeks were viewed as
viable candidates for our catalog, though their quiescent levels would typically 
be faint. To be as complete as possible, we adopted an iterative approach, first selecting sources for the catalog that were detected with BATSE using Earth occultation \citep{Harmon2004} plus bright sources ($> 50$ mCrab) detected with \swift/BAT, \maxi, or \integralsc/IBIS. This initial catalog was then expanded by adding 14 sources detected using GBM Earth occultation imaging \citep{Rodi2011}. To prepare for this catalog paper, we did a more thorough effort of source selection, selecting sources detected above $\sim 10$ mCrab with \swift/BAT in the transient monitor\footnote{\url{http://swift.gsfc.nasa.gov/docs/swift/results/transients/}} or in the \swift/BAT 58 month survey\footnote{\url{http://swift.gsfc.nasa.gov/docs/swift/results/bs58mon/}}, or detected above $\sim 20$ mCrab in public light curves from either  \rxte/ASM\footnote{\url{http://xte.mit.edu/}} or \maxi\footnote{\url{http://maxi.riken.jp/top/}}. To further broaden our inclusion of source types, we also explored cataclysmic variables (CVs) detected with {\it INTEGRAL}, Supergiant Fast X-ray Transients (SFXTs) detected with {\it INTEGRAL} and \swift/BAT, pulsar wind nebulae and gamma-ray binaries detected with \fermi/LAT \citep{Abdo11}. For active galaxies (AGN), we selected sources detected with the Oriented Scintillation Spectrometer Experiment (OSSE) on CGRO and sources detected in the LAT 2-year AGN catalog \citep{Ackermann11} that met the following criteria: (1) $\gtrsim 10\arcdeg$ from the galactic plane, (2) flagged as in the LAT Clean sample, (3) classified as either a BL Lac with a High Synchrotron Peak (HSP) or classified as a Flat Spectrum Radio Quiet source (FSRQ) with a Low Synchrotron peak. The reasoning is that for the BL Lacs, the GBM energy range will sample the fall-off of the $\nu F_{\nu}$ synchrotron peak while the LAT is sampling the IC peak, and for FSRQs, the GBM band will sample the rise of the IC peak while the LAT samples the fall of the IC peak (assuming Leptonic processes), making these sources more likely to be detectable with GBM, and (4) a test statistic of $> 600$ in the full LAT energy range in the 2-year catalog, to limit this initial sample to the brightest LAT sources. These selections resulted in a catalog, shown in Figure~\ref{fig:capmap}, containing 209 sources, including 71 active galaxies (AGN), 52 low-mass X-ray binary/neutron star (LMXB/NS) systems, 40 high-mass x-ray binary/neutron star (HMXB/NS) systems (including 5 supergiant fast X-ray transients (SFXT)),  19 black hole candidates (BHC), 8 pulsars (PSR)/pulsar wind nebulae (PWN)/supernova remnants (SNR), 6 CVs, 4 gamma ray binaries, 2 galaxy clusters, 3 stars including the Sun, 2 SGRs,  1 globular cluster, and 1 tidal disruption event (TDE).

\section{Systematic Effects}
\subsection{Description of known effects and Mitigation Strategies}

From our experience using the Earth occultation technique with BATSE and with GBM, we have identified the following systematic effects that affect Earth occultation flux measurements: (1) accuracy of the assumed source spectra, (2) large variations in the background, (3) duration of the occultation transition, (4) inaccuracies in the detector response matrices, (5) occultation limb geometry, and (6) nearby sources. In this section we describe our mitigation strategies and our efforts to reduce, account for, or quantify these effects. 

We explored the effects of incorrectly assumed source spectra using multiple runs of the Earth occultation software for the Crab assuming (1) a canonical spectrum \citep{Toor1974}, (2) an exponential cut-off spectrum (with a cutoff energy of 30 keV and e-folding energy of 13.6 keV), (3) the exponential cutoff spectrum plus a 1 mCrab power law with photon index $=-2$, (4) a power law with a photon index $= -3$, and (5) a hard power law with photon index $=-1$. The measured count rates in each step were consistent within errors, indicating that the fitting process is robust, and in particular that the treatment of interfering source steps in the fit window is little affected by the assumed source spectrum. Further we found that the statistical significance of the average flux in each energy channel was consistent for all five assumed spectra. The flux values showed systematic effects, however. For the ``incorrect" spectra,
models 3 and 4, fluxes were consistent with the canonical values within 5-10\%, while model 2, which changes rapidly at higher energies, and the very hard model 5 showed large flux deviations. This test was repeated using models 2-5 for Vela X-1 and the results were the same as for the Crab. From 12 to 100 keV, where Vela X-1 is detected with GBM, count rates and significances were consistent for all models. Model 2 is based upon Vela X-1 BATSE data \citep{Bildsten97}. Models 3 and 4 fluxes were within 5-10\% of model 2 fluxes, while model 5 again showed large deviations. Therefore, the assumed spectrum does not appear to significantly affect detection significance or fits to other sources. Its primary impact is on the reported flux values. For this paper, we have done an extensive literature search to obtain the best published spectra for the sources in our catalog. Where appropriate spectra in our energy range are not available, these tests show that model 4, a power law with photon index $=-3$, gives fluxes consistent to 5-10\% with expected models for both a harder source (Crab) and a softer source (Vela X-1). We have adopted model 4 as a generic flux model for sources lacking spectra in the GBM energy range or for sources with rapidly changing spectra.  As described in Section~\ref{sec:bat_gbm} we have also compared our results with \swift/BAT in overlapping energy bands, where we generally found good agreement.

Because the background and sources are fit simultaneously, variations in the background that cannot be fitted with a quadratic function will be absorbed into the source terms and can potentially affect source measurements. For all of these background effects, our mitigation strategy is to exclude these data from the occultation flux measurements. We performed an investigation where we examined the CTIME count rates in the 12-25 keV band for times when we saw large outliers for Earth occultation flux measurements for the Crab, Cen A, and a few additional sources. These outliers frequently corresponded to times with large variations in the background, such as large slopes due to SAA entry or exit, bright (X or M class) solar flares, or intervals where \fermi\ is slewing or rotating rapidly. To mitigate these effects, we pre-filter the data, fitting a spline model to the 12-25 keV rates and excluding intervals with large deviations from this smooth model from the GTIs, so that these data are not used for occultation step fitting. This pre-filtering eliminates SAA entries and exits and bright solar flares from further analysis and is believed effective in eliminating similar effects that we have not yet identified. The spacecraft spin-rate is given by
\begin{equation}
\left(wsj_1^2+wsj_2^2+wsj_3^2\right)^{1/2}
\end{equation}
where $wsj_i$ are the instantaneous angular velocity components for the spacecraft, available in public spacecraft position history files from the \fermi\ Science Support Center\footnote{\url{http://fermi.gsfc.nasa.gov/ssc/data/access/}}. For this catalog, fit windows with spin-rates $>4 \times 10^{-3}$ rad s$^{-1}$ were flagged and excluded from further analysis. This is an adjustable parameter in the occultation software, which is currently set
conservatively to eliminate steps where the high-spin rate adversely affects the fitting.  

An Earth occultation has a finite transition time due to the effect of absorption in the Earth's atmosphere. Since the orbital period of the spacecraft is $\sim 96$ minutes, the individual occultation steps last for $\sim8/\cos\beta$ seconds, where $\beta$  is defined as the elevation angle of the source being occulted with respect to the plane of the \fermi\ orbit. Since the \fermi\ orbit has an inclination of 25.6$^{\circ}$, sources with declinations larger than $\approx \pm 40^{\circ}$ undergo intervals where they are no longer occulted. When $\beta \sim 66^{\circ}$ these sources are still being occulted, but the occultations are very broad, lasting $> 20$ seconds. Our analysis has shown that our flux measurements become unreliable if the transition lasts longer than about 20 seconds, so we flag these long duration steps and exclude them from further analysis.

The most noticeable systematic effect arising from the detector response matrices are solar panel occultations of sources. In the mass model for \fermi\ used to derive the responses, the solar panels are included at a fixed orientation, with the top rotated in the $+X$ direction by $20^{\circ}$ (where $0^{\circ}$ is vertical alignment in the Y-Z plane). The true solar panel orientation is not fixed, and may block a source, an effect that is particularly severe at low energies. The discrepancy between the modeled response assuming a fixed solar panel orientation and the true response to the source yields predicted blockages that do not occur or real blockages that are not predicted. To account for either of these issues, we have defined potential solar panel blockage regions in spacecraft coordinates for each detector based upon the full-range of motion of the solar panels. If the source of interest or an interfering source is in that region in spacecraft coordinates, that detector is excluded from occultation step fitting. Typically this blockage only occurs in 1-2 NaI detectors, and a flux measurement based on unblocked detectors is calculated where possible. In addition to the solar panel blockage regions, we have identified two small regions where Crab, Sco X-1, or Cyg X-1 occultations were obviously blocked in NaI 11, based upon low-flux measurements only in NaI 11. These blockage regions are believed to be due to small structures on the spacecraft that were not included in the response model, so NaI 11 data is excluded from fits if any sources included in the fit are in this region. Photopeak efficiencies measured during the \fermi\ source survey (with the solar panels folded) agree with the mass model simulations to within 5\% rms for detectors with source angles $<90^{\circ}$ confirming the accuracy of the responses for larger spacecraft structures \citep{Meegan2009}.

Every $\sim 52$ days, the \fermi\ orbit precesses so that the Earth occultation limb geometry (i.e. the projection of the Earth's limb on the sky) repeats at this period. For a particular geometry, Earth occultation flux measurements may be systematically low or high due to unmodelled sources or non-point source backgrounds such as galactic ridge emission. These systematic effects can be very difficult to identify. To mitigate these effects, we use \swift/BAT data to populate a flare database listing times and brightness levels for potentially interfering sources. This database is updated with current flares and if past flares were missed, they can be added and the occultation code rerun. Using Earth occultation imaging \citep{Rodi2011}, combined with existing catalogs of known source locations, we identified 14 sources detected with $>10\sigma$ significance that were not initially included in the Earth occultation catalog. These sources were added to the catalog. The catalog and flare database continue to be iteratively updated to extract our best results. 

Nearby sources, especially bright or highly variable sources, can make occultation measurements very difficult. If two occultation steps occur within 8 seconds of each other, the step fitting breaks down, so we automatically flag these steps and exclude them from further analysis. The angular resolution of the occultation technique is 360$^{\circ} \times \Delta t/P$ where $\Delta t$ is the occultation duration and $P$ is \fermi's orbital period ($\sim 96$ minutes). The occultation duration $\Delta t$ varies as $8/cos (\beta)$ seconds, where $\beta$ is the elevation angle between the source and the orbital plane of \fermi. The angular resolution ranges from $\sim 0.5^{\circ}$ at $\beta = 0^{\circ}$ to $\sim 1.25^{\circ}$ at $\beta = 66^{\circ}$. When $\beta > 66^{\circ}$, occultations no longer occur. Some remaining nearby source effects are relatively easy to identify, e.g the bright outburst of A0535+26 in December 2009 that is visible in the raw Crab light curve.  When these effects can be easily identified, we manually flag these data and exclude them from further analysis. However in crowded regions such as the galactic center, such simple approaches can break down, resulting in a flux measurement that is the sum of multiple sources within $\sim 1^{\circ}$, if uncatalogued sources become active.

In the next section, we describe our analysis of ``blank" sky positions (where no source is present) where we quantify observed systematic effects.

\subsection{``Ghost" source analysis}

\begin{figure}[!h]
\center{\includegraphics[width=4.5cm,height=3cm]{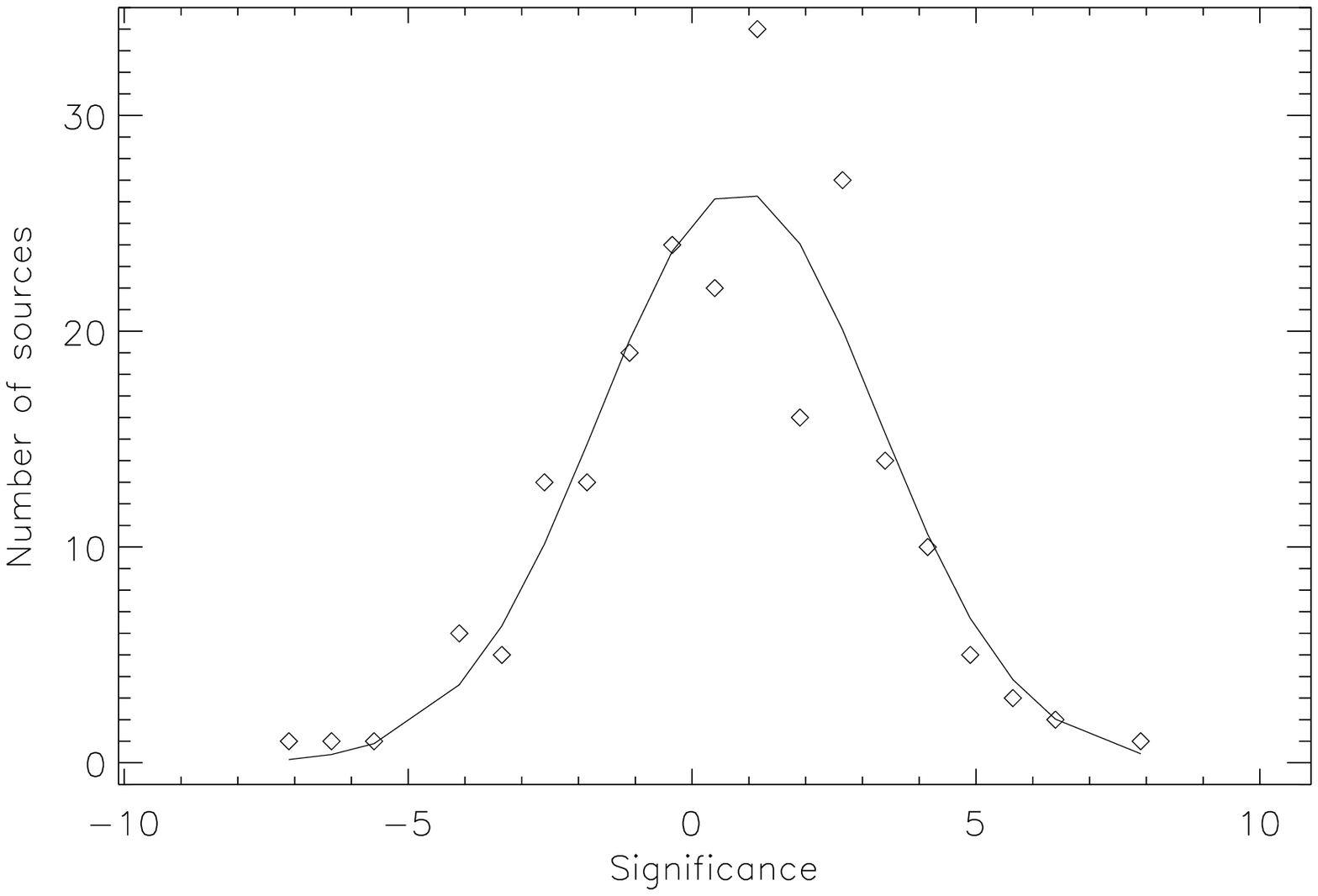}}
\vspace{-0.1in}
\center{\includegraphics[width=4.5cm,height=3cm]{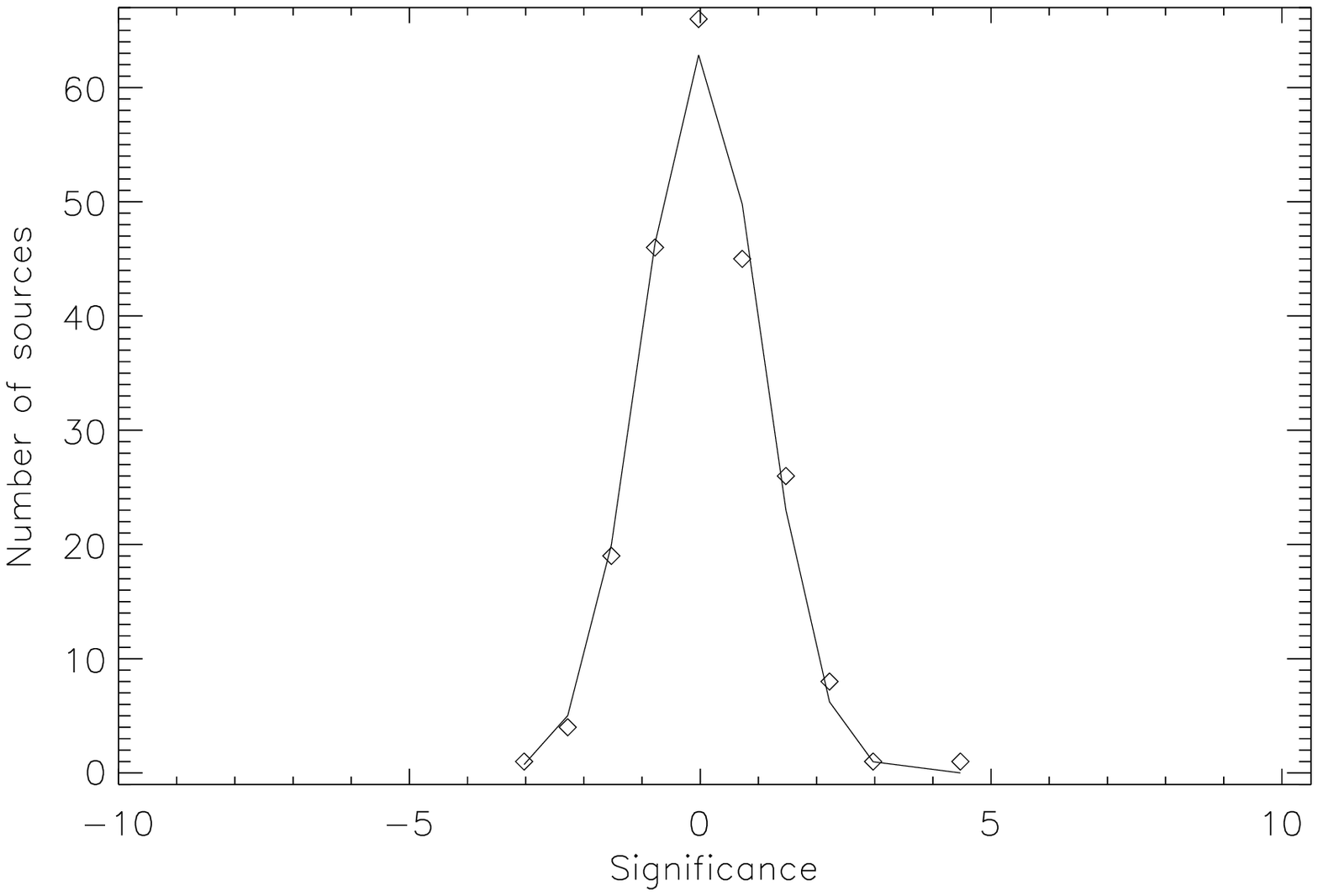}}
\vspace{-0.1in}
\center{\includegraphics[width=4.5cm,height=3cm]{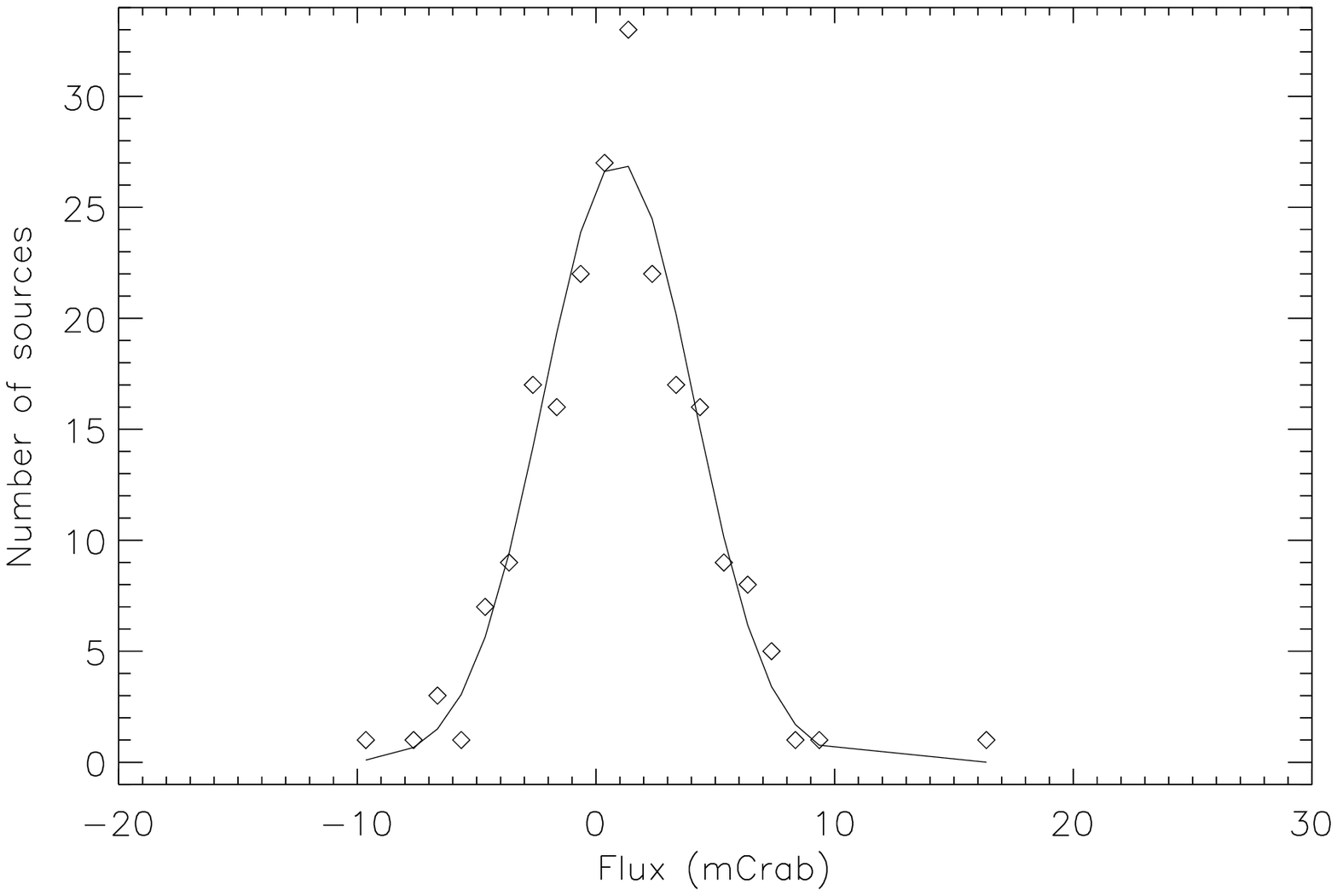}}
\vspace{-0.1in}
\caption{Top: Flux significance for the ``ghost" sources in the 12-25 keV band before systematic error correction. The standard deviation is 2.46. Center: Flux significance for the ``ghost" sources in the 12-25 keV band after systematic error correction. The new standard deviation is 1.02. Bottom: The distribution of ``ghost" source fluxes in the 12-25 keV band has a mean flux of 0.94 mCrab and a standard deviation of 3.15 mCrab.\label{fig:blank_sources}}
\end{figure}

\begin{table}
\begin{scriptsize}
\caption{\scriptsize{Systematic error estimates for GBM Earth occultation analysis} \label{tab:sys}}
\begin{tabular}{lc}
\hline\noalign{\smallskip} \hline\noalign{\smallskip}
Energy band (keV) & Systematic error (mCrab) \\
\hline\noalign{\smallskip}
8-12 & 3.4 \\
12-25 & 2.8 \\
25-50 & 2.2  \\
50-100 & 1.5 \\
100-300 & 3.1 \\
300-500 & 3.4 \\
\hline\noalign{\smallskip}
\end{tabular}
\end{scriptsize}
\end{table}

To examine the remaining systematic effects, 3-year light curves were selected from $\sim 512$ ``ghost" sources run through the occultation software in conjunction with GBM imaging analyses \citep{Rodi2011}. The initial list was reduced by excluding any ``ghost" sources within $\pm 10\arcdeg $ in longitude and latitude of the galactic center. Any ``ghost" sources within $2\arcdeg $ of a source in the GBM catalog were also removed, resulting in a sample of about 200 ``ghost" sources distributed over the whole sky. We investigated scatter plots of various combinations of parameters including location on the sky, flux, flux significance, flux statistical error, and flux standard deviation in all eight energy bands. 

These distributions were used to estimate overall systematic errors on the flux measurements. Since no source was expected at these locations, the distribution of the flux significance in each channel is expected to be centered on zero with a standard distribution of 1.0. The distributions were centered on zero, but were broader than expected.  In the flux distributions, the expected standard deviation gives a measure of the total error. To find the systematic error, we used the significance distributions to calculate a scale factor for each energy band by dividing the expected standard deviation (1.0) by the measured standard deviation. These scale factors were then multiplied by the measured flux distribution standard deviation to calculate the expected statistical flux standard deviation for each energy band. From there, we calculated the difference between the squares of the measured flux standard deviation and the calculated statistical standard deviation to get the square of the systematic standard deviation. This method gives systematic errors shown in Table~\ref{tab:sys}. These errors added in quadrature to the statistical errors for the ``ghost" sources result in standard deviations of $\sim 1$ for the flux significance distributions. As an example of the process, Figure~\ref{fig:blank_sources} shows the flux significance distributions, before and after systematic error correction, and the flux distribution for the 12-25 keV band.

\section{Sensitivity}

\begin{figure}[!h]
\includegraphics[width=3in]{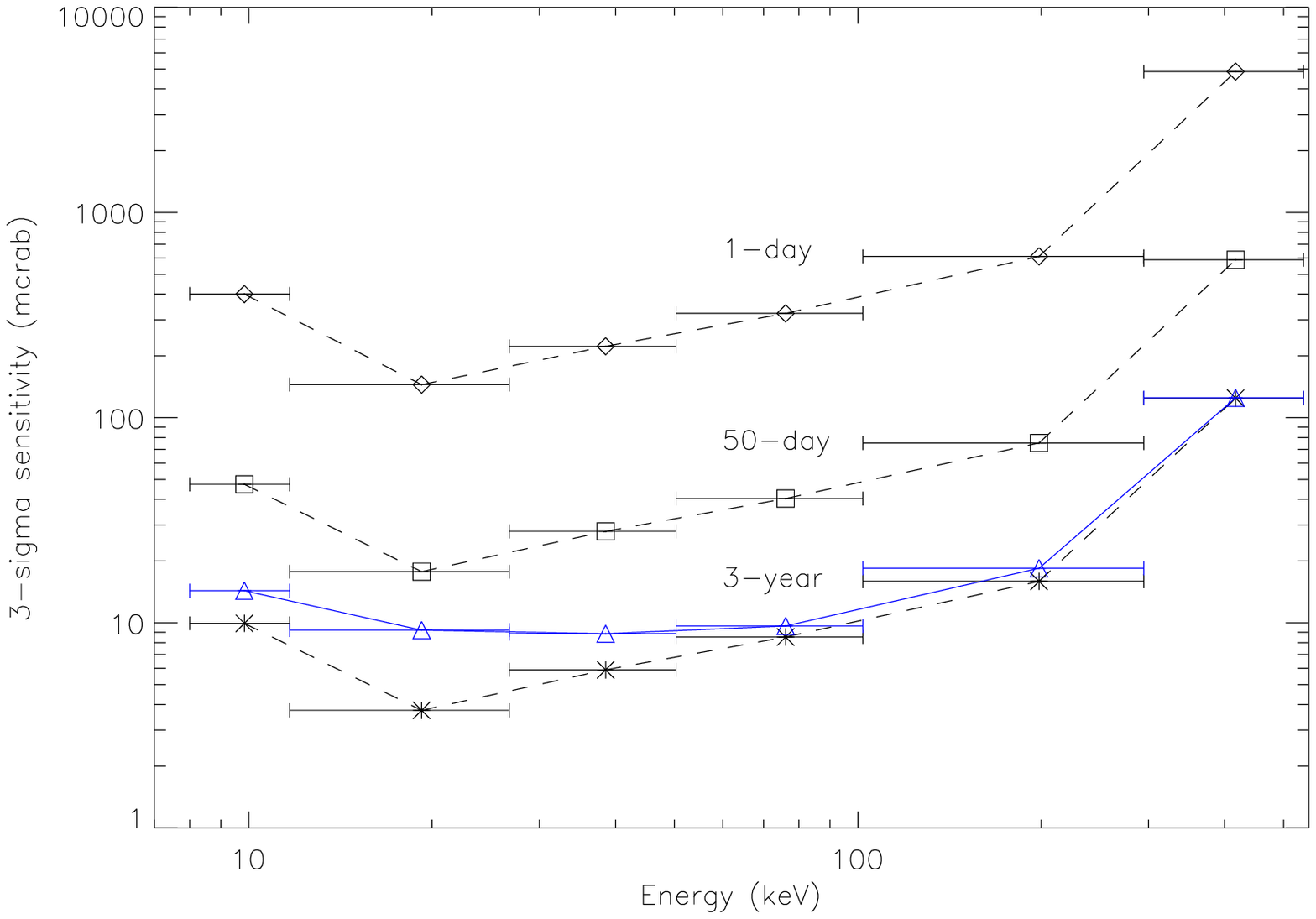}
\caption{Estimated 3-$\sigma$ sensitivities for the GBM Earth occultation technique from 8-500 keV. Statistical estimates are shown for 1-day (diamonds), 50-day (squares), and 3-year (asterisks) intervals. Systematic errors from Table~\ref{tab:sys} have been added in quadrature to the 3-year statistical errors and are plotted as triangles connected with solid lines.
\label{sens}}
\end{figure} 

Calculating the sensitivity for the Earth occultation technique is made challenging by the constantly changing spacecraft source geometry, detector response, and the constantly changing hard X-ray sky. To estimate the statistical sensitivity of the technique, we used four detected sources, the Crab (a galactic source in a relatively uncrowded region), Centaurus A (an active galaxy at high galactic latitude and high declination to account for steps lost when the source was at high beta angles), NGC 4151 (another active galaxy at high galactic latitude and moderate declination), and GRS 1915+105 (a persistent black hole system near but not at the galactic center). For each of these sources, 1-day, 50-day, and 3-year average measured fluxes and statistical errors were computed. The average error in each energy channel for each source was then computed. The approximate statistical sensitivity for each source was computed as three times the average error in mCrab units. For the three year averages only the systematic errors from Table~\ref{tab:sys} were added in quadrature. Figure~\ref{sens} shows the estimated sensitivities, averaged for the four sources.

As a check of our sensitivity estimate, we followed the \citet{Harmon2002} sensitivity estimate for BATSE where they used fitted count rates and count rate errors to estimate the flux sensitivity, specifically
\begin{equation}
F_{\rm min} = \delta r_{\rm Crab} \frac{N_{\sigma} F_{\rm Crab}}{r_{\rm Crab}}
\end{equation}
where $F_{\rm min}$ is the minimum detectable flux for $N_{\sigma}$ significance, $\delta r_{\rm Crab}$ is the uncertainty on the fitted Crab count rate $r_{\rm Crab}$, and $F_{\rm Crab}$ is the Crab Nebula flux. To implement this method, we computed the weighted mean count rate for each occultation step across the detectors in each fit and then computed weighted mean count rates with time. The 3-year sensitivity estimate from this method was consistent within 1-2\% with our simple approach using the flux errors. Based on the way the GBM fluxes are computed, using the weighted average of the scale factors between detectors, one would expect these two approaches to be approximately equivalent.

\section{Results}

\subsection{Fitting the Crab energy spectrum\label{sec:crabspec}}

\begin{figure}[!h]
\includegraphics[width=3in]{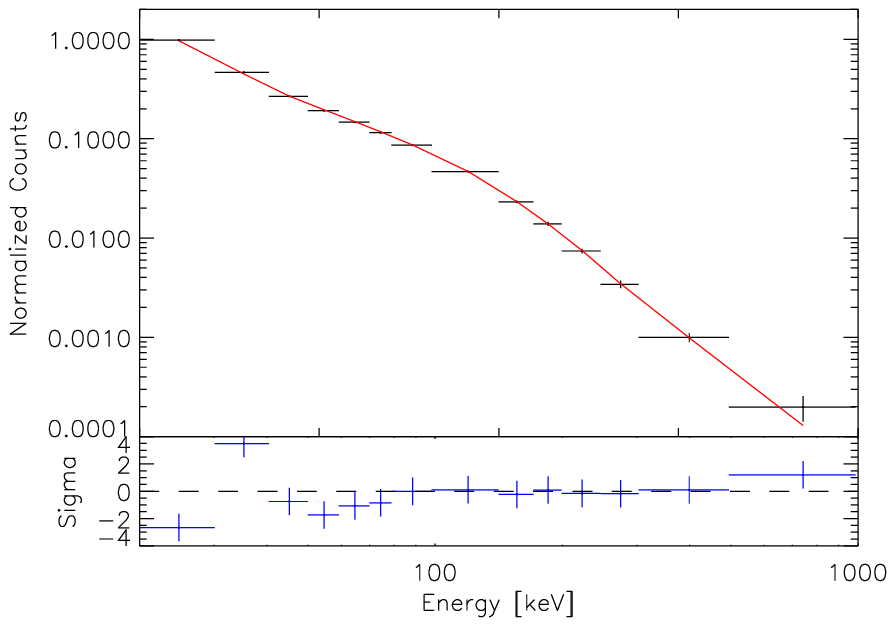}  
\caption{Fit to the Crab spectrum using 14,620 Earth occultation steps, approximately 1.7 Ms of on source time, and 42 position bins. An average spectrum, averaged over the 69 spectra simultaneously fit, and the fit model are shown. The best fit broken power-law model had a lower photon index of $2.057 \pm 0.01$, break energy of $98 \pm 9$ keV, and an upper photon index of $2.36 \pm 0.05$.}
\label{fig:crabspec}
\end{figure}

Spectral analysis with the EOT requires more than the eight energy channels available with CTIME data. However the current EOT software is designed for 8-channel data. The simplest approach to increasing the number of energy channels is to combine CSPEC channels to create a custom 8-channel CSPEC data set.  Rerunning the EOT code with different custom 8-channel CSPEC data sets results in a final spectrum with a multiple of eight channels.  We can customize the energy bins (to within the constraints of CSPEC energy edges) to the source of interest.  The resulting rates calculated from the step fits for each detector are then binned according to the position of the source in the space-craft coordinate frame.  A detector response is then generated for each position bin.  The spectra and responses for the position bins are fit to a photon model in XSPEC.  Due to the limited exposure time in individual occultation steps for a source,  spectral analysis with EOT will usually be performed with time averaged spectra.  To demonstrate this technique and to add confidence in the EOT we have performed this analysis using occultation rates calculated from observations of the Crab.  Since the pulsar and the nebula can not be distinguished with the EOT we expect an averaged spectrum for the pulsar and nebula.  Two different runs with custom CSPEC data were used to create the spectrum; a low energy run with bins: (10-15 keV, 15-20 keV, 20-30 keV, 30-40 keV, 40-50 keV, 50-60 keV,60-70 keV,70-80 keV), and a high energy run with bins: (80-100 keV, 100-140 keV, 140-170 keV, 170-200 keV, 200-250 keV, 250-300 keV,300-500 keV, 500-1000 keV). The data used are for the entire three year interval from 2008 August 12 to 2011 August 11. Position bins with fewer than 80 steps were excluded from the analysis. Each position bin was $6^{\circ} \times 8^{\circ}$. A total of 42 position bins were used in this analysis. These bins contained a total of 14,620 steps for an exposure time of approximately 1.7Ms (assuming $\sim 120$s of exposure per step). Each sky bin typically was favorably viewed by 1-4 detectors (typically one detector), resulting in a total of 10 different NaI detectors. The sky bins and detectors resulted in 69 spectra, each with 14 energy bins from 20-1000.0 keV, that were simultaneously fitted with a broken power-law model in XSPEC. Below 19.4 keV, the data contained considerable scatter that reduced the quality of the fit.  Due to the numerous spectra simultaneously fit, an average spectrum and the fit model are shown in Figure~\ref{fig:crabspec}. A broken power-law with a lower index of $2.057 \pm 0.009$, a break energy of $98 \pm 9$ keV, and an upper index of $2.36 \pm 0.05$ fit the data with $\chi^2$ = 1280.44 with 962 degrees of freedom ($\chi^2_{\nu} = 1.33$). 
This result is consistent with \citet{Jourdain_2009} who fit the \integralsc/SPI Crab data to a broken power-law with an index of $2.07 \pm 0.01$ below 100 keV and and 2.23 $\pm$0.05 above 100 keV.

\subsection{3-year catalog}

\begin{figure}[!h]
\center{\includegraphics[width=3.75cm,height=2.68cm]{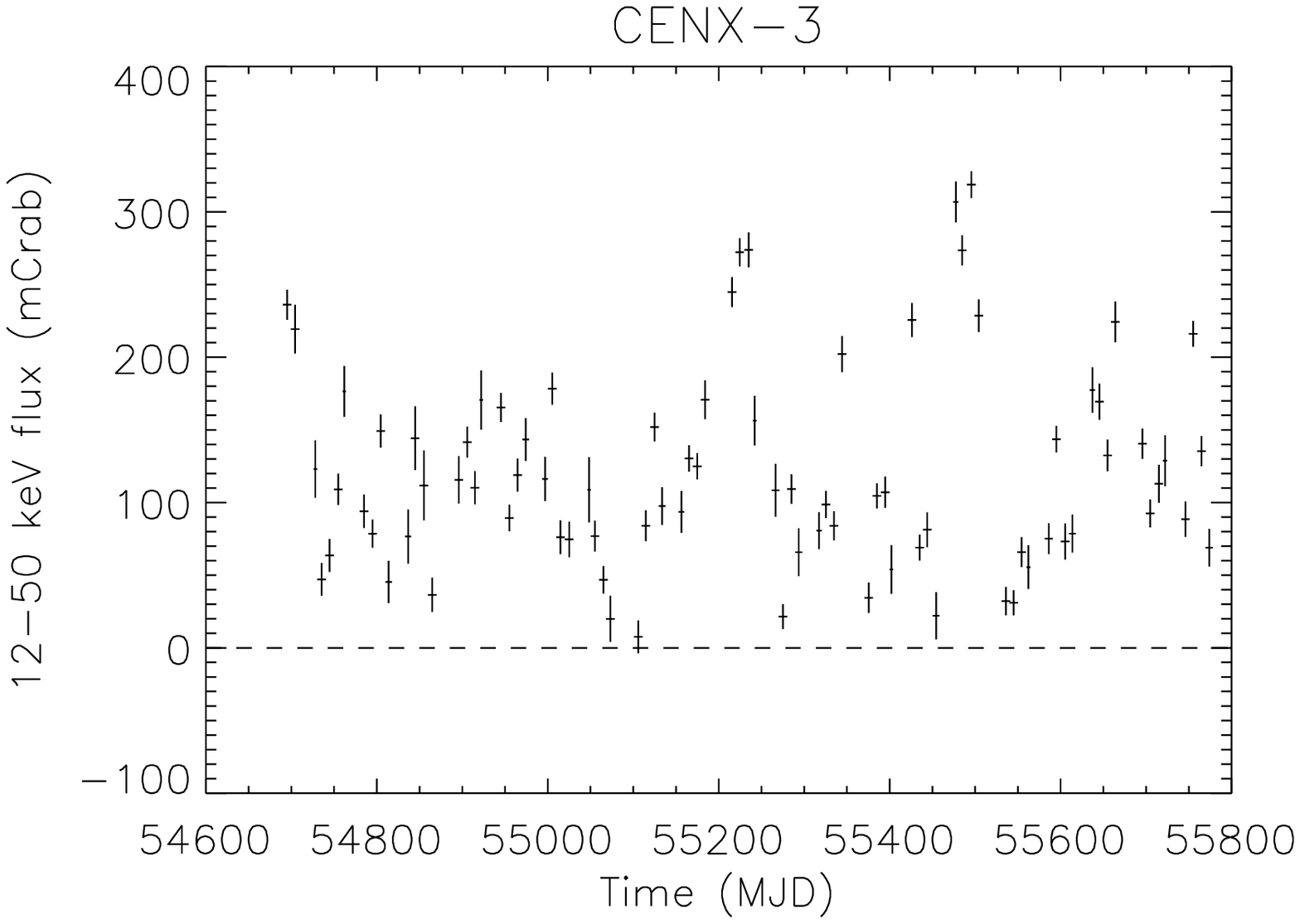}
\includegraphics[width=3.75cm,height=2.68cm]{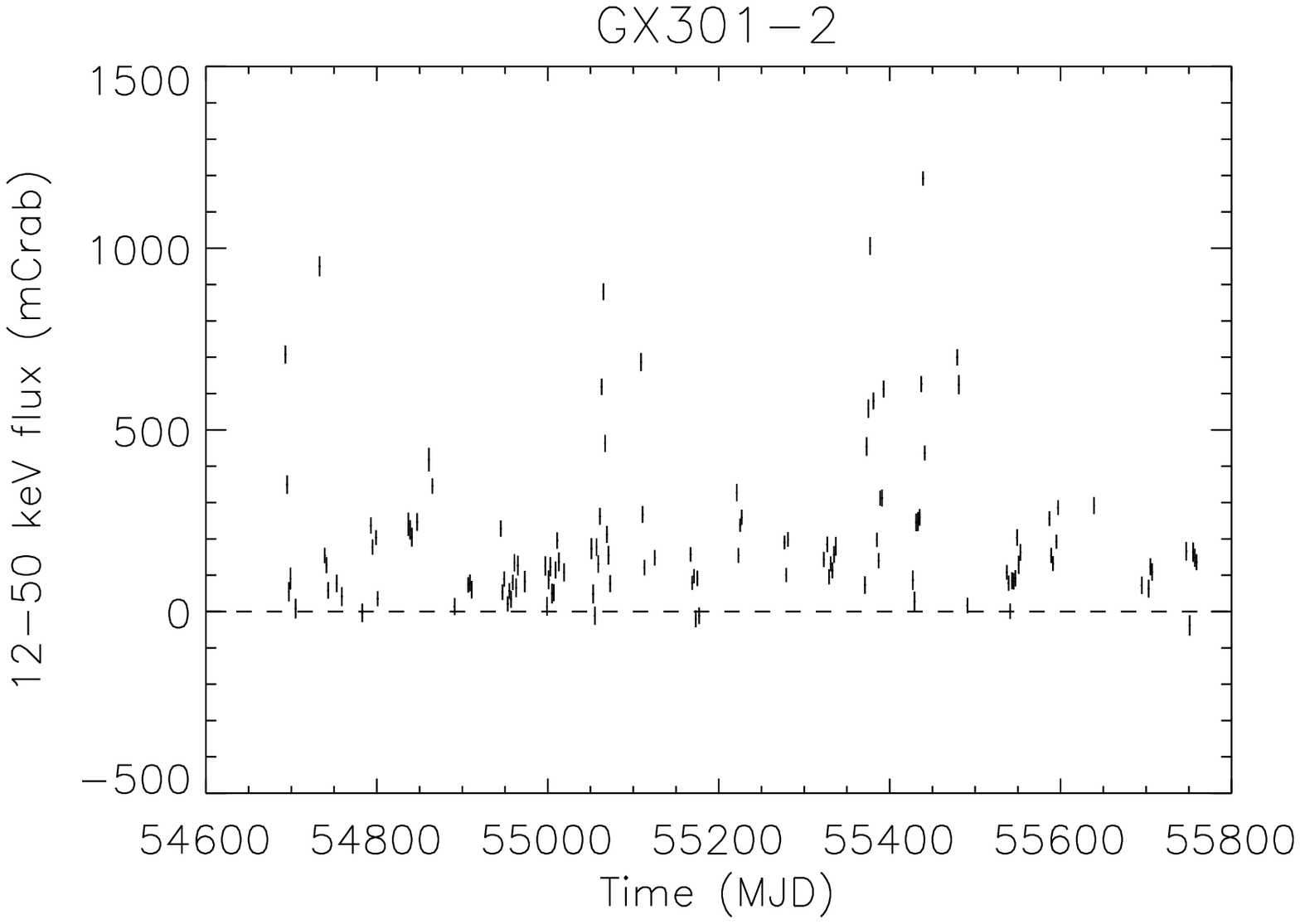}}
\vspace{-0.1in}
\center{\includegraphics[width=3.75cm,height=2.68cm]{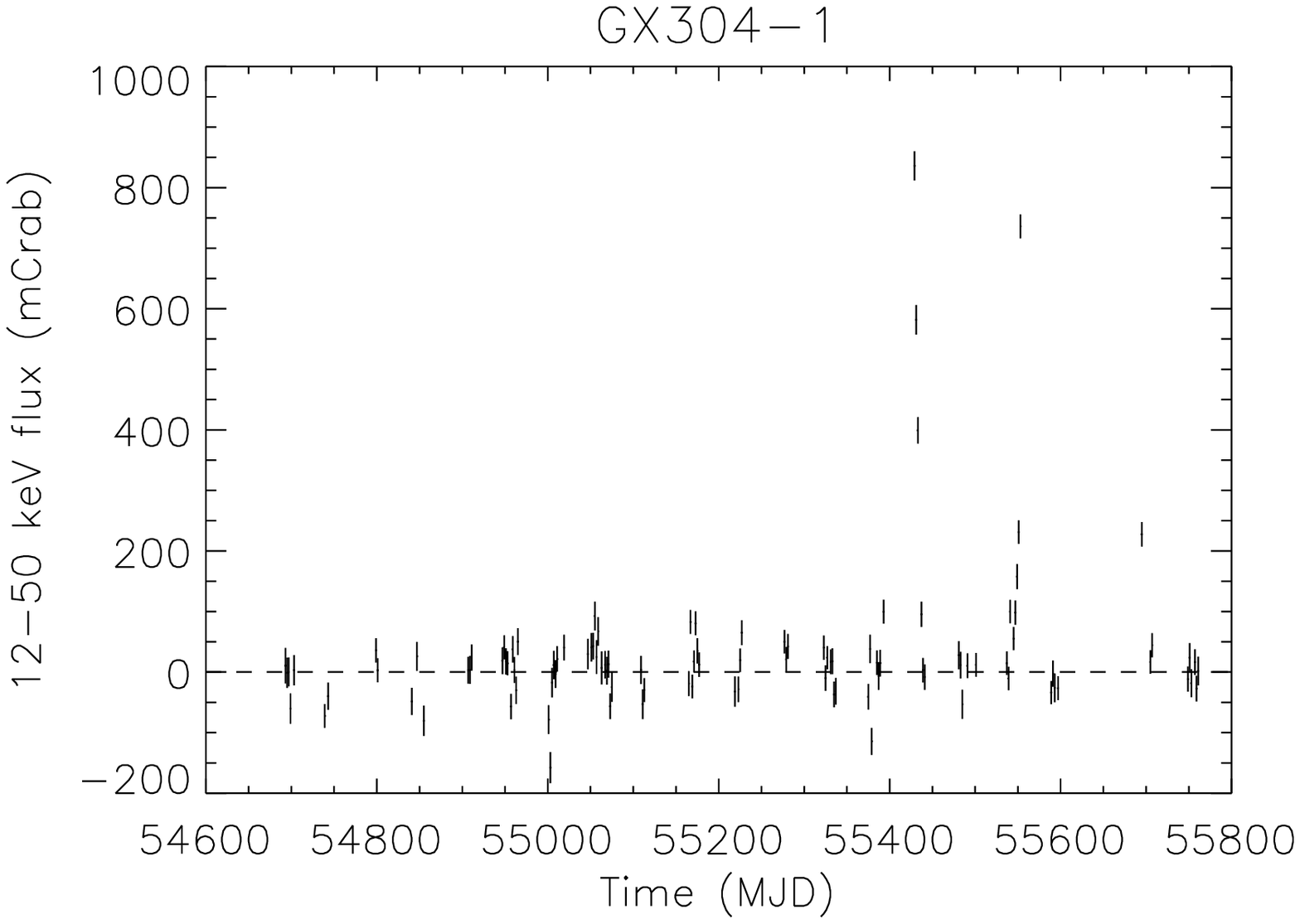}
\includegraphics[width=3.75cm,height=2.68cm]{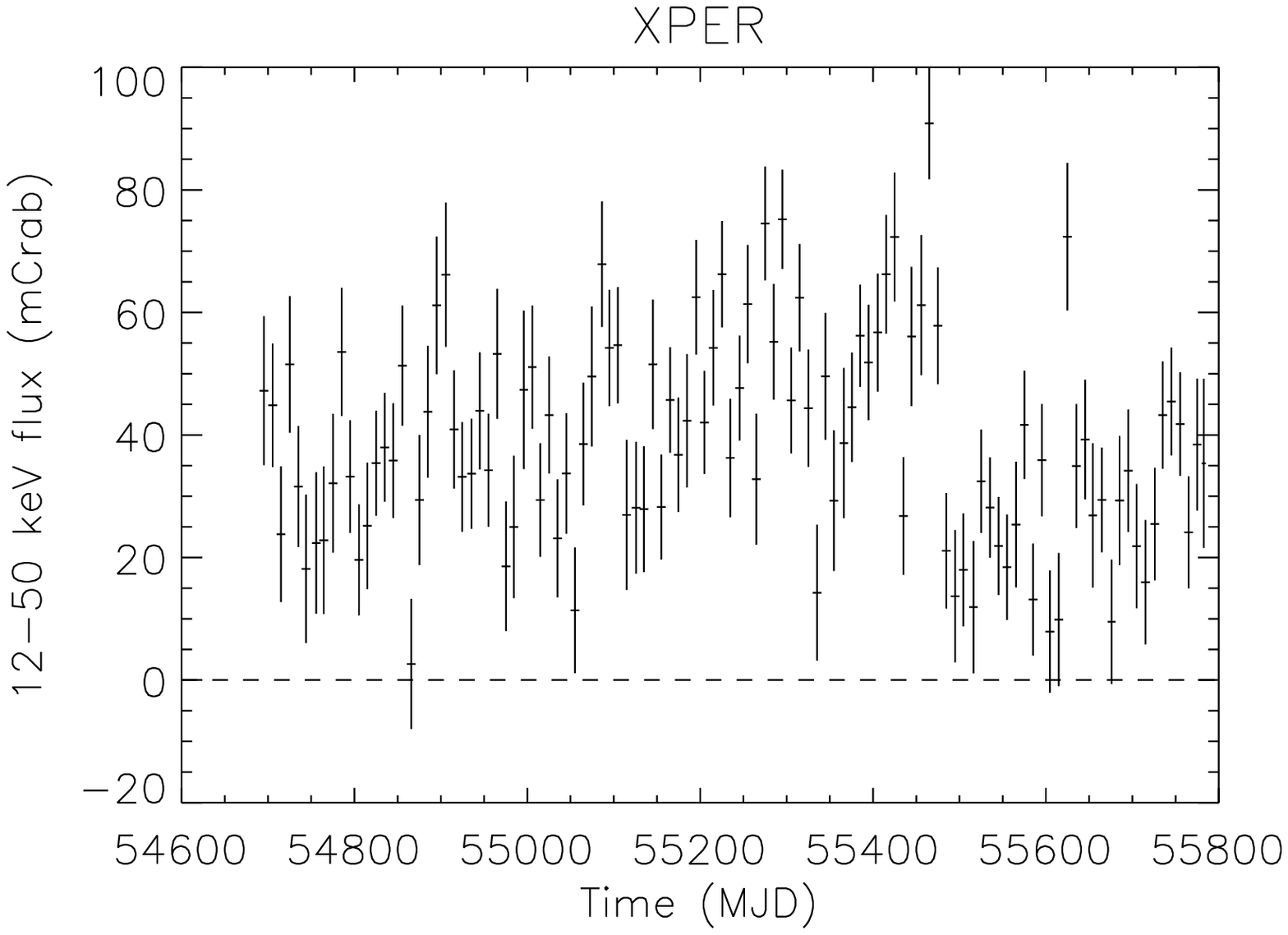}}
\vspace{-0.1in}
\center{\includegraphics[width=3.75cm,height=2.68cm]{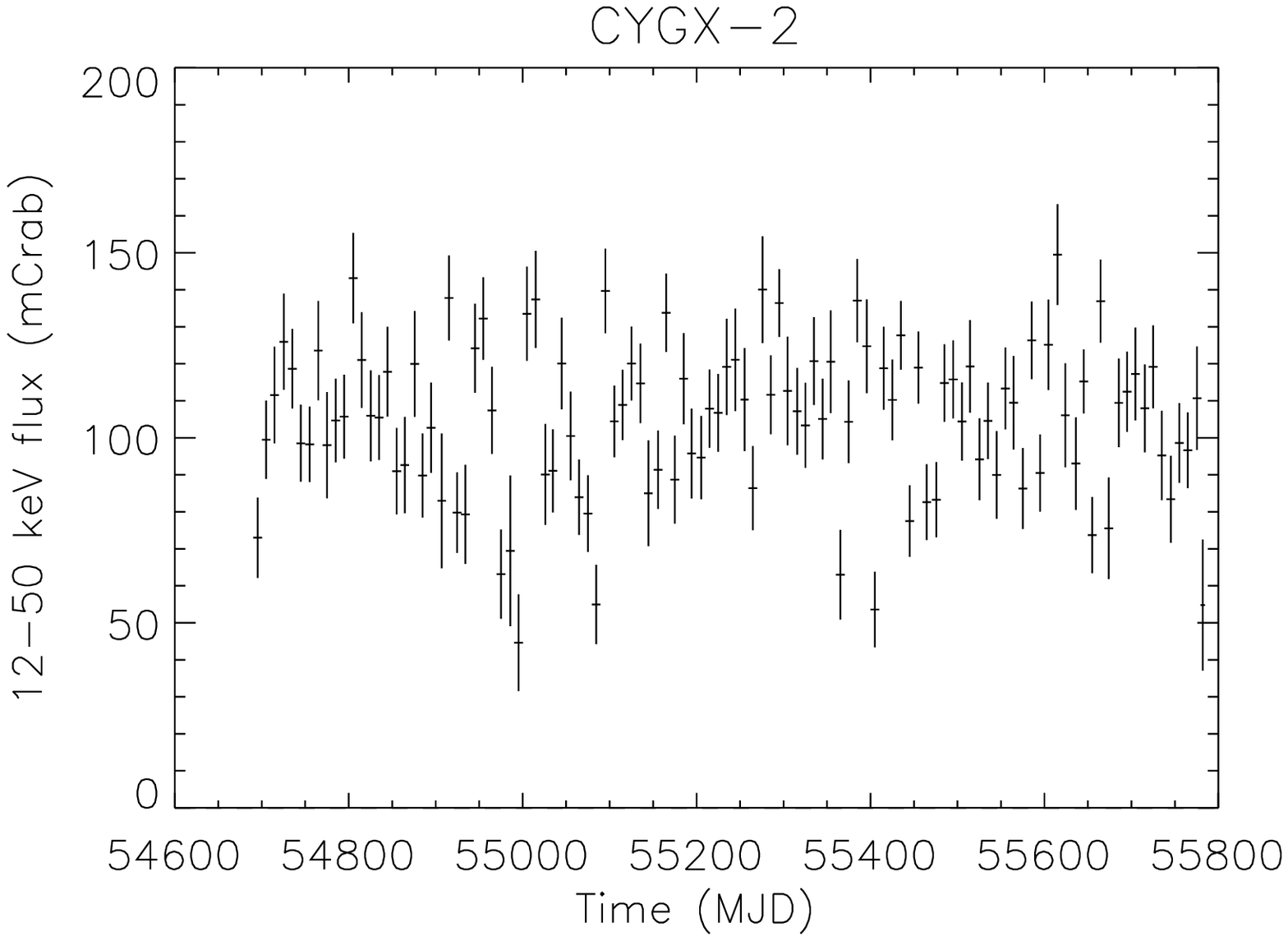}
\includegraphics[width=3.75cm,height=2.68cm]{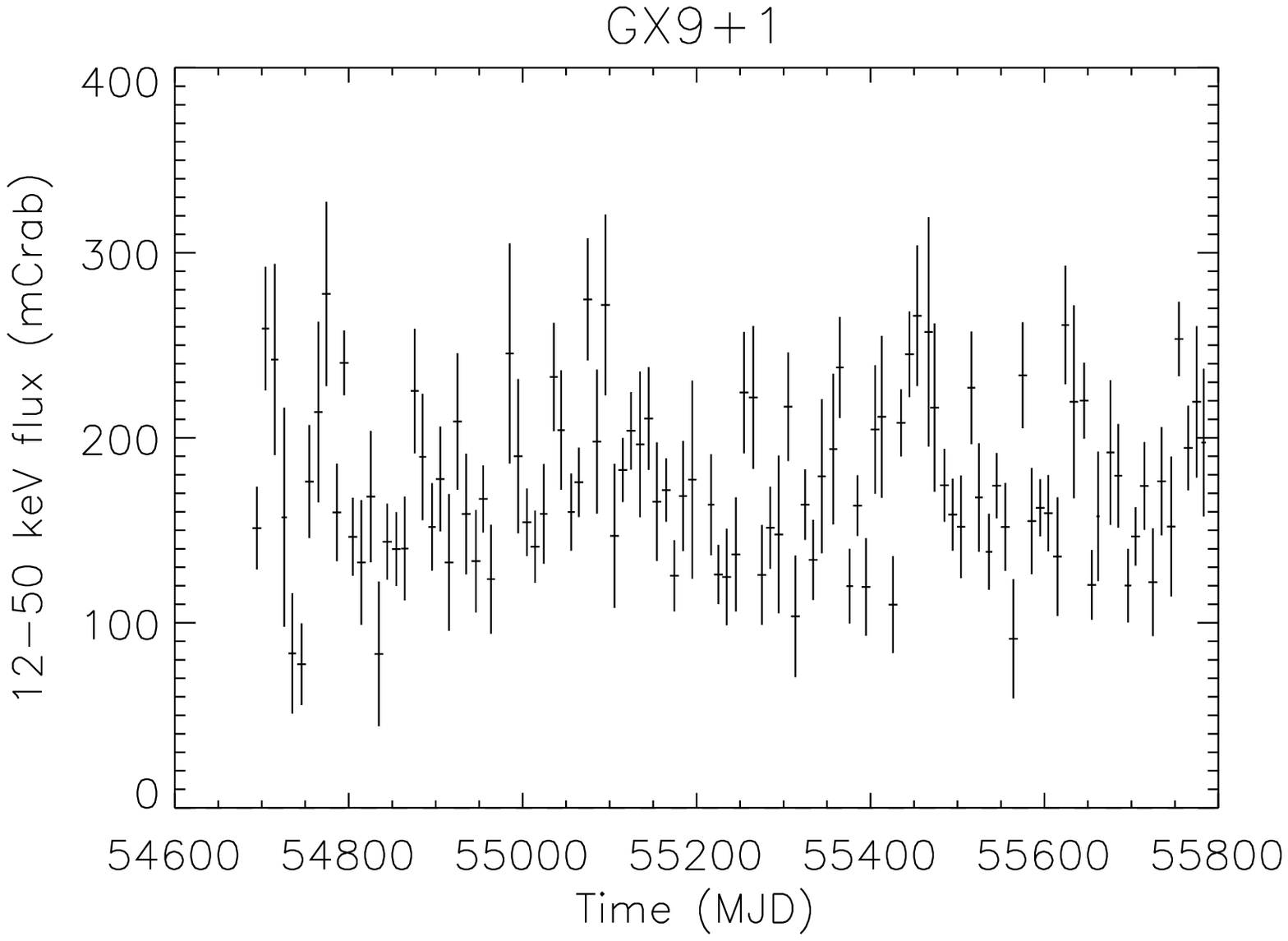}}
\vspace{-0.1in}
\center{\includegraphics[width=3.75cm,height=2.68cm]{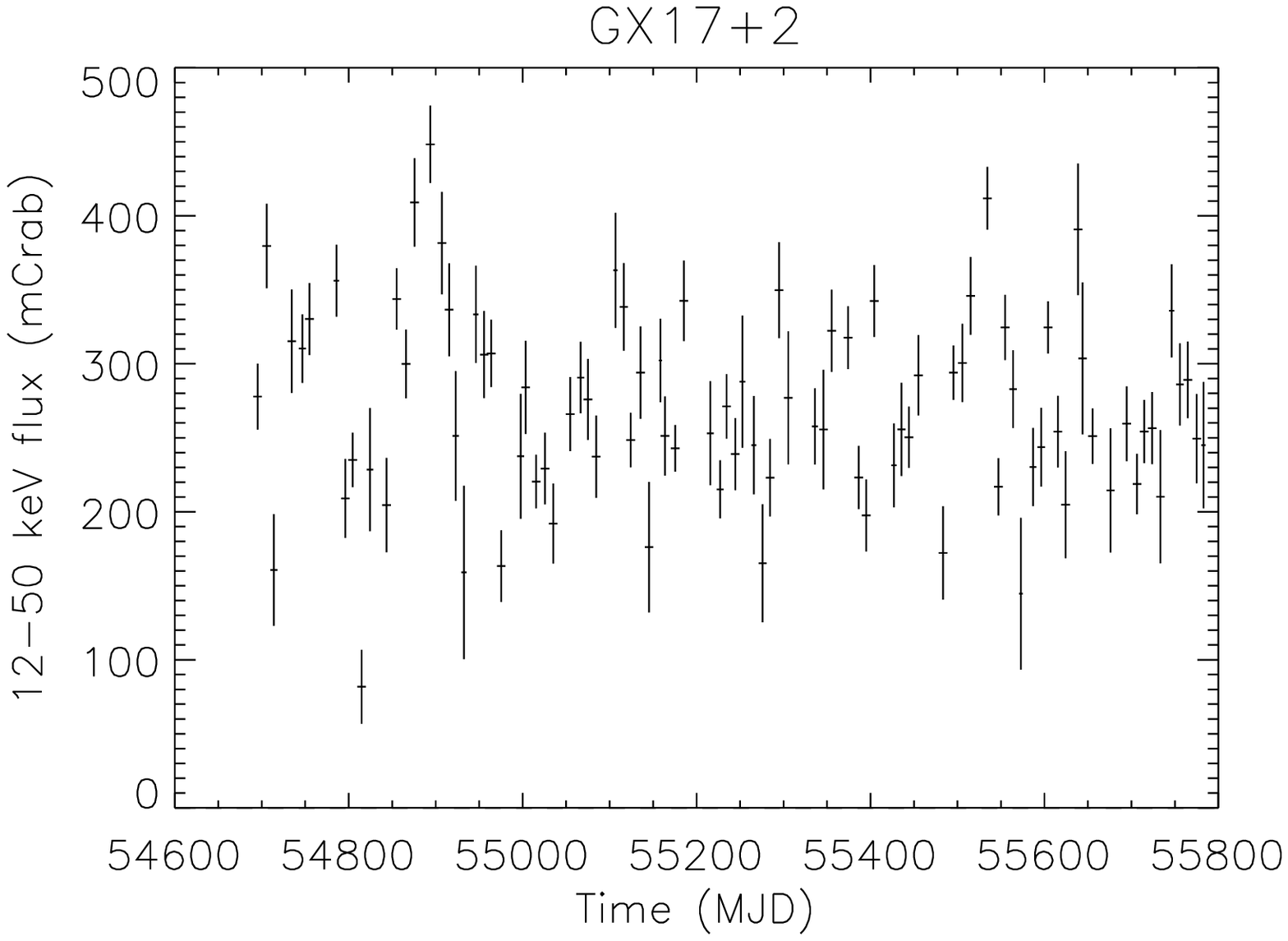}
\includegraphics[width=3.75cm,height=2.68cm]{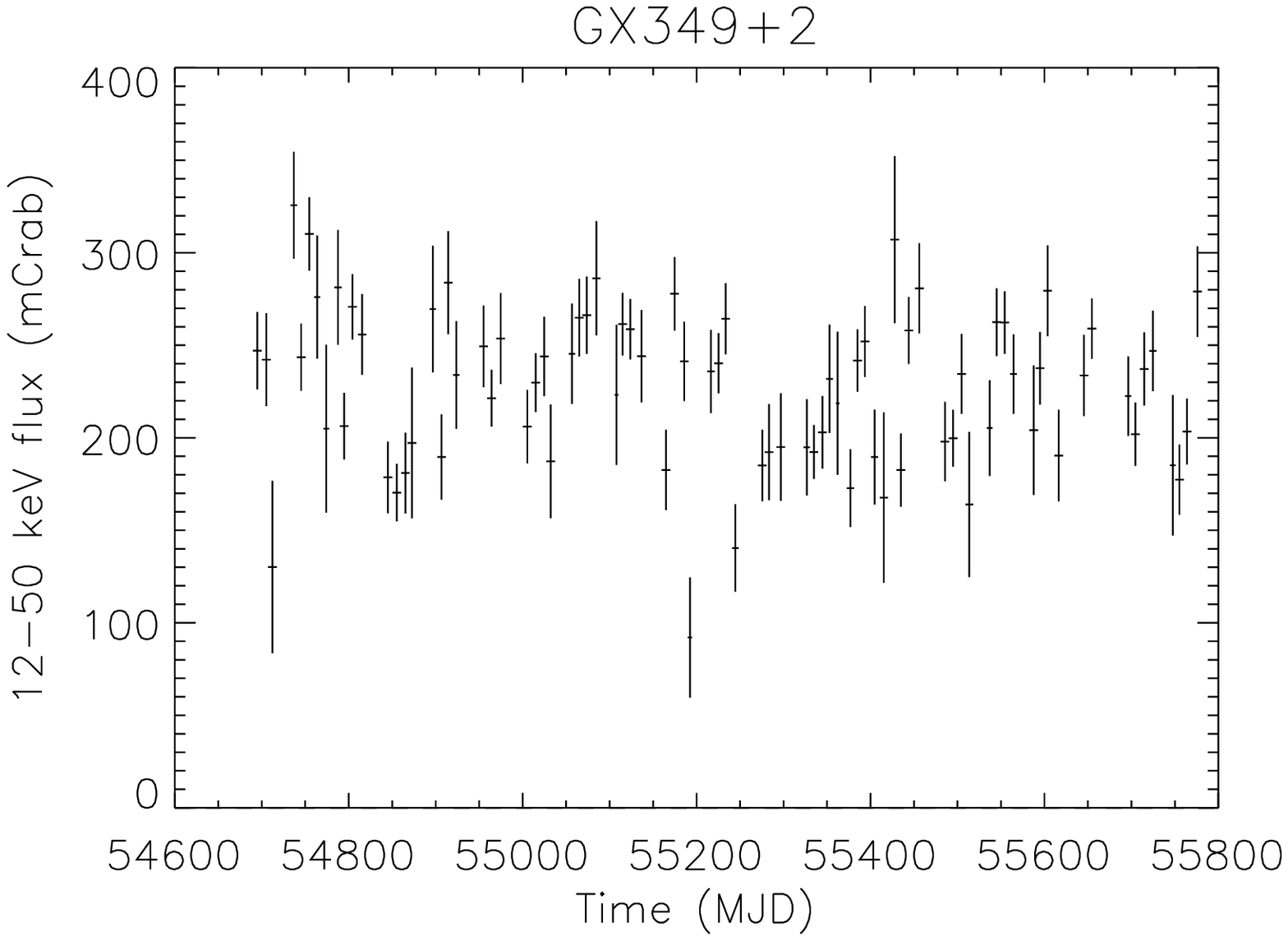}}
\vspace{-0.1in}
\caption{Sample 12-50 keV GBM light curves for four HMXB/NS systems containing accreting pulsars and four LMXB/NS systems, including two X-ray bursters, Cygnus X-2 and GX 17+2. Points plotted are 2-day average fluxes for GX 301-2 and GX 304-1 and are 10-day averages for the other six objects.}
\label{fig:hmxb_lmxb}
\end{figure}

Tables \ref{tbl:main}, \ref{tbl:marginal}, and \ref{tbl:uplim} contain the three year (August 12, 2008 - August 11, 2011) GBM Occultation catalog results for 209 sources. Table~\ref{tbl:main} lists detected sources (as defined below), Table~\ref{tbl:marginal} lists marginal detections ($3-5\sigma$), and Table~\ref{tbl:uplim} lists $3\sigma$ upper-limits for non-detected sources in the catalog. The average flux in mCrabs for each source over three years is listed for 12-25 keV, 25-50 keV, 50-100 keV and 100-300 keV energy bands as well as the significance in the 12-50 keV and 12-300 keV energy bands. Downloadable up to date postscript plots and plain text files of the 4-band light curves for all sources are available from our website \url{http://heastro.phys.lsu.edu/gbm/}. All errors include statistical and the estimated systematic errors given in Table~\ref{tab:sys}.  If the significance for any source in any of the above energy bands equals or exceeds 5 sigma or the source is detected in the transient search (section~\ref{sec:transient}) or with the orbit period folding technique (section~\ref{sec:period}) then the information on that source is listed in Table~\ref{tbl:main} and the source is considered a detection. In this catalog,  99 sources are considered detections. Detection categories, column four in Table~\ref{tbl:main}-\ref{tbl:uplim}, are defined as: A = $> 5\sigma$ in any of the bands 12-25, 25-50, 50-100, 100-300, 12-50, or 12-300 keV, B=3-5 $\sigma$ in those bands, N = non-detection = $> -3 \sigma$ and $<+3 \sigma$ in those bands, I = indeterminate $<-3 \sigma$ in any of the bands, which is indicative of source interference problems, T = transient detected in the transient search (section~\ref{sec:transient}), P = detected in the orbit period folding (section~\ref{sec:period}).  The most numerous source classification detected in categories A and/or T are LMXB/NS (40 sources) followed by HMXB/NS (31 sources), including one SFXT. Sample HMXB/NS and LMXB/NS light curves are shown in Figure~\ref{fig:hmxb_lmxb}. The 12 black hole candidates (BHC) are the next most numerous binary detected with 7 BHCs being detected above 100 keV and with Cyg X-1 detected up to and above 300 keV.  We also have 12 AGNs detected, most of which are Seyfert 2 galaxies (NGC 1275, NGC 2110, NGC 4388, Cen A, NGC 5252, Circinus Galaxy, NGC 5506) with two Seyfert 1 galaxies (NGC 4151, IC 4329A), a radio galaxy (IGR J21247+5058), and two quasars (3C 273 and 3C 454.3), all of which are also seen with the Fermi/LAT. Furthermore, Cen A is detected up to and above 100 keV.  The Crab pulsar together with its wind nebula is our second most significant source behind the LMXB/NS Sco X-1. The Ophiuchus Cluster is also detected as category A but only in the 12-25 keV band. The Sun (also discussed in Section~\ref{sec:fitting}) is also detected in the 3-year average to 96 sigma in the 8-12 keV band and numerous flares are detected to $>$10 sigma up to $\sim 50$ keV using the Earth Occultation Technique. GBM detects solar flares in triggered observations up to $\sim 8$ MeV with detections to higher energies in \fermi/LAT \citep{Ackermann2011b}.

Category B (3-5 $\sigma$) contains  23 sources, listed in Table~\ref{tbl:marginal}, including six AGN, three BL Lac objects (MRK421, H 1517+656, B2 1732+38A), two Seyfert 1 (NGC3783 and TXS 1700+685) and one Seyfert 2 (MCG-05-23-016), four LMXB/NS (4U1323--62,  GRS1724-308, SAXJ1806.5-2115, 4U2127+119), five HMXB/NS (3A0114+650, IGRJ16418--4532, SAX J1818.6-1703,XTEJ1855-026, 4U2206+54),two PSR/PWN (MSH 15-52, Vela-X), one CV (GK Per), the Vela pulsar, one BHC (LMX X-3), the Coma Cluster, a gamma ray binary (1FGLJ1018.6-5856), and a supernova remnant (Cas A). These sources may become detectable above $5\sigma$ with additional observation time.

\begin{figure}[!h]
\center{\includegraphics[width=3.75cm,height=2.68cm]{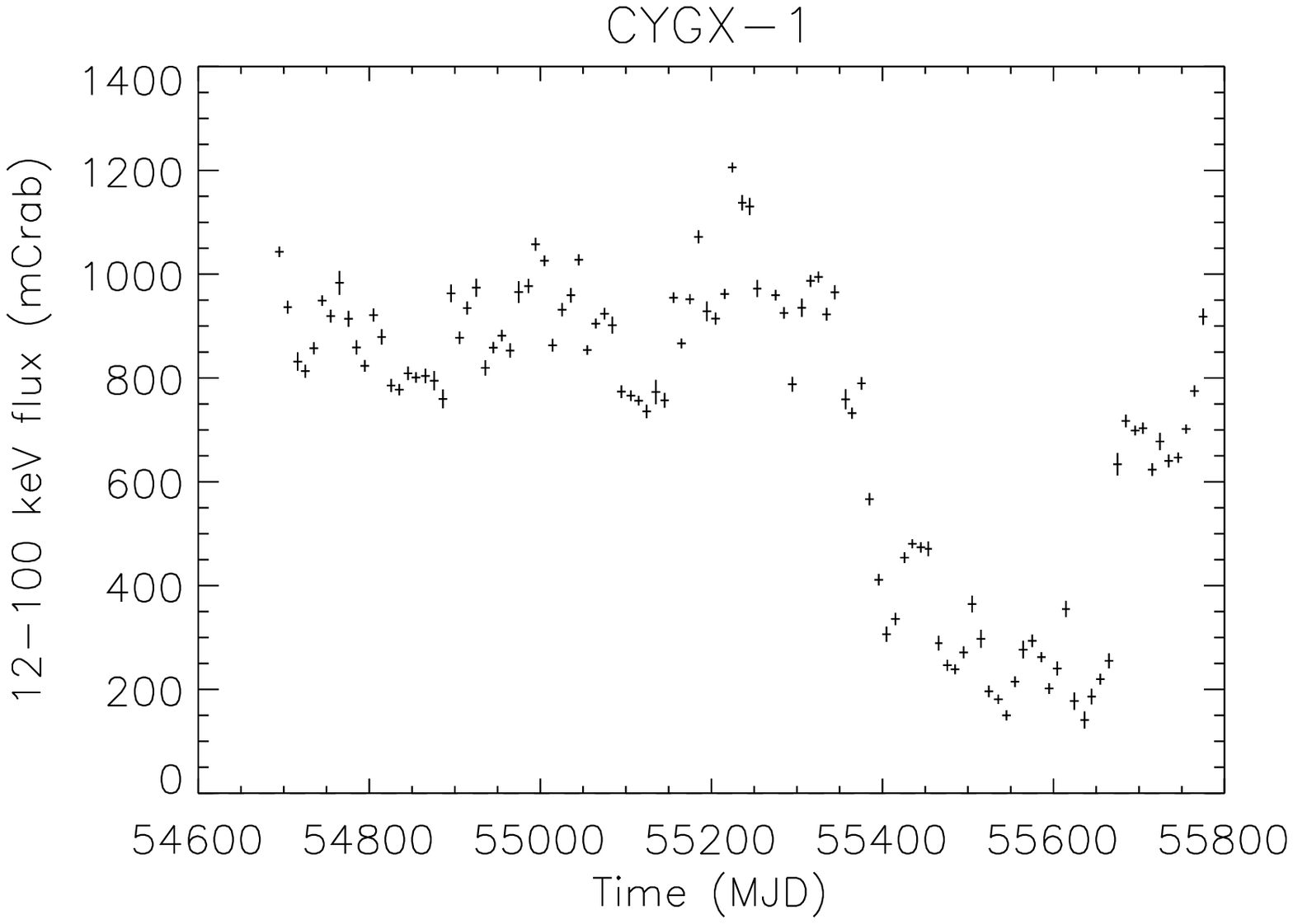}
\includegraphics[width=3.75cm,height=2.68cm]{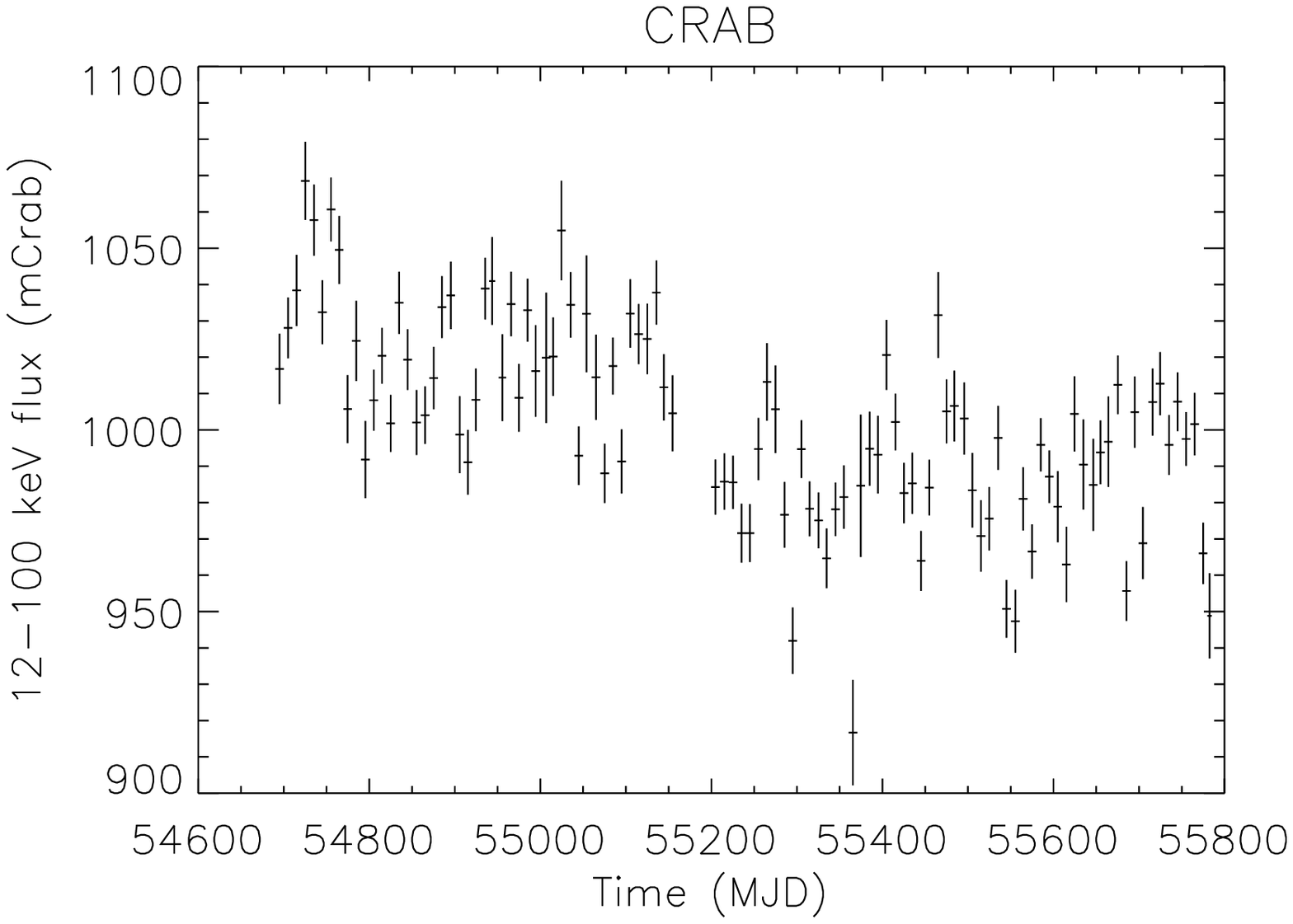}}
\vspace{-0.1in}
\center{\includegraphics[width=3.75cm,height=2.68cm]{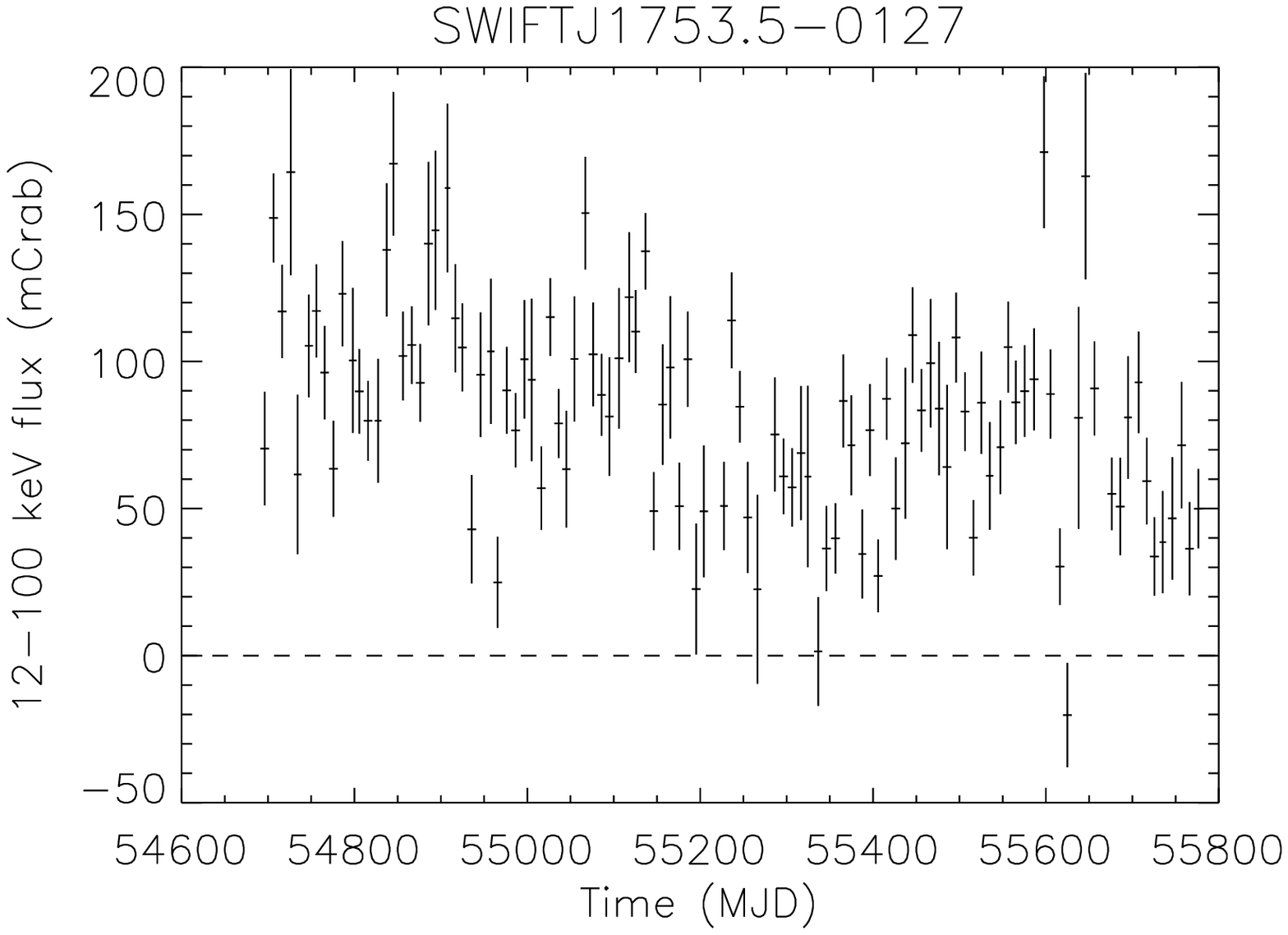}
\includegraphics[width=3.75cm,height=2.68cm]{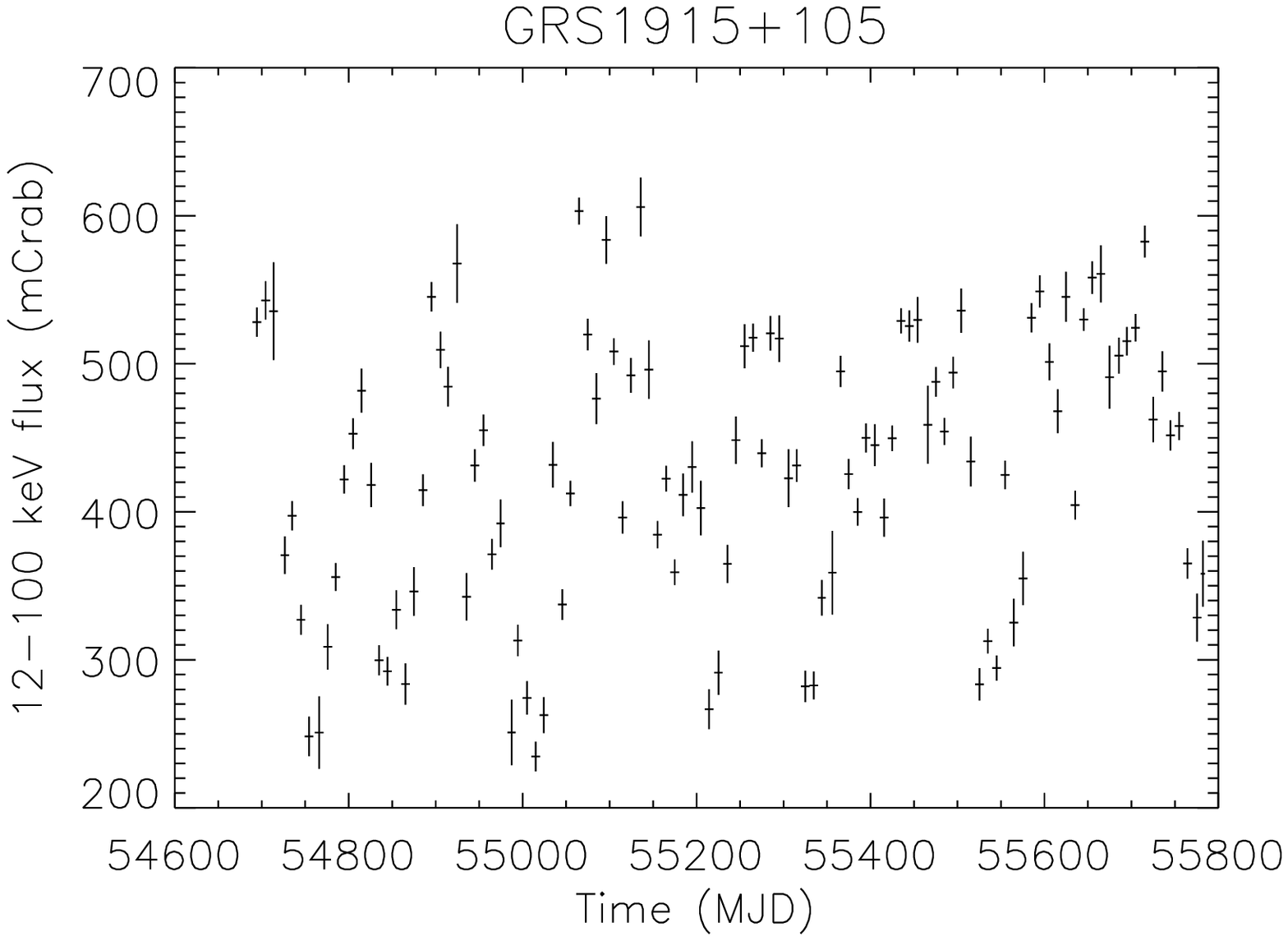}}
\vspace{-0.1in}
\center{\includegraphics[width=3.75cm,height=2.68cm]{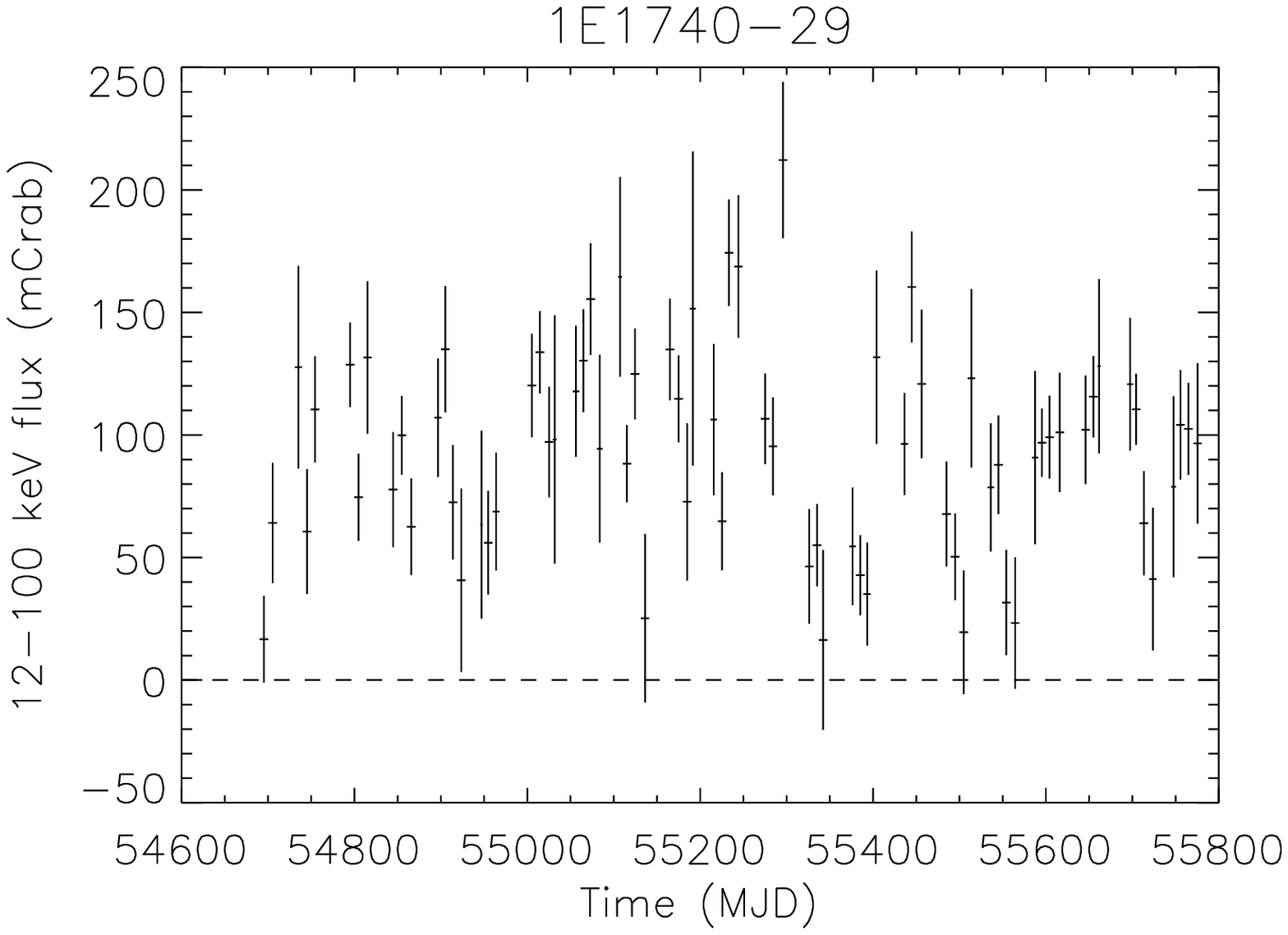}
\includegraphics[width=3.75cm,height=2.68cm]{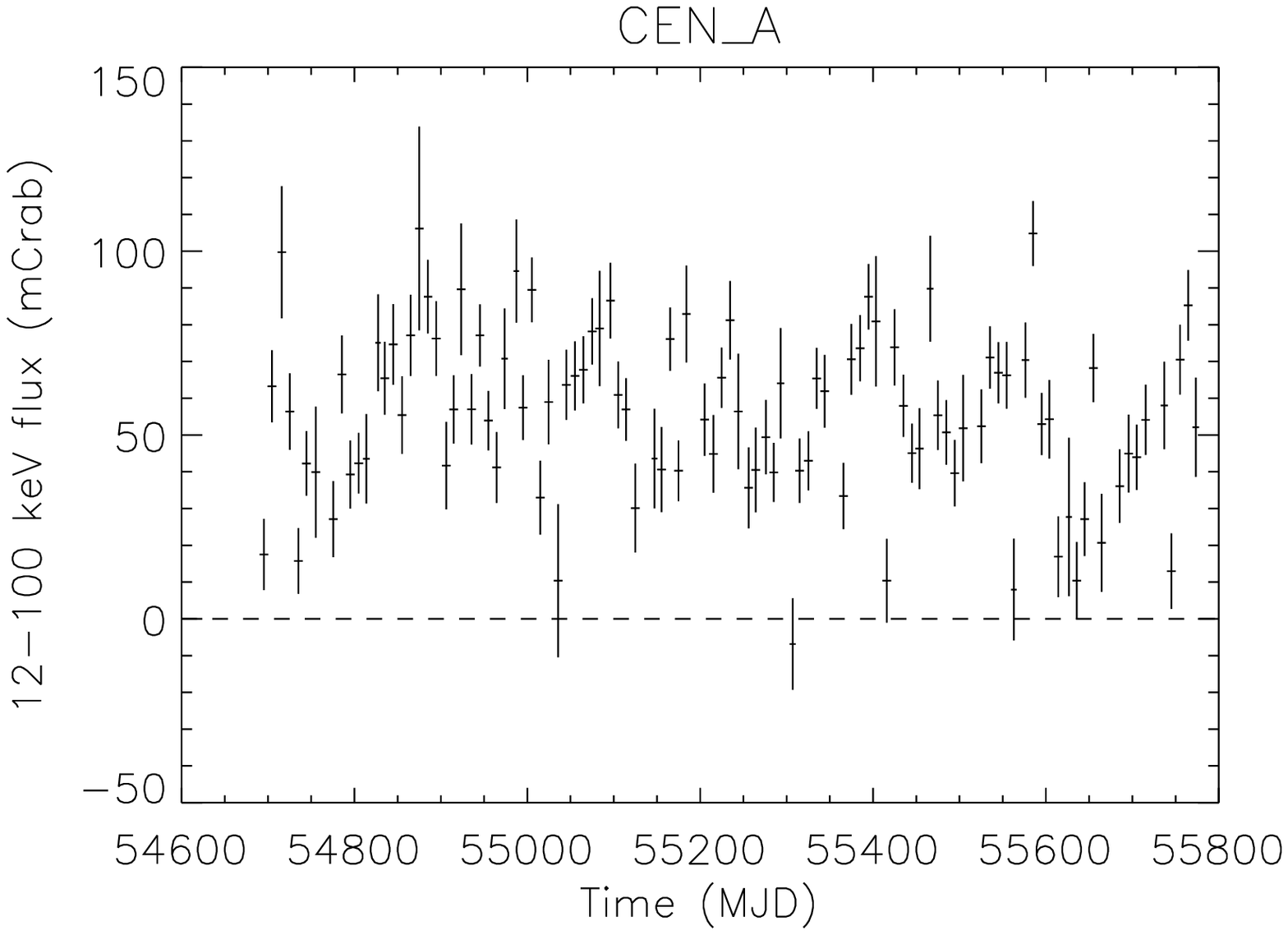}}
\vspace{-0.1in}
\center{\includegraphics[width=3.75cm,height=2.68cm]{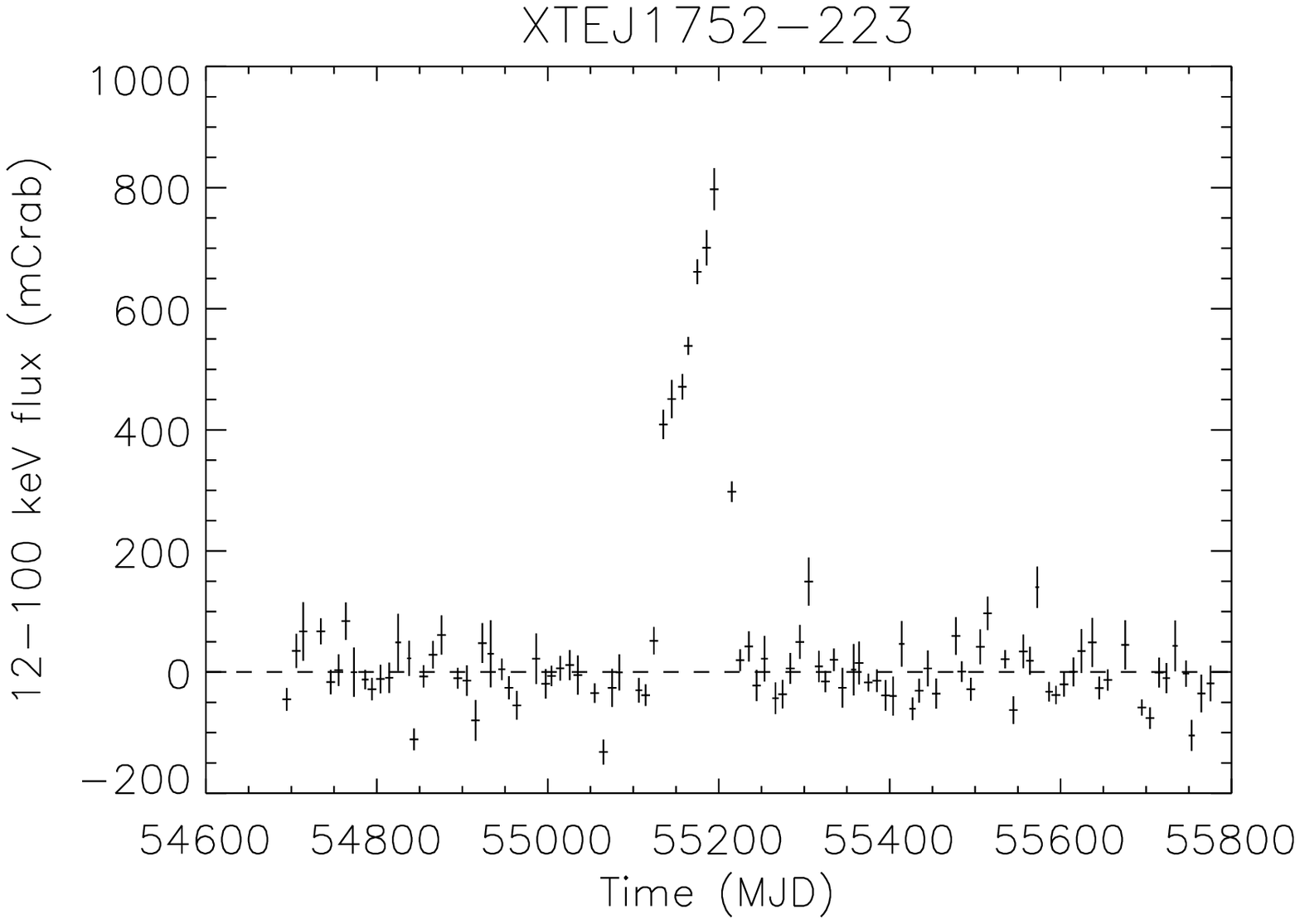}
\includegraphics[width=3.75cm,height=2.68cm]{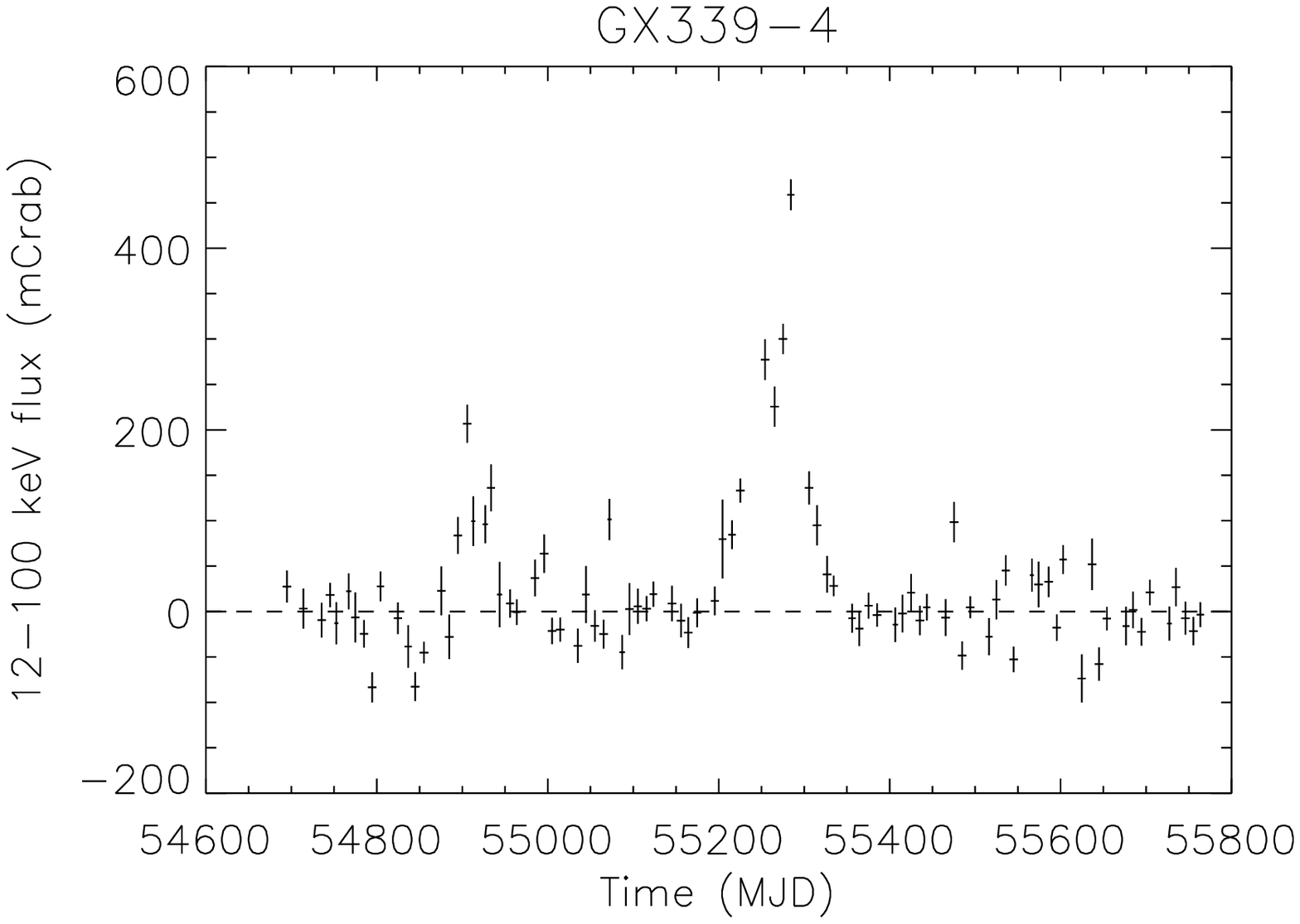}}
\vspace{-0.1in}
\caption{GBM lower energy (12-100 keV) light curves for the eight objects detected above 100 keV in this 3-year catalog. Each data point is a 10-day average flux. The dashed line denotes zero flux.}
\label{fig:12to100keV}
\end{figure}

\begin{figure}[!h]
\center{\includegraphics[width=3.75cm,height=2.68cm]{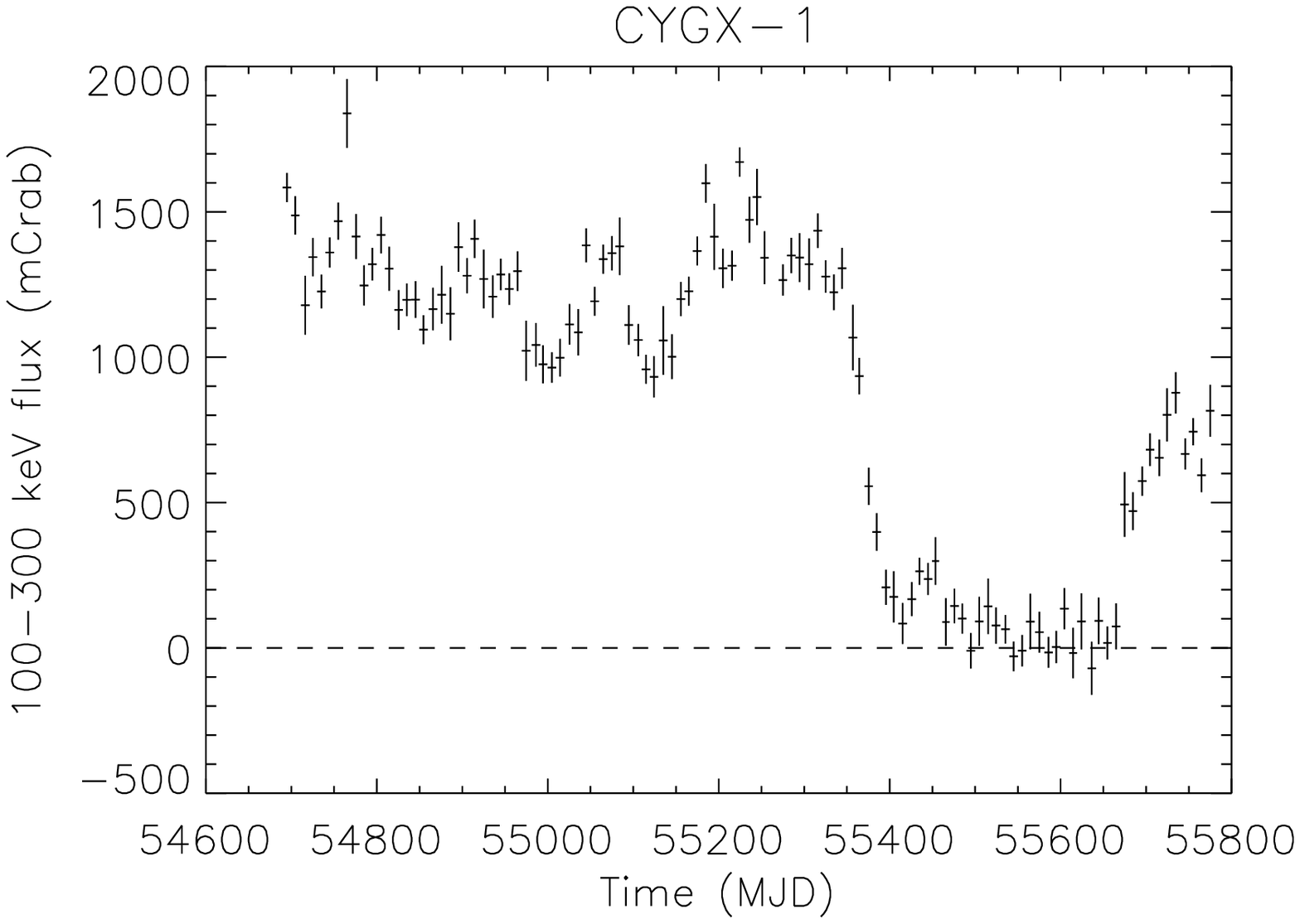}
\includegraphics[width=3.75cm,height=2.68cm]{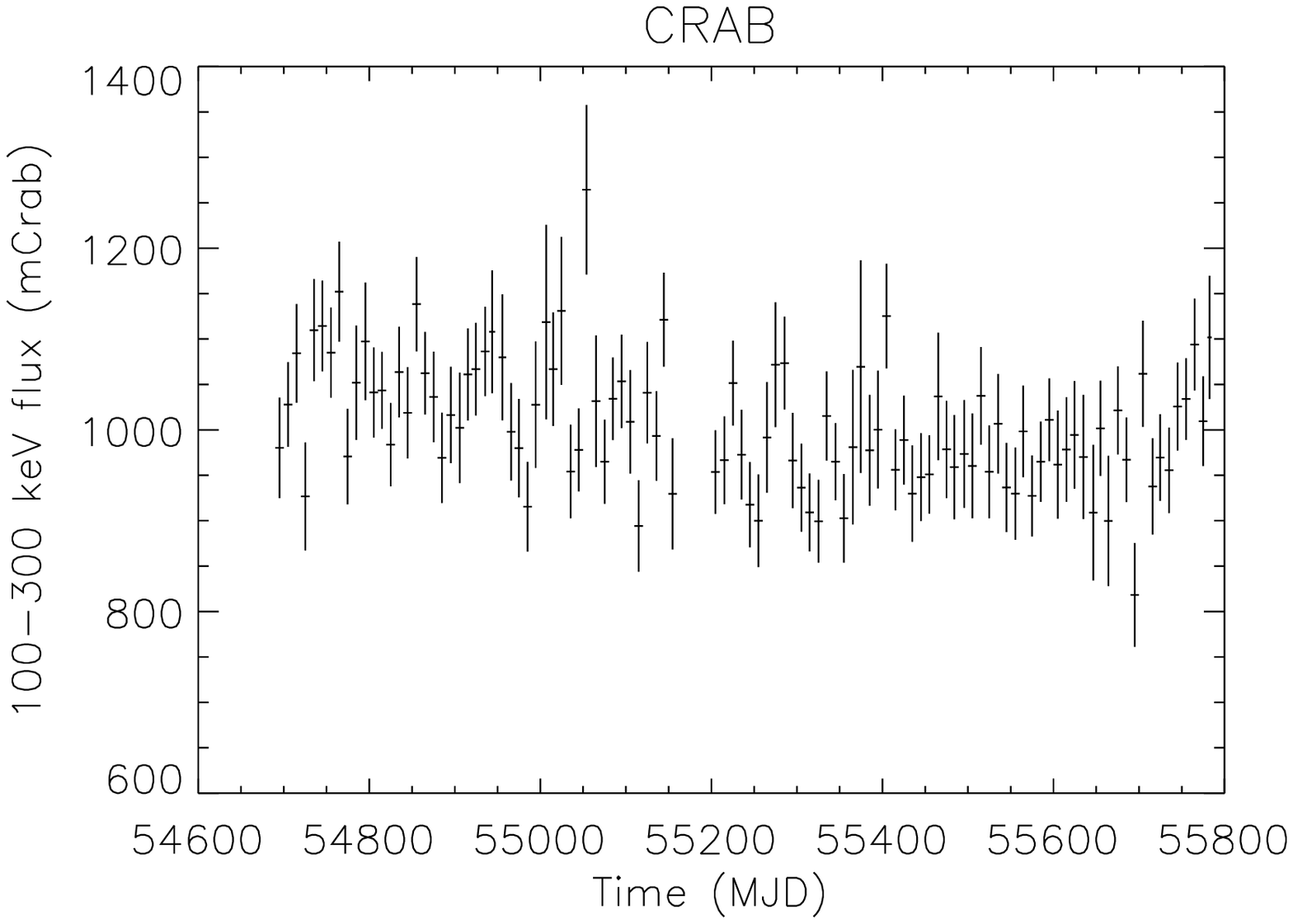}}
\vspace{-0.1in}
\center{\includegraphics[width=3.75cm,height=2.68cm]{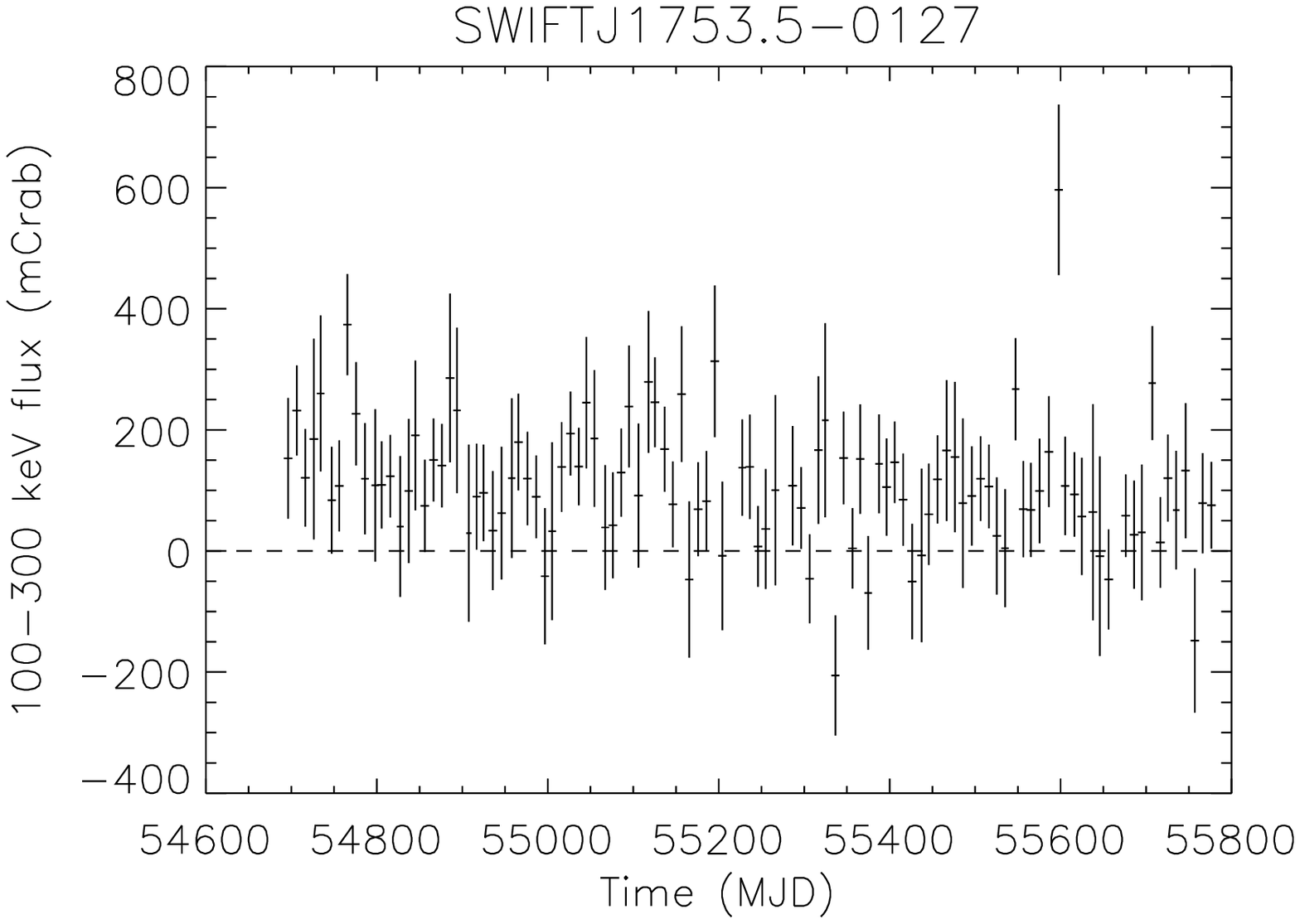}
\includegraphics[width=3.75cm,height=2.68cm]{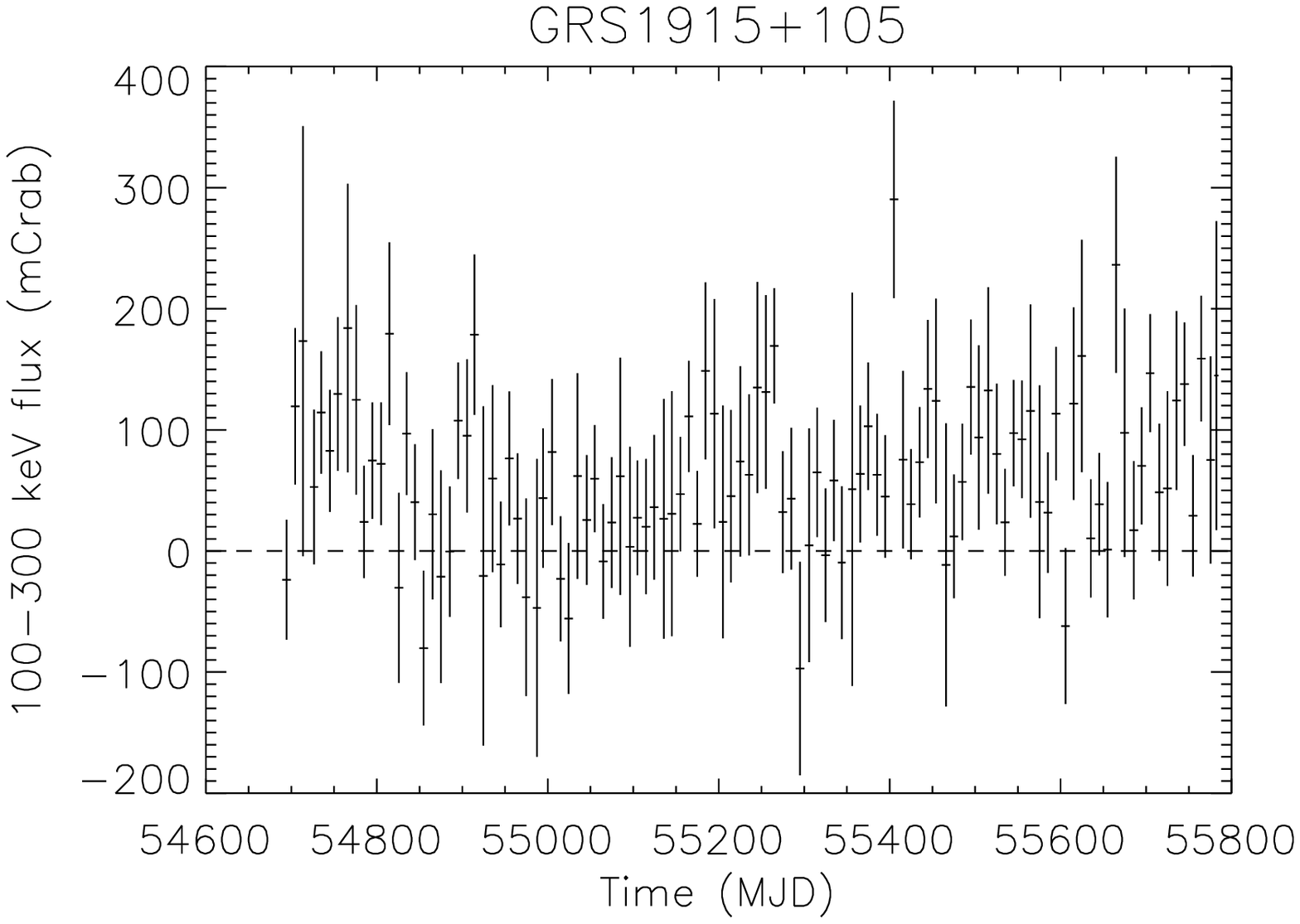}}
\vspace{-0.1in}
\center{\includegraphics[width=3.75cm,height=2.68cm]{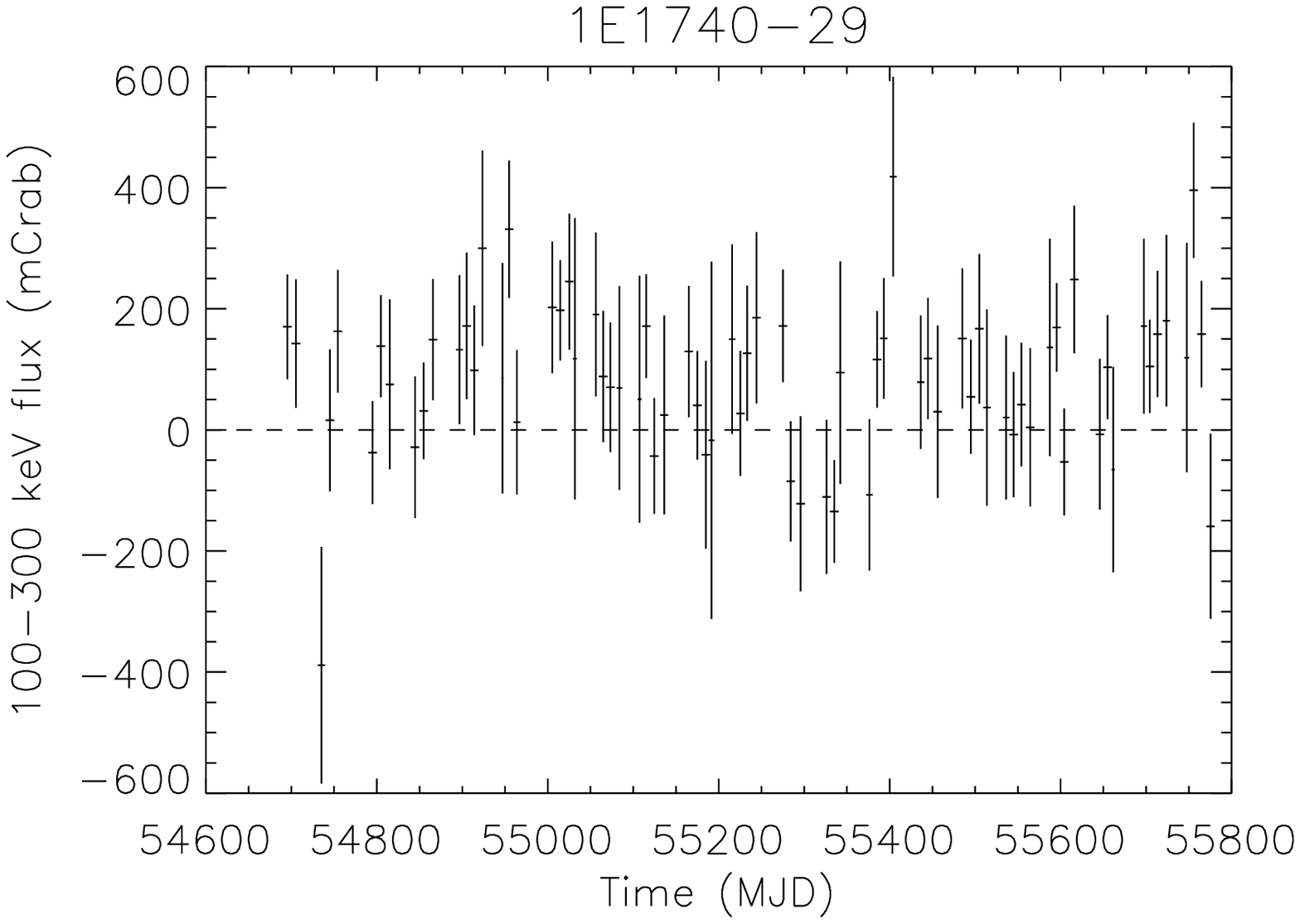}
\includegraphics[width=3.75cm,height=2.68cm]{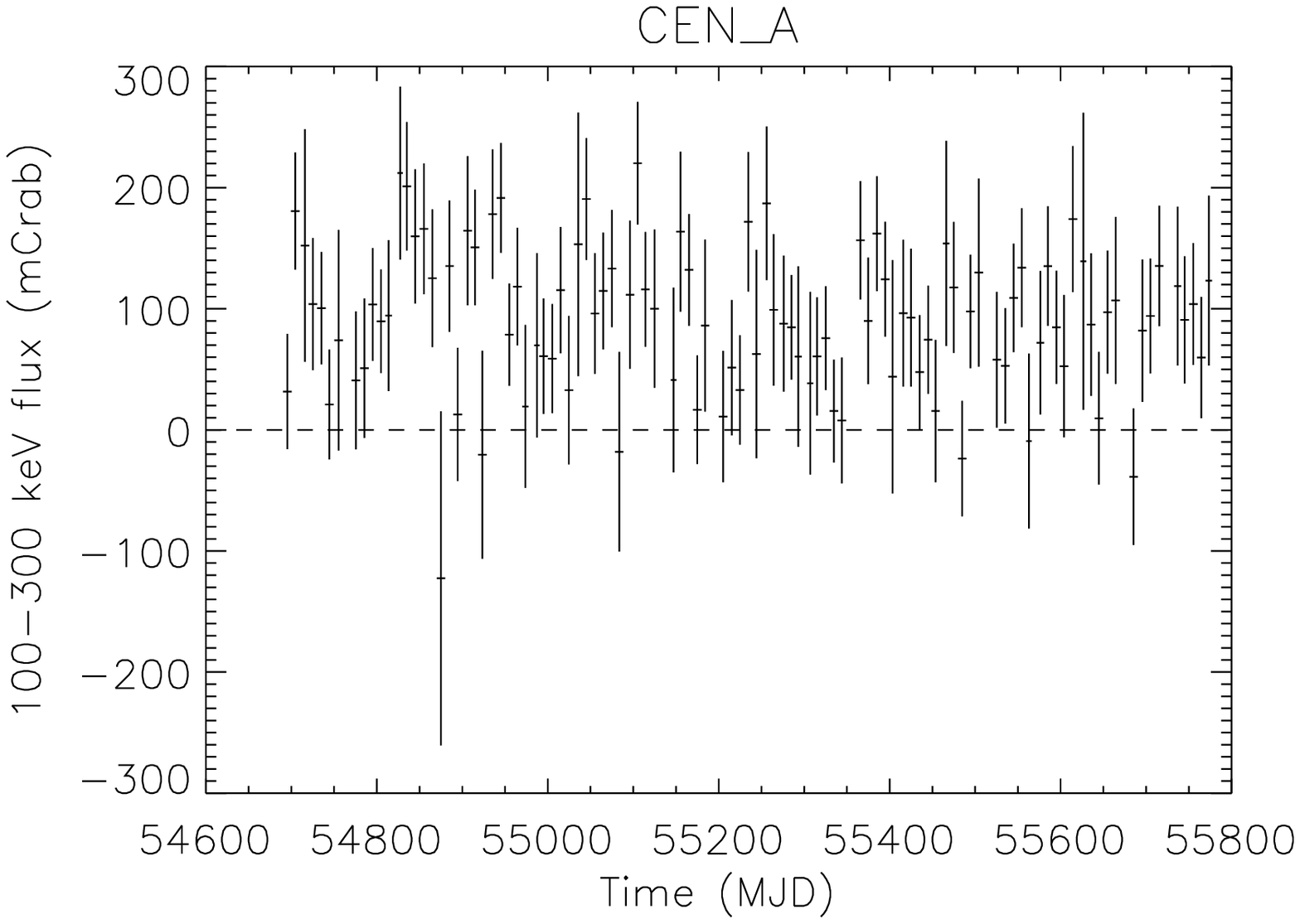}}
\vspace{-0.1in}
\center{\includegraphics[width=3.75cm,height=2.68cm]{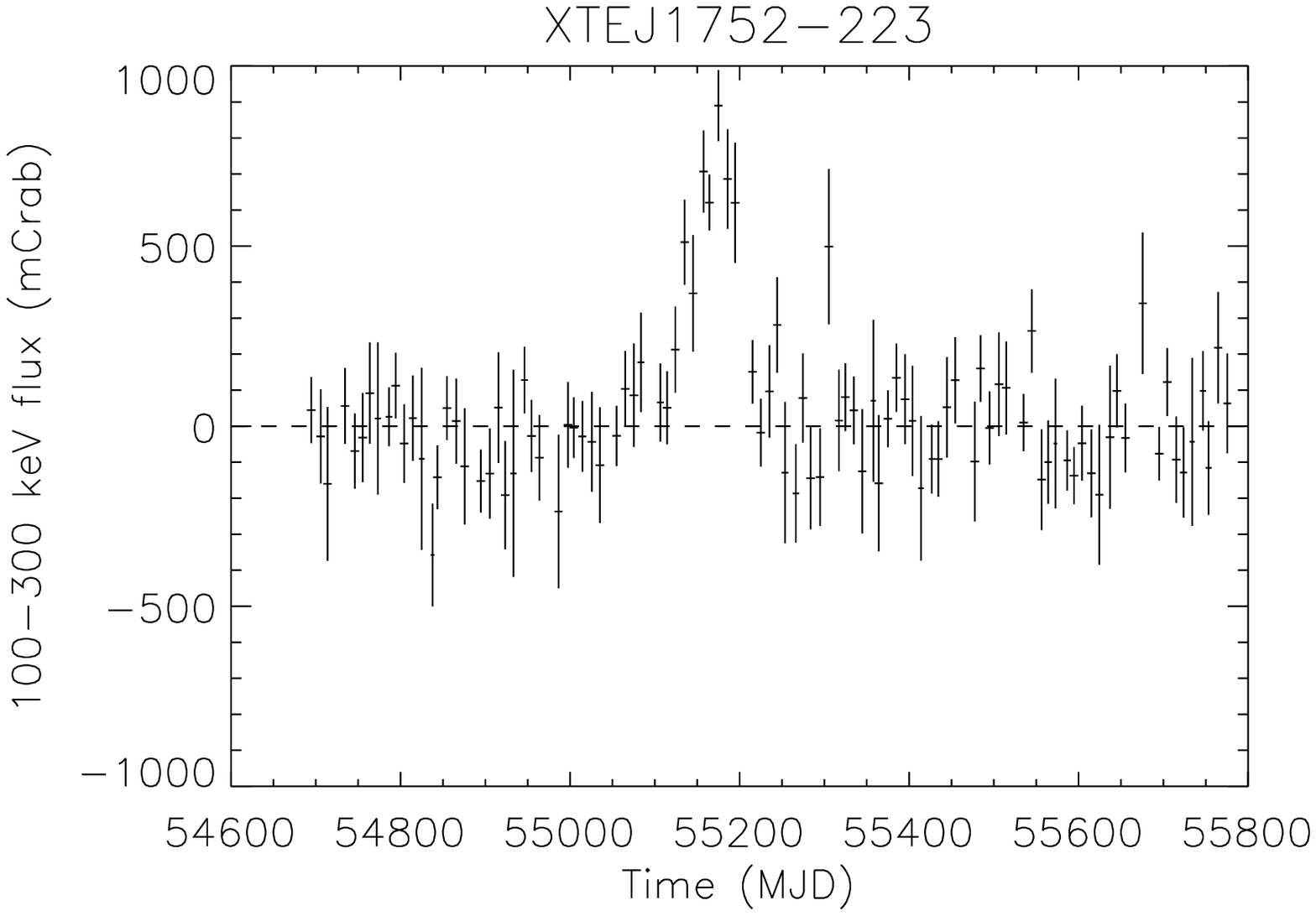}
\includegraphics[width=3.75cm,height=2.68cm]{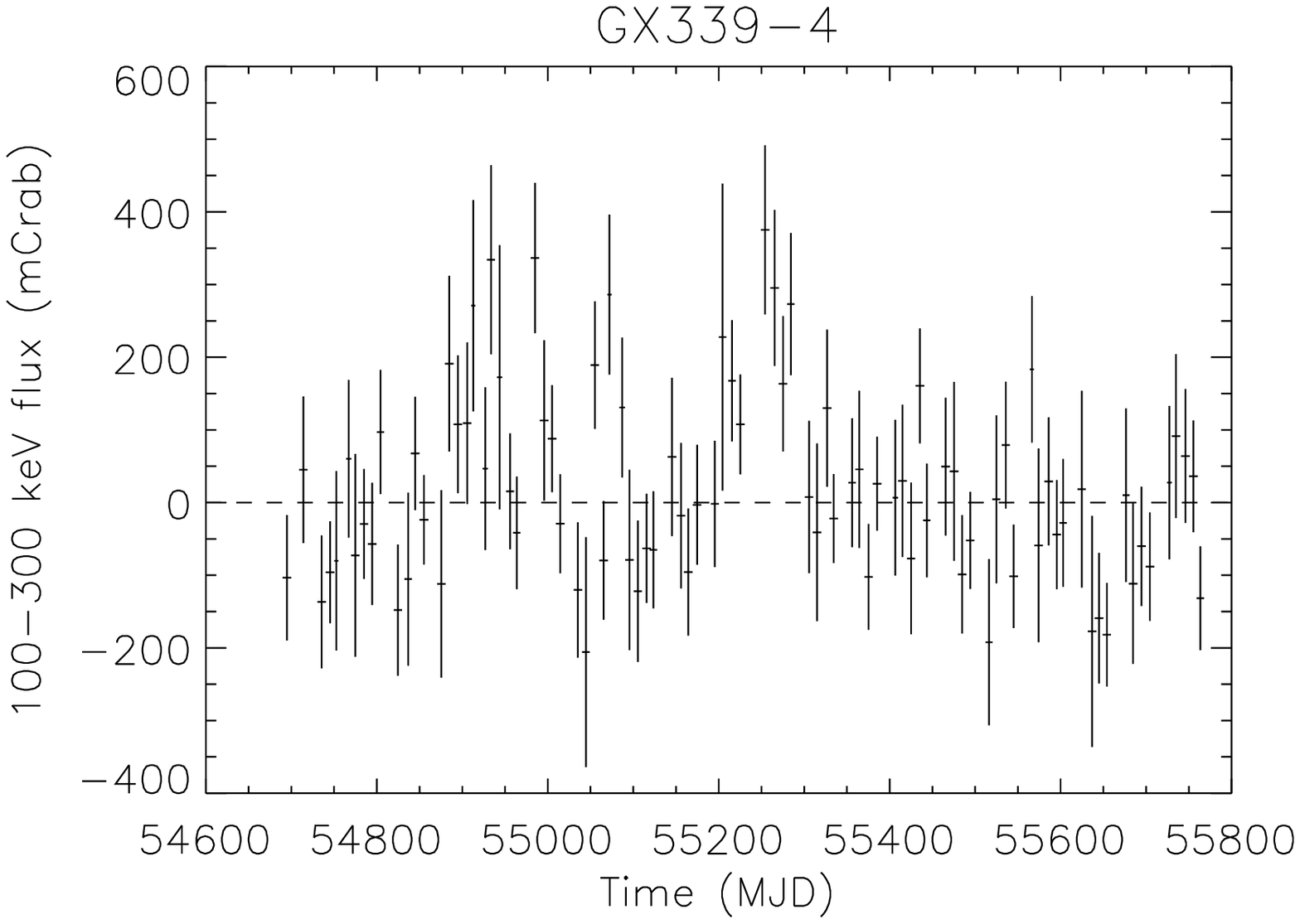}}
\vspace{-0.1in}
\caption{GBM higher energy (100-300 keV) light curves for the eight objects detected above 100 keV in this 3-year catalog. Each data point is a 10-day average flux. The dashed line denotes zero flux.}
\label{fig:100to300keV}
\end{figure}

The Earth occultation technique allows us to do important monitoring of the
hard X-ray/low energy gamma-ray sky.  This is especially important as GBM is
the only instrument that can provide nearly continuous monitoring of sources above 100 keV.  For example, Figures~\ref{fig:12to100keV} and \ref{fig:100to300keV} show that Cyg X-1 has made two hard-to-soft state
transitions between 2010 July and 2011 December \citep{Negoro10, Rushton10, Wilsonhodge10, Negoro11, Grinberg11, Case2011b}.  We have used the GBM spectral extraction tools (see Section 7.1)  to generate spectra in the 10-400
keV energy range for the hard and soft states.  We were able to verify that
Cyg X-1 was in an unusually hard state from at least the beginning of the
{\it Fermi} mission up to the first hard-to-soft transition \citep{Nowak11, Case2011d}.  After Cyg X-1 made the transition back
to the hard state, the spectrum more closely resembled the typical hard state
spectrum.  Cyg X-1 returned to the hard state in 2011 December following a
second hard-to-soft transition, and we were able to alert the community that
the transition back to the hard state had occurred \citep{Case2011c}. The
spectrum in the hard state is again similar to the canonical hard
state spectrum.  A more detailed study of Cyg X-1 will be published soon \citep{Case2012}.

In 2010 we made the surprising discovery using GBM Earth occultation data that the Crab Nebula flux had declined by 7\% in the 15-100 keV band over the first two years of the \fermi\ mission, and confirmed this discovery with \integralsc, \rxte, and \swift\ \citep{Wilson2011}. In the third year, the GBM data showed that this decline has leveled off and appears to have started a gradual recovery, so far less dramatic than the two-year decline (See Figures~\ref{fig:12to100keV} \& \ref{fig:100to300keV} and \citet{Wilson2011b}) 

Analysis of the first two years of data looking only above 100 keV resulted
in 8 sources (6 persistent, 2 transient) detected in the 100--300 keV band
and 2 sources detected above 300 keV \citep{Case2011a}.  The persistent
sources detected are the Crab, the BHCs Cyg X-1, SWIFT J1753.5-0127, GRS
1915+105, and 1E 1740-29, and the AGN Cen A.  The transient sources detected
are the BHCs XTE J1752-223 and GX 339-4.  These results were generated using
our previous version of the occultation code with a smaller input catalog.
The detection threshold was $7\sigma$ and no systematic errors were
included.  Using three years of data with the catalog presented here, Figures~\ref{fig:12to100keV} and \ref{fig:100to300keV} show that all 8 of the sources are detected at greater than $5\sigma$ including the systematic errors (see Table~\ref{tbl:main}).  The Crab and Cyg X-1 remain the only two sources detected above 300 keV, with 3-year average fluxes of $1000 \pm 40$ mCrab ($25\sigma$) and $500 \pm 50$ mCrab ($10 \sigma$), respectively. 

Our imaging analysis \citep{Rodi2011} has also indicated the presence of a significant source
in the 100-300 keV band near the BHC GRS 1758-258.  However, since GRS 1758-258 sits less than
$1^{\circ}$ from the strong soft source GX 5-1, the source confusion
filtering removes most of the steps from both of these sources.  We have
reprocessed the data with GX 5-1 classified as a weak source (class 4),
assuming then that all of the $>100$ keV emission is from GRS 1758-258.  The
3-yr average flux for GRS 1758-258 in the 100--300 keV band is $52.6\pm8.2$
mCrab and is detected at a significance of $6.4\sigma$.

Several other sources are at less than $5\sigma$ (including systematic errors) above 100 keV but
current measurements suggest that they may reach the detection threshold as more data are accumulated.  These
include the AGN 3C273 ($3.9\sigma$), NGC 4151 ($2.9\sigma$), 3C454.3 ($2.7\sigma$), and NGC 2110 ($2.9\sigma$), as well as the LMXB/NS systems GS 1826-238 ($4.4\sigma$) and 1A 1742-292 ($3.2\sigma$).  Either of the last two sources would be the first neutron
star system detected in the GBM occultation data in the 100-300 keV energy band.

\subsection{Comparisons with \swift/BAT \label{sec:bat_gbm}}

\begin{figure}[!h]
\center{\includegraphics[width=3.8cm,height=2.71cm]{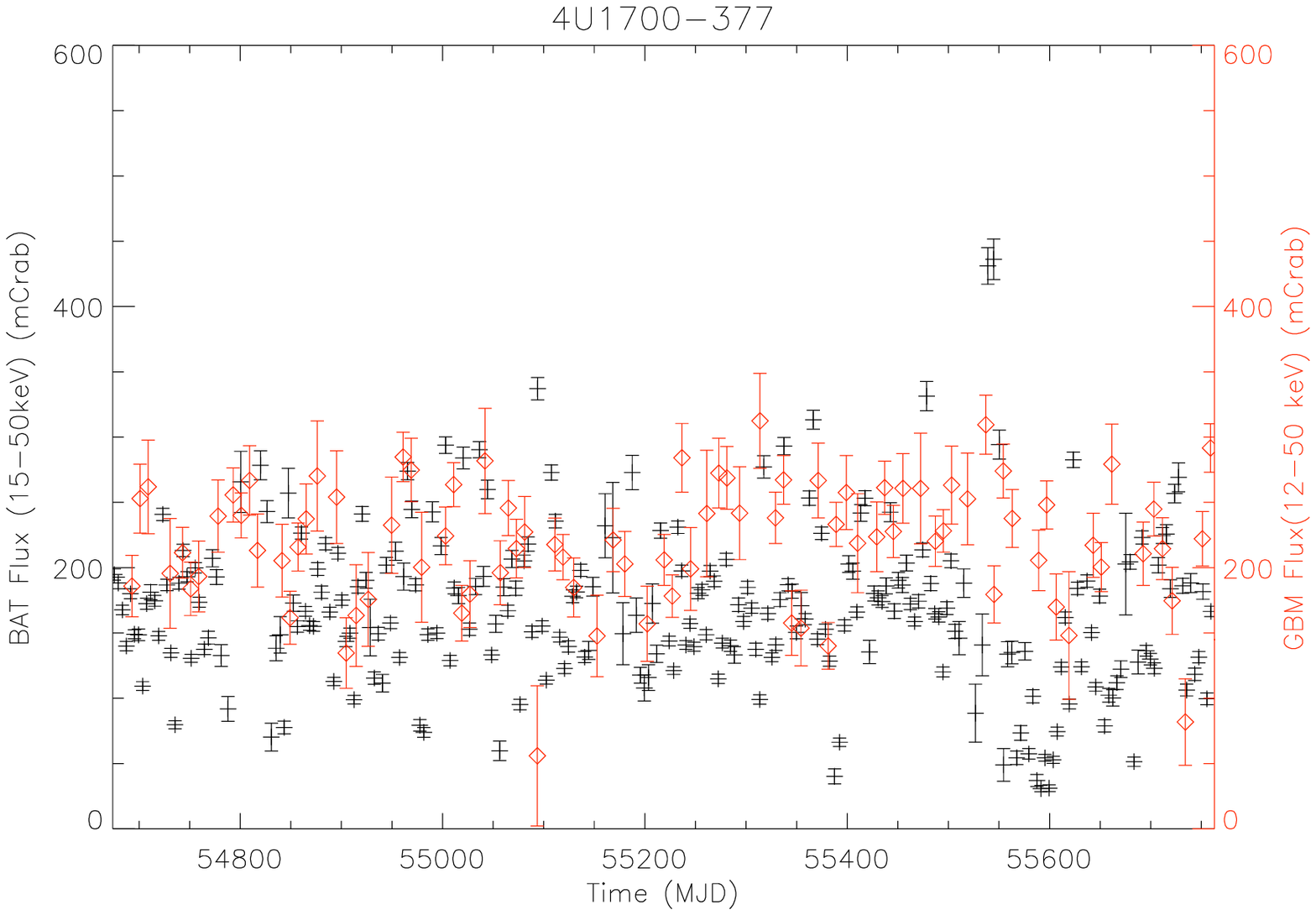}
\includegraphics[width=3.8cm,height=2.71cm]{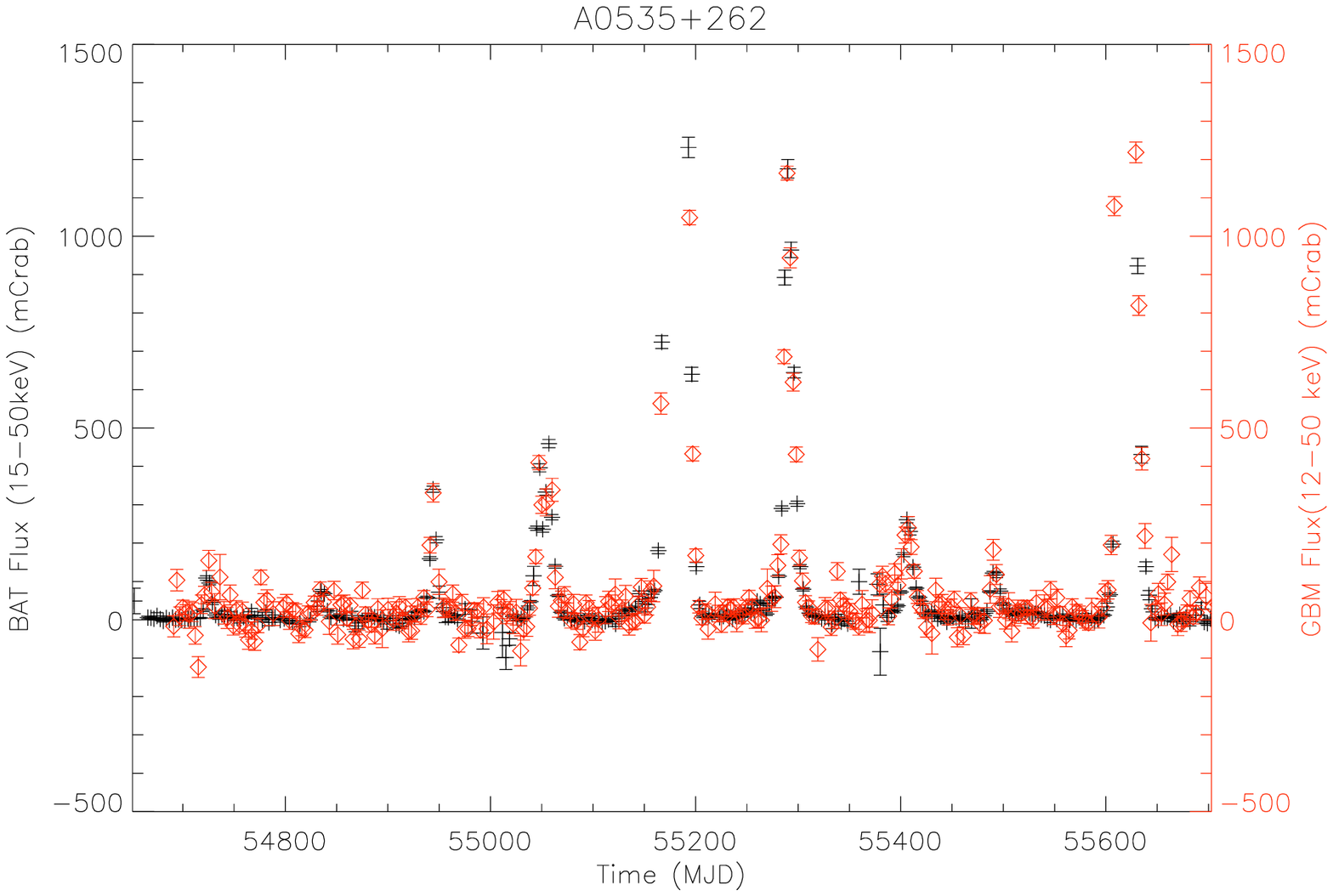}}
\vspace{-0.1in}
\center{\includegraphics[width=3.8cm,height=2.71cm]{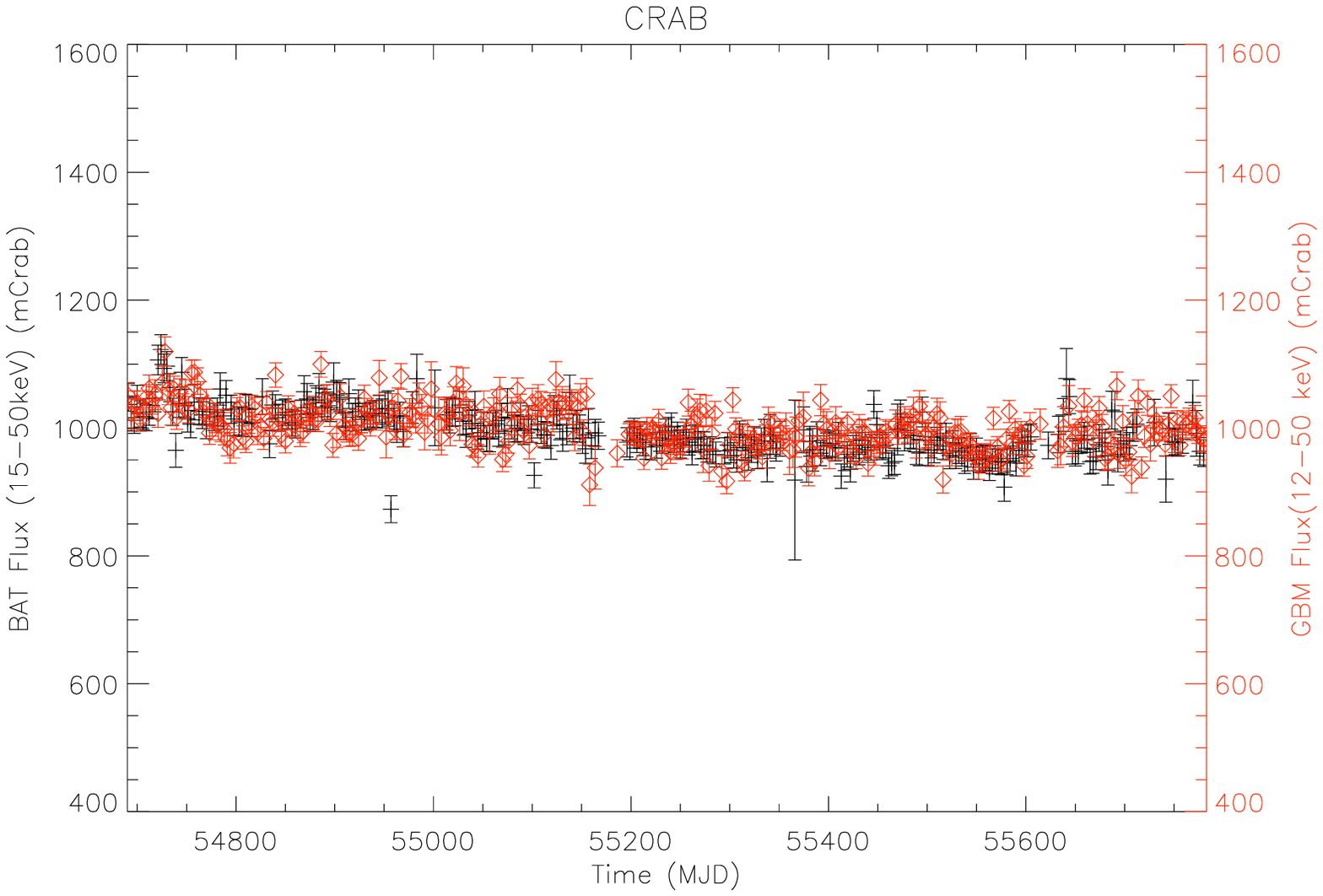}
\includegraphics[width=3.8cm,height=2.71cm]{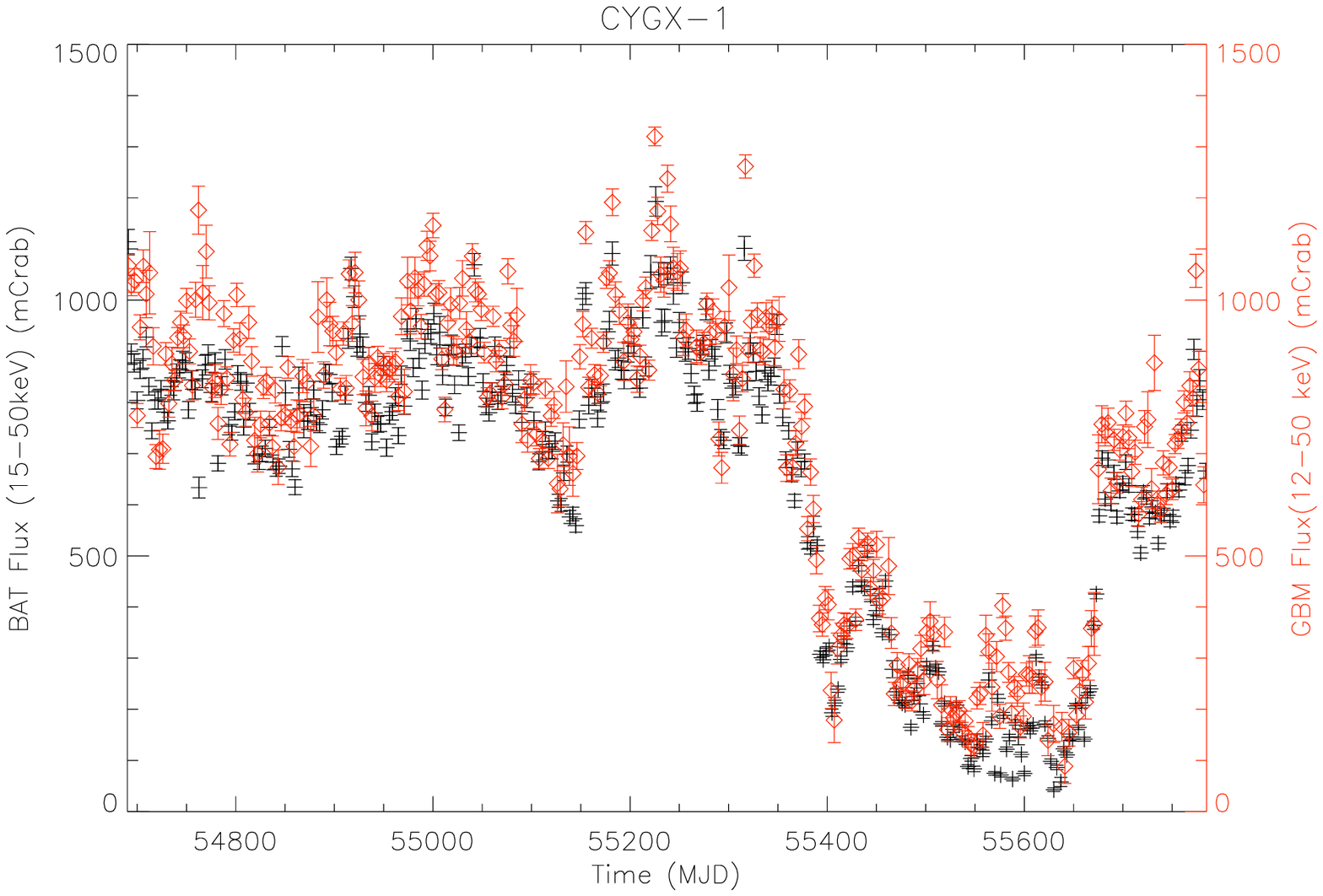}}
\vspace{-0.1in}
\center{\includegraphics[width=3.8cm,height=2.71cm]{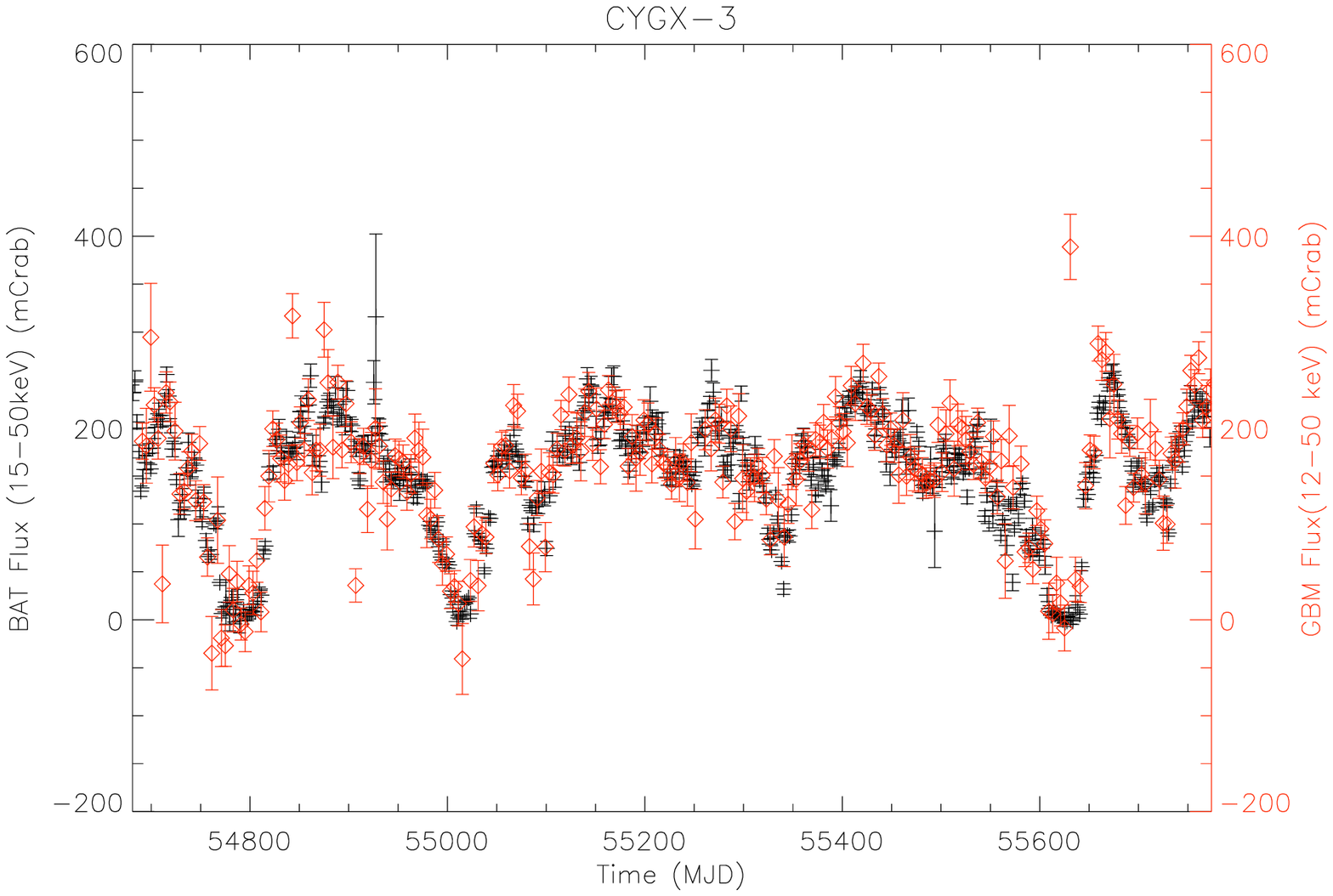}
\includegraphics[width=3.8cm,height=2.71cm]{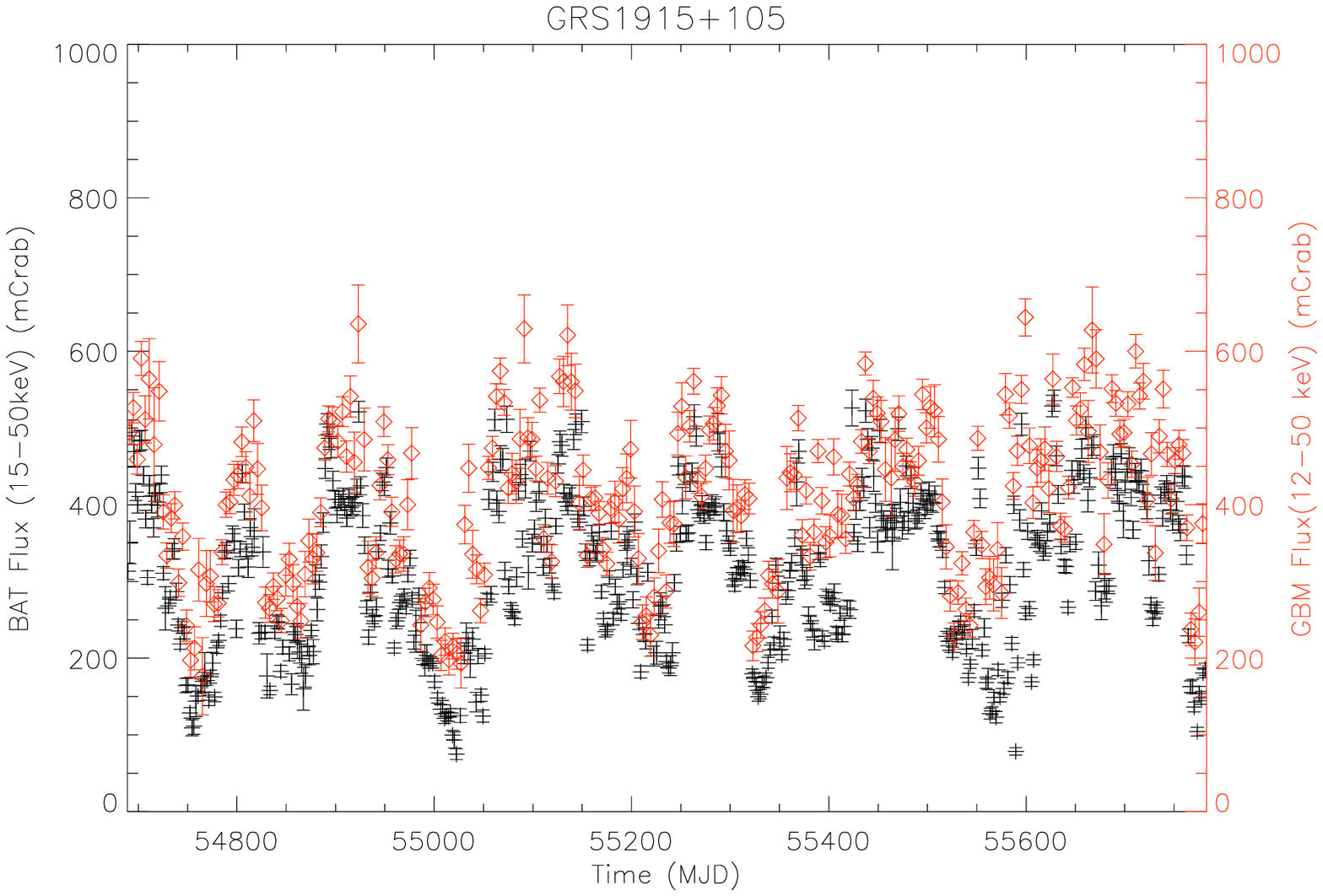}}
\vspace{-0.1in}
\center{\includegraphics[width=3.8cm,height=2.71cm]{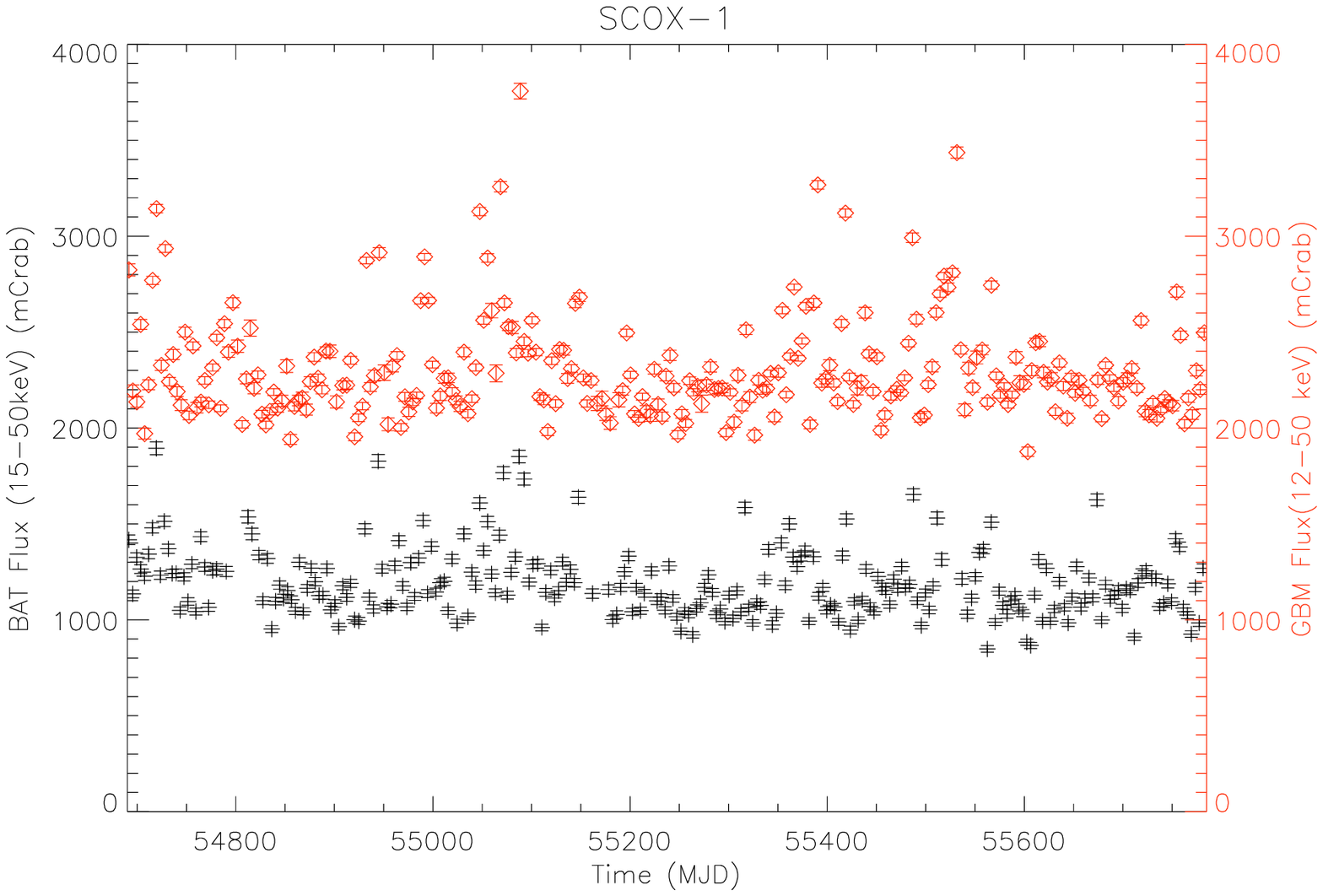}
\includegraphics[width=3.8cm,height=2.71cm]{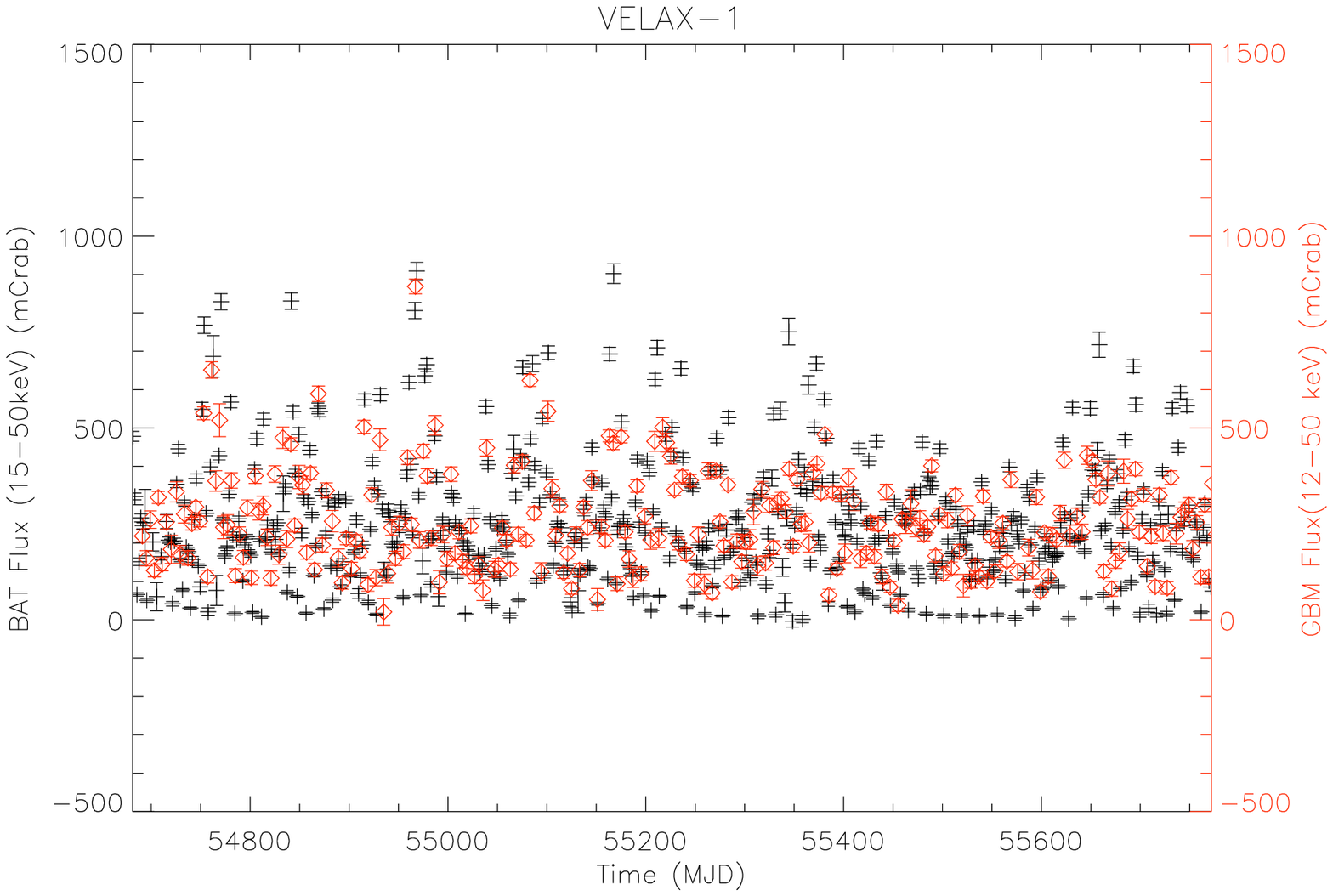}}
\vspace{-0.1in}
\caption{Comparisons between GBM 12-50 keV flux (diamonds) and BAT 15-50 keV flux (crosses). Plotted points are 2-4 day averages. Start and stop times for these averages are not precisely aligned between the two instruments.} 
\label{gbmbat}
\vspace{-0.25in}
\end{figure}

The GBM CTIME data channels 1-2 cover the 12-50 keV energy band, while \swift/BAT public transient monitor data\footnote{\url{http://swift.gsfc.nasa.gov/docs/swift/results/transients/}} spans the 15-50 keV range. As a simple check that the GBM EOT results are reasonable, we have plotted 12-50 keV GBM data for the three year interval from 2008 Aug 12 to 2011 Aug 12  along with the same time interval for \swift/BAT. Figure~\ref{gbmbat} shows comparisons between GBM and \swift/BAT data for eight representative sources. Each dataset is normalized to its own three-year average value for the Crab flux. The data are binned into 2-4 day averages but the bins are not exactly aligned in time between the two instruments. The two datasets agree quite well for these sources with the exception being Sco X-1. Because Sco X-1 has a very soft spectrum, the differing energy bands (12-50 keV for GBM and 15-50 keV for \swift/BAT) become important.

\subsection{Transient outbursts\label{sec:transient}}

\begin{landscape}
\begin{figure}
\includegraphics[width=6.5in,angle=90]{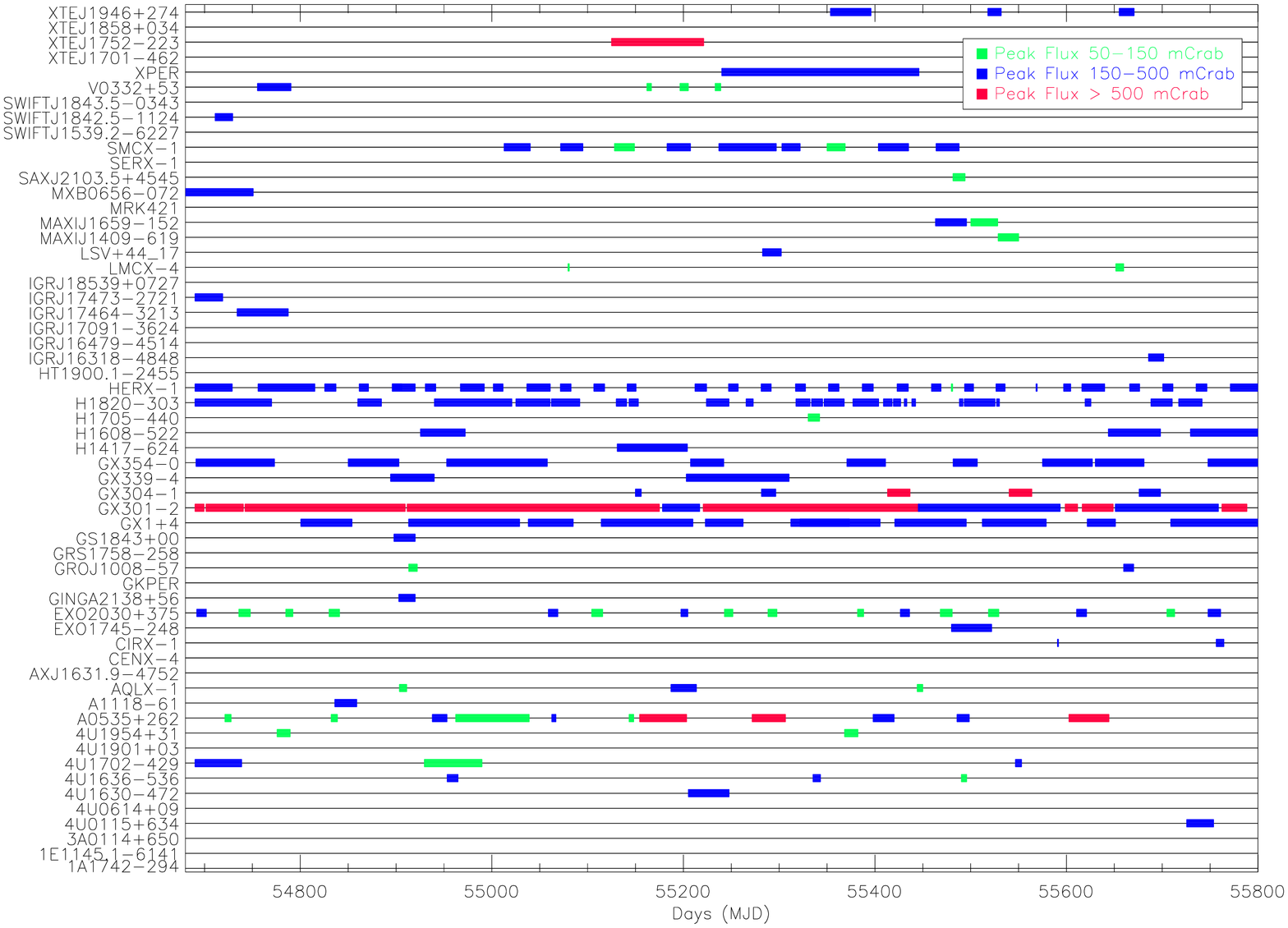}
\caption{Bar plot of transient outbursts detected with GBM. \label{transients}}
\end{figure}
\end{landscape}

Figure~\ref{transients} shows all transient outbursts detected to date with GBM. The time intervals for each transient outburst were defined using the flare database, described in Section~\ref{sec:fitting}, with the outburst time intervals defined using \swift/BAT data. All sources from the flare database are included in the plot, with the exception of Cyg X-1 and Cyg X-3, which are persistent sources that were listed in the flare database because they changed flux class during the course of the mission. The color bars represent outbursts with $> 5\sigma$ statistical significance in the 12-25 keV energy band, integrated over the full outburst interval. Three peak flux categories are shown: 50-150 mCrab (green), 150-500 mCrab (blue), and $> 500$ mCrab (red). Eight sources (V0332+53, LSV+44 17, MXB 0656-072, MAXI J1659-152, IGRJ17473-2721, EXO 1745-248, and SAX J2103.5+4545) had 3-year flux averages below our catalog detection limit in all energy bands, but were detectable at $>5 \sigma$ in the 12-25 keV band for relatively brief transient outbursts. 

GBM is able to detect flares on timescales as short as a few days and $> 50$ mCrab. These detections are consistent with the sensitivity plot (Figure~\ref{sens}) in which the 1-day sensitivity is $\sim 150$ mCrab in the 12-25 keV band. Thus flares with a low peak flux require a longer duration to reach the detection limit. The results from the transient source plot are consistent with this in that the outbursts that are not detected often lasted only a few days. The other outbursts not detected occurred during times when data were not available because of source confusion or no occultations. Because outbursts were added to the flare database only when the peak flux exceeded 50 mCrab in BAT, there are no events with peak fluxes less than 50 mCrab. Consequently outbursts with peak fluxes less than 50 mCrab may be detectable but have not been included in this initial catalog, along with events that were missed by \swift/BAT due to solar constraints. A more detailed analysis of transient outbursts is planned for a future publication.

\onecolumn
\begin{landscape}
\begin{tiny}
%
\begin{center}
\begin{longtable}{lcccllllccl}

\caption[Sources]{GBM Occultation Catalog Detections } \\\hline
  \centering
 \label{tbl:main}
 Source Name & ra & dec & Category & Flux & Flux & Flux & Flux & Sig & Sig & Type \\
 &       &       &  &12-25 keV & 25-50 keV & 50-100 keV & 100-300 keV & 12-50 keV & 12-300 keV & \\
& (deg) & (deg) & &  (mCrabs) & (mCrabs) & (mCrabs) &  (mCrabs) &  ($\sigma$)&  ($\sigma$)  &\\
\hline
 \endfirsthead
\caption{continued from previous page}\\\hline  
  Source Name & ra & dec & Category & Flux & Flux & Flux & Flux & Sig & Sig & type\\
  &       &       &  &12-25 keV & 25-50 keV & 50-100 keV &100-300 keV & 12-50 keV & 12-300 keV  &\\
& (deg) & (deg) &  &(mCrabs) & (mCrabs) & (mCrabs) &  (mCrabs) &  ($\sigma$) &  ($\sigma$)  &\\
\hline
\endhead
SUN &  \nodata & \nodata & T & 5.72$\pm$ 3.11 &  -0.03$\pm$2.84 &  0.76$\pm$3.28 &  2.15$\pm$6.13 &     1.4 &    1.1 &  Star\\
SMCX-1 &   19.27 &  -73.43 & ATP & 40.07$\pm$ 3.23 &  18.14$\pm$3.55 &  -8.97$\pm$4.30 &  -0.47$\pm$5.52 &    12.1 &    5.7 &  HMXB/NS\\
4U0115+634 &   19.63 &   63.74 & AT & 13.06$\pm$ 3.07 &  12.24$\pm$3.48 &  6.88$\pm$4.34 &  6.89$\pm$6.54 &     5.5 &    4.3 &  HMXB/NS\\
NGC1275 &   49.95 &   41.51 & A & 17.13$\pm$ 3.04 &  2.03$\pm$2.91 &  4.54$\pm$3.21 &  2.89$\pm$6.18 &     4.6 &    3.3 &  Seyfert 2\\
V0332+53 &   53.75 &   53.17 & T & 11.27$\pm$ 3.02 &  1.83$\pm$3.09 &  -1.52$\pm$3.60 &  4.11$\pm$6.07 &     3.0 &    1.9 &  HMXB/NS\\
XPER &   58.85 &   31.05 & AT & 41.97$\pm$ 3.01 &  32.46$\pm$2.79 &  37.79$\pm$3.05 &  10.29$\pm$5.21 &    18.1 &   16.8 &  HMXB/NS\\
LSV+44 17 &   70.25 &   44.53 & T & 4.29$\pm$ 3.10 &  6.72$\pm$2.92 &  7.79$\pm$3.34 &  0.07$\pm$6.42 &     2.6 &    2.2 &  HMXB/NS\\
LMCX-4 &   83.20 &  -66.37 & ATP & 24.75$\pm$ 3.10 &  19.11$\pm$3.26 &  0.95$\pm$3.80 &  -8.23$\pm$5.55 &     9.8 &    4.5 &  HMXB/NS\\
CRAB &   83.63 &   22.01 & A & 1000.00$\pm$ 3.00 &  1000.00$\pm$2.88 &  1000.00$\pm$3.05 &  1000.00$\pm$6.02 &   481.1 &  504.8 &  Pulsar/PWN\\
A0535+262 &   84.73 &   26.32 & ATP & 188.89$\pm$ 3.04 &  190.64$\pm$3.08 &  75.12$\pm$3.59 &  9.83$\pm$5.70 &    87.7 &   58.0 &  HMXB/NS\\
NGC2110 &   88.04 &   -7.46 & A & 12.21$\pm$ 3.03 &  15.15$\pm$2.89 &  15.82$\pm$3.15 &  17.74$\pm$6.08 &     6.5 &    7.6 &  Seyfert 2\\
4U0614+09 &   94.28 &    9.14 & A & 29.96$\pm$ 3.00 &  24.58$\pm$2.88 &  20.12$\pm$3.06 &  8.90$\pm$5.96 &    13.1 &   10.6 &  LMXB/NS\\
MXB0656-072 &  104.61 &   -7.26 & T & 3.05$\pm$ 2.99 &  -1.26$\pm$2.79 &  1.69$\pm$3.02 &  -7.59$\pm$5.04 &     0.4 &   -0.6 &  HMXB/NS\\
VELAX-1 &  135.53 &  -40.56 & AP & 274.83$\pm$ 3.03 &  213.39$\pm$2.94 &  28.06$\pm$3.29 &  2.53$\pm$5.53 &   115.7 &   67.4 &  HMXB/NS\\
GROJ1008-57 &  152.44 &  -58.29 & T & 11.77$\pm$ 3.09 &  5.60$\pm$3.12 &  0.47$\pm$3.62 &  -17.07$\pm$6.30 &     4.0 &    0.1 &  HMXB/NS\\
A1118-61 &  170.24 &  -61.92 & T & 15.88$\pm$ 3.57 &  4.02$\pm$4.39 &  -4.91$\pm$5.16 &  11.34$\pm$9.01 &     3.5 &    2.2 &  HMXB/NS\\
CENX-3 &  170.31 &  -60.62 & AP & 153.47$\pm$ 3.18 &  31.48$\pm$3.10 &  4.74$\pm$3.56 &  -5.26$\pm$5.33 &    41.7 &   23.7 &  HMXB/NS\\
1E1145.1-6141 &  176.87 &  -61.95 & AP & 24.27$\pm$ 3.22 &  20.19$\pm$3.53 &  19.29$\pm$4.33 &  -28.10$\pm$7.65 &     9.3 &    3.6 &  HMXB/NS\\
NGC4151 &  182.65 &   39.42 & A & 31.62$\pm$ 3.11 &  32.08$\pm$2.97 &  36.82$\pm$3.38 &  17.93$\pm$6.23 &    14.8 &   14.3 &  Seyfert 1\\
NGC4388 &  186.45 &   12.66 & A & 11.54$\pm$ 3.02 &  13.77$\pm$2.84 &  13.21$\pm$3.07 &  14.63$\pm$5.97 &     6.1 &    6.7 &  Seyfert 2\\
GX301-2 &  186.66 &  -62.77 & ATP & 260.11$\pm$ 3.25 &  110.28$\pm$3.19 &  -1.83$\pm$3.86 &  -13.15$\pm$6.50 &    81.3 &   40.3 &  HMXB/NS\\
3C273 &  187.27 &    2.05 & A & 15.31$\pm$ 2.99 &  19.75$\pm$2.86 &  24.30$\pm$2.99 &  22.56$\pm$5.84 &     8.5 &   10.6 &  Quasar\\
4U1254-690 &  194.40 &  -69.29 & A & 17.35$\pm$ 3.42 &  2.94$\pm$3.35 &  2.08$\pm$3.92 &  -6.57$\pm$5.04 &     4.2 &    2.0 &  LMXB/NS\\
GX304-1 &  195.32 &  -61.60 & ATP & 38.21$\pm$ 3.17 &  35.38$\pm$3.39 &  9.05$\pm$3.82 &  -5.52$\pm$7.20 &    15.8 &    8.2 &  HMXB/NS\\
CEN A &  201.37 &  -43.02 & A & 49.92$\pm$ 3.06 &  61.50$\pm$3.03 &  70.40$\pm$3.23 &  92.69$\pm$6.19 &    25.9 &   33.5 &  Seyfert 2\\
NGC5252 &  204.57 &    4.54 & A & 6.54$\pm$ 3.03 &  11.62$\pm$2.86 &  11.87$\pm$3.10 &  11.54$\pm$6.02 &     4.4 &    5.2 &  Seyfert 2\\
IC4329A &  207.33 &  -30.31 & A & 10.51$\pm$ 3.08 &  10.89$\pm$2.95 &  13.04$\pm$3.25 &  8.13$\pm$6.26 &     5.0 &    5.2 &  Seyfert 1\\
MAXIJ1409-619 &  212.01 &  -61.98 & T & 10.87$\pm$ 3.17 &  10.25$\pm$3.35 &  3.29$\pm$4.02 &  1.01$\pm$7.15 &     4.6 &    2.7 &  HMXB/NS\\
CIRCINUS GALAXY &  213.29 &  -65.34 & A & 12.57$\pm$ 3.12 &  19.01$\pm$3.41 &  0.03$\pm$1.50 &  -5.87$\pm$6.51 &     6.8 &    3.2 &  Seyfert 2\\
NGC5506 &  213.31 &   -3.21 & A & 11.32$\pm$ 3.03 &  12.18$\pm$2.87 &  4.85$\pm$3.06 &  11.31$\pm$5.96 &     5.6 &    5.0 &  Seyfert 2\\
H1417-624 &  215.30 &  -62.70 & ATP & 19.95$\pm$ 3.17 &  19.44$\pm$3.24 &  -0.04$\pm$3.70 &  -9.96$\pm$7.05 &     8.7 &    3.2 &  HMXB/NS\\
CIRX-1 &  230.17 &  -57.17 & T & 14.13$\pm$ 3.44 &  2.31$\pm$3.09 &  4.04$\pm$2.81 &  -3.08$\pm$7.28 &     3.6 &    1.9 &  LMXB/NS\\
4U1538-52 &  235.60 &  -52.39 & AP & 23.62$\pm$ 3.10 &  12.97$\pm$3.19 &  2.89$\pm$3.41 &  1.20$\pm$6.37 &     8.2 &    4.8 &  HMXB/NS\\
H1608-522 &  243.18 &  -52.42 & T & 11.35$\pm$ 3.51 &  9.53$\pm$3.10 &  5.33$\pm$2.97 &  2.09$\pm$7.45 &     4.5 &    3.0 &  LMXB/NS\\
SCOX-1 &  244.98 &  -15.64 & AP & 4364.73$\pm$ 3.31 &  222.51$\pm$2.99 &  19.49$\pm$3.38 &  12.48$\pm$6.69 &  1028.1 &  529.8 &  LMXB/NS\\
X1624-490 &  247.01 &  -49.20 & A & 28.22$\pm$ 3.54 &  8.43$\pm$2.81 &  1.73$\pm$2.41 &  6.82$\pm$7.90 &     8.1 &    4.8 &  LMXB/NS\\
IGRJ16318-4848 &  247.97 &  -48.80 & AT & 39.93$\pm$ 3.38 &  39.55$\pm$3.58 &  4.90$\pm$4.40 &  1.15$\pm$8.32 &    16.1 &    8.1 &  HMXB/NS\\
AXJ1631.9-4752 &  248.01 &  -47.87 & A & 32.94$\pm$ 3.24 &  39.43$\pm$3.52 &  8.42$\pm$4.01 &  3.41$\pm$7.59 &    15.1 &    8.6 &  HMXB/NS\\
4U1626-67 &  248.07 &  -67.46 & A & 61.79$\pm$ 3.27 &  26.10$\pm$3.64 &  -4.62$\pm$4.58 &  -3.47$\pm$7.09 &    18.0 &    8.2 &  LMXB/NS\\
4U1630-472 &  248.50 &  -47.39 & AT & 26.76$\pm$ 3.22 &  32.90$\pm$3.48 &  9.15$\pm$3.90 &  -1.36$\pm$7.32 &    12.6 &    7.1 &  BHC\\
4U1636-536 &  250.23 &  -53.75 & AT & 47.15$\pm$ 3.49 &  31.88$\pm$3.98 &  6.34$\pm$5.09 &  -0.09$\pm$9.19 &    14.9 &    7.3 &  LMXB/NS\\
SWIFTJ164449.3P573451 &  251.20 &   57.60 & A & 12.72$\pm$ 3.23 &  13.84$\pm$3.38 &  3.91$\pm$4.08 &  -5.12$\pm$7.68 &     5.7 &    2.6 &  TDE\\
GX340+0 &  251.45 &  -45.61 & A & 173.17$\pm$ 3.63 &  -0.07$\pm$3.32 &  -6.20$\pm$4.27 &  -6.86$\pm$8.23 &    35.2 &   15.2 &  LMXB/NS\\
HERX-1 &  254.46 &   35.34 & ATP & 86.36$\pm$ 3.04 &  39.68$\pm$2.97 &  5.87$\pm$3.43 &  6.67$\pm$5.31 &    29.6 &   18.2 &  LMXB/NS\\
MAXIJ1659-152 &  254.76 &  -15.26 & T & 3.29$\pm$ 3.19 &  1.44$\pm$3.24 &  0.55$\pm$3.67 &  -7.14$\pm$7.11 &     1.0 &   -0.2 &  BHC\\
OAO1657-415 &  255.20 &  -41.67 & AP & 62.30$\pm$ 3.42 &  66.70$\pm$4.02 &  31.40$\pm$5.27 &  17.88$\pm$9.08 &    24.4 &   15.2 &  HMXB/NS\\
GX339-4 &  255.71 &  -48.79 & AT & 18.39$\pm$ 3.61 &  23.83$\pm$4.05 &  42.49$\pm$5.21 &  1.58$\pm$9.88 &     7.8 &    6.9 &  BHC\\
4U1700-377 &  255.99 &  -37.84 & AP & 230.57$\pm$ 3.73 &  209.48$\pm$4.59 &  112.86$\pm$6.21 &  20.67$\pm$10.68 &    74.5 &   41.9 &  HMXB/NS\\
GX349+2 &  256.45 &  -36.42 & A & 304.96$\pm$ 4.08 &  12.58$\pm$3.84 &  -0.22$\pm$4.03 &  18.63$\pm$10.20 &    56.7 &   27.3 &  LMXB/NS\\
4U1702-429 &  256.56 &  -43.04 & AT & 36.72$\pm$ 3.26 &  14.77$\pm$3.52 &  9.26$\pm$3.43 &  6.24$\pm$5.77 &    10.7 &    8.1 &  LMXB/NS\\
H1705-440 &  257.23 &  -44.10 & AT & 40.89$\pm$ 3.30 &  15.51$\pm$3.69 &  5.03$\pm$3.75 &  4.13$\pm$6.33 &    11.4 &    7.4 &  LMXB/NS\\
4U1708-407 &  258.10 &  -40.84 & A & 19.83$\pm$ 3.84 &  -2.96$\pm$4.44 &  8.12$\pm$5.84 &  19.46$\pm$11.16 &     2.9 &    3.2 &  LMXB/NS\\
OPH CLUSTER &  258.11 &  -23.38 & A & 20.51$\pm$ 3.49 &  2.62$\pm$3.71 &  -4.82$\pm$4.61 &  -4.72$\pm$8.78 &     4.5 &    1.2 &  Cluster of Galaxies\\
GX9+9 &  262.93 &  -16.96 & A & 70.93$\pm$ 3.86 &  2.49$\pm$3.70 &  -0.75$\pm$4.53 &  6.68$\pm$5.74 &    13.7 &    8.8 &  LMXB/NS\\
GX354-0 &  263.00 &  -33.83 & AT & 87.00$\pm$ 3.57 &  53.10$\pm$3.91 &  20.86$\pm$4.99 &  -3.96$\pm$9.53 &    26.5 &   13.1 &  LMXB/NS\\
GX1+4 &  263.01 &  -24.75 & AT & 79.52$\pm$ 3.57 &  95.81$\pm$3.86 &  73.61$\pm$4.87 &  21.01$\pm$9.27 &    33.3 &   23.0 &  HMXB/NS\\
H1730-333 &  263.35 &  -33.39 & AT & 40.75$\pm$ 3.96 &  13.69$\pm$4.58 &  6.40$\pm$6.09 &  -23.93$\pm$11.56 &     9.0 &    2.6 &  LMXB/NS\\
KS1731-260 &  263.55 &  -26.09 & A & 38.82$\pm$ 4.89 &  27.46$\pm$4.71 &  21.48$\pm$5.54 &  15.28$\pm$13.46 &     9.8 &    6.4 &  LMXB/NS\\
SLX1735-269 &  264.57 &  -27.00 & A & 23.57$\pm$ 4.43 &  28.77$\pm$5.31 &  20.71$\pm$7.14 &  -9.99$\pm$13.40 &     7.6 &    3.8 &  LMXB/NS\\
X1735-444 &  264.74 &  -44.45 & A & 83.80$\pm$ 3.90 &  5.86$\pm$3.85 &  5.31$\pm$4.16 &  22.00$\pm$9.98 &    16.4 &    9.7 &  LMXB/NS\\
1E1740-29 &  265.97 &  -29.74 & A & 91.94$\pm$ 4.37 &  79.68$\pm$5.19 &  117.22$\pm$7.03 &  90.98$\pm$13.22 &    25.3 &   23.1 &  BHC\\
MAXIJ1745-288 &  266.46 &  -28.82 & A & 85.05$\pm$ 4.59 &  49.91$\pm$5.56 &  36.26$\pm$7.60 &  28.82$\pm$14.24 &    18.7 &   11.3 &  LMXB/NS\\
1A1742-294 &  266.52 &  -29.51 & A & 78.37$\pm$ 7.22 &  44.36$\pm$9.65 &  61.89$\pm$12.78 &  73.63$\pm$23.06 &    10.2 &    8.9 &  LMXB/NS\\
IGRJ17464-3213 &  266.58 &  -32.24 & AT & 38.54$\pm$ 4.62 &  31.07$\pm$5.89 &  44.59$\pm$7.73 &  16.86$\pm$14.67 &     9.3 &    7.2 &  BHC\\
IGRJ17473-2721 &  266.83 &  -27.34 & T & 27.32$\pm$ 5.62 &  8.04$\pm$6.95 &  11.99$\pm$9.74 &  16.80$\pm$18.42 &     4.0 &    2.8 &  LMXB/NS\\
GX3+1 &  266.98 &  -26.56 & A & 94.94$\pm$ 4.51 &  7.69$\pm$4.44 &  -1.13$\pm$4.70 &  -8.96$\pm$11.95 &    16.2 &    6.5 &  LMXB/NS\\
EXO1745-248 &  267.23 &  -24.89 & T & 0.71$\pm$ 4.42 &  -2.59$\pm$4.64 &  -5.43$\pm$6.78 &  9.90$\pm$12.40 &    -0.3 &    0.2 &  LMXB/NS\\
4U1746-370 &  267.55 &  -37.05 & A & 24.03$\pm$ 3.94 &  3.79$\pm$4.62 &  7.78$\pm$6.19 &  18.48$\pm$11.81 &     4.6 &    3.7 &  LMXB/NS\\
XTEJ1752-223 &  268.04 &  -22.33 & AT & 25.98$\pm$ 4.03 &  51.77$\pm$4.73 &  53.43$\pm$6.29 &  52.12$\pm$11.94 &    12.5 &   12.3 &  BHC\\
SWIFTJ1753.5-0127 &  268.37 &   -1.45 & A & 67.78$\pm$ 3.46 &  90.51$\pm$3.87 &  111.28$\pm$4.95 &  110.43$\pm$9.09 &    30.5 &   32.8 &  BHC\\
GX5-1 &  270.27 &  -25.08 & A & 310.11$\pm$ 12.34 &  30.15$\pm$17.01 &  -28.81$\pm$24.87 &  73.29$\pm$46.80 &    16.2 &    6.7 &  LMXB/NS\\
GRS1758-258 &  270.30 &  -25.73 & A & 65.30$\pm$ 9.95 &  28.17$\pm$15.74 &  93.54$\pm$22.21 &  21.72$\pm$43.99 &     5.0 &    4.0 &  BHC\\
GX9+1 &  270.38 &  -20.53 & A & 223.96$\pm$ 4.26 &  9.08$\pm$3.44 &  -0.62$\pm$4.03 &  -18.80$\pm$10.38 &    42.5 &   17.2 &  LMXB/NS\\
GX13+1 &  273.63 &  -17.16 & A & 77.54$\pm$ 4.46 &  2.13$\pm$4.43 &  -3.04$\pm$4.72 &  11.35$\pm$12.03 &    12.7 &    6.1 &  LMXB/NS\\
4U1812-12 &  273.80 &  -12.08 & A & 26.80$\pm$ 4.84 &  36.51$\pm$6.71 &  35.61$\pm$9.53 &  -7.49$\pm$17.27 &     7.7 &    4.3 &  LMXB/NS\\
GX17+2 &  274.01 &  -14.04 & A & 353.23$\pm$ 4.38 &  48.03$\pm$4.52 &  40.39$\pm$6.06 &  16.99$\pm$11.58 &    63.7 &   31.6 &  LMXB/NS\\
H1820-303 &  275.92 &  -30.37 & A & 159.62$\pm$ 3.57 &  15.83$\pm$4.03 &  -0.37$\pm$5.04 &  -0.62$\pm$9.12 &    32.6 &   14.9 &  LMXB/NS\\
3A1822-371 &  276.45 &  -37.10 & A & 49.06$\pm$ 3.49 &  5.64$\pm$3.90 &  -5.13$\pm$4.69 &  -2.14$\pm$8.57 &    10.4 &    4.3 &  LMXB/NS\\
GS1826-238 &  277.37 &  -23.80 & A & 77.18$\pm$ 3.40 &  69.19$\pm$3.82 &  58.57$\pm$4.74 &  40.42$\pm$9.24 &    28.6 &   21.2 &  LMXB/NS\\
SERX-1 &  279.99 &    5.04 & A & 50.29$\pm$ 3.21 &  7.10$\pm$3.10 &  3.35$\pm$3.63 &  -8.90$\pm$5.79 &    12.9 &    6.3 &  LMXB/NS\\
SWIFTJ1842.5-1124 &  280.50 &  -11.40 & T & 4.18$\pm$ 3.12 &  6.86$\pm$3.20 &  10.25$\pm$3.50 &  7.06$\pm$6.70 &     2.5 &    3.2 &  BHC\\
SWIFTJ1843.5-0343 &  280.90 &   -3.73 & A & 12.72$\pm$ 3.19 &  10.25$\pm$3.18 &  6.01$\pm$3.69 &  -1.02$\pm$7.06 &     5.1 &    3.1 &  LMXB/NS\\
GS1843+00 &  281.41 &    0.89 & AT & 31.17$\pm$ 3.10 &  14.84$\pm$3.21 &  5.86$\pm$3.76 &  -1.78$\pm$6.69 &    10.3 &    5.6 &  HMXB/NS\\
IGRJ18483-0311 &  282.08 &   -3.16 & A & 18.91$\pm$ 3.18 &  11.99$\pm$3.16 &  9.05$\pm$3.65 &  11.66$\pm$6.99 &     6.9 &    5.7 &  HMXB/SFXT\\
HT1900.1-2455 &  285.04 &  -24.92 & A & 19.31$\pm$ 3.09 &  26.65$\pm$3.09 &  19.92$\pm$3.50 &  8.04$\pm$6.69 &    10.5 &    8.5 &  LMXB/NS\\
H1907+097 &  287.40 &    9.83 & A & 46.21$\pm$ 3.14 &  25.31$\pm$3.03 &  16.11$\pm$3.40 &  -2.56$\pm$4.91 &    16.4 &   11.5 &  HMXB/NS\\
4U1909+07 &  287.70 &    7.60 & A & 44.01$\pm$ 3.17 &  34.83$\pm$3.27 &  19.95$\pm$3.91 &  -8.29$\pm$6.96 &    17.3 &    9.9 &  HMXB/NS\\
AQLX-1 &  287.82 &    0.58 & T & 6.61$\pm$ 3.14 &  9.92$\pm$3.22 &  4.70$\pm$3.60 &  -0.19$\pm$6.93 &     3.7 &    2.3 &  LMXB/NS\\
GRS1915+105 &  288.82 &   10.97 & A & 540.46$\pm$ 3.14 &  271.96$\pm$3.02 &  138.32$\pm$3.40 &  60.65$\pm$6.51 &   186.4 &  118.4 &  BHC\\
XTEJ1946+274 &  296.41 &   27.36 & ATP & 19.04$\pm$ 3.16 &  10.93$\pm$3.19 &  4.80$\pm$3.73 &  -8.98$\pm$7.12 &     6.7 &    2.8 &  HMXB/NS\\
4U1954+31 &  298.93 &   32.10 & AT & 18.00$\pm$ 3.26 &  15.46$\pm$3.56 &  3.04$\pm$4.50 &  -5.53$\pm$7.64 &     6.9 &    3.1 &  LMXB/NS\\
CYGX-1 &  299.59 &   35.20 & AP & 640.41$\pm$ 3.15 &  779.52$\pm$3.17 &  944.20$\pm$3.64 &  881.27$\pm$6.93 &   317.6 &  360.0 &  BHC\\
EXO2030+375 &  308.06 &   37.64 & ATP & 29.88$\pm$ 3.13 &  23.37$\pm$3.34 &  9.05$\pm$4.07 &  -2.26$\pm$7.40 &    11.6 &    6.2 &  HMXB/NS\\
CYGX-3 &  308.11 &   40.96 & AP & 196.13$\pm$ 3.04 &  126.56$\pm$3.05 &  45.43$\pm$3.30 &  13.83$\pm$6.34 &    74.9 &   45.8 &  BHC\\
SAXJ2103.5+4545 &  315.90 &   45.76 & T & 5.60$\pm$ 3.05 &  5.19$\pm$3.12 &  2.97$\pm$3.32 &  -1.37$\pm$6.21 &     2.5 &    1.5 &  HMXB/NS\\
IGRJ21247+5058 &  321.18 &   50.98 & A & 13.76$\pm$ 3.19 &  10.93$\pm$3.29 &  8.20$\pm$3.95 &  6.78$\pm$7.50 &     5.4 &    4.1 &  Radio Galaxy\\
GINGA2138+56 &  324.88 &   56.99 & T & 9.15$\pm$ 3.18 &  7.42$\pm$3.28 &  -1.60$\pm$4.02 &  -1.44$\pm$6.33 &     3.6 &    1.5 &  HMXB/NS\\
CYGX-2 &  326.17 &   38.32 & A & 138.01$\pm$ 3.10 &  14.04$\pm$2.85 &  -2.24$\pm$3.08 &  5.60$\pm$5.97 &    36.1 &   19.6 &  LMXB/NS\\
3C454.3 &  343.49 &   16.15 & A & 5.57$\pm$ 2.95 &  6.19$\pm$2.78 &  11.73$\pm$2.76 &  14.56$\pm$5.32 &     2.9 &    5.3 &  Quasar\\
\hline
\end{longtable}
\end{center}
\end{tiny}


\begin{tiny}
%
\begin{center}
\begin{longtable}{lcccllllccl}

\caption[Sources]{GBM Occultation Catalog Marginal Detections } \\\hline
  \centering
 \label{tbl:marginal}
 Source Name & ra & dec & Category & Flux & Flux & Flux & Flux & Sig & Sig & Type \\
 &       &       &  &12-25 keV & 25-50 keV & 50-100 keV & 100-300 keV & 12-50 keV & 12-300 keV & \\
& (deg) & (deg) & &  (mCrabs) & (mCrabs) & (mCrabs) &  (mCrabs) &  ($\sigma$)&  ($\sigma$)  &\\
\hline
 \endfirsthead
\caption{continued from previous page}\\\hline  
  Source Name & ra & dec & Category & Flux & Flux & Flux & Flux & Sig & Sig & type\\
  &       &       &  &12-25 keV & 25-50 keV & 50-100 keV &100-300 keV & 12-50 keV & 12-300 keV  &\\
& (deg) & (deg) &  &(mCrabs) & (mCrabs) & (mCrabs) &  (mCrabs) &  ($\sigma$) &  ($\sigma$)  &\\
\hline
\endhead
3A0114+650 &   19.51 &   65.29 & B & 12.96$\pm$ 3.15 &  5.66$\pm$3.39 &  10.81$\pm$4.18 &  14.41$\pm$7.16 &     4.0 &    4.6 & HMXB/NS\\
GKPER &   52.80 &   43.90 & B & 8.58$\pm$ 3.05 &  6.75$\pm$2.96 &  -1.20$\pm$3.28 &  6.67$\pm$6.34 &     3.6 &    2.5 & CV\\
LMCX-3 &   84.79 &  -64.08 & B & 12.36$\pm$ 3.18 &  8.76$\pm$3.27 &  9.08$\pm$3.84 &  -4.86$\pm$7.20 &     4.6 &    2.7 & BHC\\
VELA-X &  128.29 &  -45.19 & B & 12.42$\pm$ 3.12 &  6.38$\pm$3.11 &  12.16$\pm$3.57 &  6.32$\pm$6.78 &     4.3 &    4.2 & Pulsar/PWN\\
VELA PSR &  128.85 &  -45.18 & B & 12.80$\pm$ 3.06 &  7.40$\pm$3.12 &  8.83$\pm$3.44 &  6.54$\pm$6.59 &     4.6 &    4.1 & Pulsar/PWN\\
MCG-05-23-016 &  146.92 &  -30.95 & B & 7.86$\pm$ 3.06 &  9.69$\pm$2.95 &  4.40$\pm$3.25 &  4.09$\pm$6.21 &     4.1 &    3.2 & Seyfert 2\\
1FGLJ1018.6-5856 &  154.73 &  -58.95 & B & 7.21$\pm$ 3.16 &  12.63$\pm$3.19 &  -5.93$\pm$3.68 &  3.18$\pm$6.93 &     4.4 &    1.9 & Gamma Ray Binary\\
MRK421 &  166.11 &   38.21 & B & 11.24$\pm$ 3.01 &  4.59$\pm$2.71 &  6.58$\pm$3.14 &  5.39$\pm$6.08 &     3.9 &    3.5 & BL Lac - type object\\
NGC3783 &  174.75 &  -37.73 & B & 7.35$\pm$ 3.09 &  11.67$\pm$3.00 &  2.44$\pm$3.36 &  -5.90$\pm$6.43 &     4.4 &    1.8 & Seyfert 1\\
COMA CLUSTER &  194.95 &   27.98 & B & 11.47$\pm$ 3.02 &  8.93$\pm$2.85 &  1.24$\pm$3.09 &  -5.56$\pm$5.97 &     4.9 &    2.0 & Cluster of Galaxies\\
4U1323-62 &  201.65 &  -62.14 & B & 15.42$\pm$ 3.27 &  6.72$\pm$3.42 &  8.99$\pm$4.10 &  -8.09$\pm$7.73 &     4.7 &    2.3 & LMXB/NS\\
MSH15-52 &  228.51 &  -59.26 & B & 2.13$\pm$ 3.24 &  10.09$\pm$3.32 &  13.85$\pm$3.92 &  8.01$\pm$7.40 &     2.6 &    3.6 & Pulsar/PWN\\
H 1517+656 &  229.45 &   65.42 & B & 7.73$\pm$ 3.23 &  7.62$\pm$3.47 &  3.70$\pm$4.18 &  -4.39$\pm$7.90 &     3.2 &    1.4 & BL Lac - type object\\
IGRJ16418-4532 &  250.45 &  -45.53 & B & 17.46$\pm$ 3.67 &  4.13$\pm$4.06 &  -4.30$\pm$5.18 &  -2.56$\pm$9.76 &     3.9 &    1.2 & HMXB/SFXT\\
TXS 1700+685 &  255.04 &   68.50 & B & 10.09$\pm$ 3.21 &  6.93$\pm$3.60 &  3.32$\pm$4.17 &  -3.92$\pm$7.73 &     3.5 &    1.6 & Seyfert 1\\
GRS1724-308 &  261.90 &  -30.80 & B & 13.86$\pm$ 3.82 &  14.19$\pm$4.31 &  -0.56$\pm$5.64 &  -5.91$\pm$10.67 &     4.9 &    1.6 & LMXB/NS\\
B2 1732+38A &  263.59 &   38.96 & B & 2.50$\pm$ 3.04 &  2.38$\pm$3.05 &  9.79$\pm$3.25 &  -4.15$\pm$6.22 &     1.1 &    1.3 & BL Lac - type object\\
SAXJ1806.5-2215 &  271.64 &  -22.24 & B & 18.21$\pm$ 4.09 &  1.80$\pm$4.91 &  5.85$\pm$6.66 &  14.30$\pm$12.74 &     3.1 &    2.6 & LMXB/NS\\
SAXJ1818.6-1703 &  274.66 &  -17.05 & B & 14.13$\pm$ 4.22 &  5.66$\pm$4.95 &  -3.01$\pm$6.63 &  4.14$\pm$12.74 &     3.0 &    1.3 & HMXB/SFXT\\
XTEJ1855-026 &  283.88 &   -2.61 & B & 14.11$\pm$ 3.17 &  7.56$\pm$3.14 &  2.02$\pm$3.63 &  -0.45$\pm$6.96 &     4.9 &    2.6 & HMXB/NS\\
4U2127+119 &  322.49 &   12.17 & B & 6.12$\pm$ 3.05 &  7.68$\pm$2.90 &  7.69$\pm$3.18 &  1.57$\pm$6.14 &     3.3 &    2.8 & LMXB/NS\\
4U2206+54 &  331.98 &   54.52 & B & 8.81$\pm$ 3.15 &  9.67$\pm$3.32 &  8.22$\pm$3.97 &  6.18$\pm$7.48 &     4.0 &    3.4 & HMXB/NS\\
CAS A &  350.86 &   58.82 & B & 7.35$\pm$ 3.15 &  8.92$\pm$3.21 &  2.20$\pm$3.81 &  15.52$\pm$7.22 &     3.6 &    3.6 & SNR\\
\hline
\end{longtable}
\end{center}
\end{tiny}

\begin{tiny}
%
\begin{center}
\begin{longtable}{lccclllll}

\caption[Sources]{GBM Occultation Catalog Upper Limits} \\\hline
  \centering
 \label{tbl:uplim}
 Source Name & ra & dec & Category & $<$ Flux $(3 \sigma)$ & $<$ Flux $(3 \sigma)$& $<$ Flux $(3 \sigma)$  & $<$ Flux $(3 \sigma)$& Type \\
 &       &       &  &12-25 keV & 25-50 keV & 50-100 keV & 100-300 keV &  \\
& (deg) & (deg) & &  (mCrabs) & (mCrabs) & (mCrabs) &  (mCrabs) & \\
\hline
 \endfirsthead
\caption{continued from previous page}\\\hline  
  Source Name & ra & dec & Category & Flux $(3 \sigma)$  & Flux $(3 \sigma)$& Flux $(3 \sigma)$  & Flux $(3 \sigma)$ & type\\
  &       &       &  &12-25 keV & 25-50 keV & 50-100 keV &100-300 keV &\\
& (deg) & (deg) &  &(mCrabs) & (mCrabs) & (mCrabs) &  (mCrabs) & \\
\hline
\endhead
IGR J00234+6141 &    5.74 &   61.69 & N &  9.51 &   9.78 &  11.70 &  22.20 & CV\\
V709 CAS &    7.20 &   59.29 & N &  9.45 &   9.60 &  11.37 &  21.57 & CV\\
BD+6270 &    9.30 &   61.38 & N &  9.48 &   9.75 &  11.61 &  22.05 & Star\\
FERMIJ0109+6134 &   17.44 &   61.56 & N &  9.12 &   9.39 &   9.96 &  18.30 & Blazar\\
PKS 0116-219 &   19.74 &  -21.69 & N &  8.85 &   8.34 &   8.31 &  16.08 & Quasar\\
PKS 0142-278 &   26.26 &  -27.56 & N &  8.85 &   8.40 &   8.43 &  16.26 & Quasar\\
PKS 0215+015 &   34.45 &    1.75 & N &  8.82 &   8.28 &   8.22 &  15.87 & Quasar\\
S4 0218+35 &   35.27 &   35.94 & N &  8.85 &   8.40 &   8.40 &  16.17 & Quasar\\
PKS 0235-618 &   39.22 &  -61.60 & N &  9.15 &   9.36 &  10.08 &  18.96 & Quasar\\
LSI+61 303 &   40.13 &   61.23 & N &  9.18 &   9.48 &  10.47 &  19.86 & Gamma Ray Binary\\
PKS 0244-470 &   41.50 &  -46.86 & N &  9.00 &   8.91 &   9.30 &  17.76 & Quasar\\
ALGOL &   47.04 &   40.96 & N &  9.09 &   8.64 &   9.48 &  18.24 & Star\\
PKS 0332-403 &   53.56 &  -40.14 & N &  9.06 &   8.79 &   9.57 &  18.57 & BL Lac - type object\\
PKS 0347-211 &   57.49 &  -21.05 & N &  8.85 &   8.43 &   8.46 &  16.29 & Quasar\\
PKS 0402-362 &   60.97 &  -36.08 & N &  8.91 &   8.61 &   8.85 &  16.95 & Quasar\\
PKS 0420-01 &   65.82 &   -1.34 & N &  8.85 &   8.37 &   8.34 &  16.11 & Quasar\\
PKS 0440-00 &   70.66 &   -0.30 & N &  8.88 &   8.49 &   8.55 &  16.44 & Quasar\\
3C129 &   72.29 &   45.01 & N &  9.21 &   8.97 &  10.14 &  19.47 & Quasar\\
LMCX-2 &   80.12 &  -71.96 & N &  9.75 &  10.20 &  12.24 &  22.86 & LMXB/NS\\
PKS 0528+134 &   82.74 &   13.53 & N &  9.09 &   8.97 &   9.69 &  18.78 & Quasar\\
MAXIJ0556-332 &   89.19 &  -33.17 & N &  9.12 &   8.79 &   9.69 &  18.60 & LMXB/NS\\
PKS 0601-70 &   90.30 &  -70.60 & N &  9.33 &   9.81 &  10.77 &  20.01 & Quasar\\
TXS 0628-240 &   97.75 &  -24.11 & N &  9.06 &   8.76 &   9.48 &  18.39 & BL Lac - type object \\
HESSJ0632+057 &   98.25 &    5.80 & N &  9.06 &   8.52 &   9.21 &  17.79 & Gamma Ray Binary\\
MG2 J071354+193 &  108.48 &   19.58 & N &  8.85 &   8.37 &   8.37 &  16.14 & Blazar\\
4C +14.23 &  111.32 &   14.42 & N &  8.85 &   8.37 &   8.37 &  16.17 & Quasar\\
PKS 0805-07 &  122.06 &   -7.85 & N &  8.88 &   8.49 &   8.55 &  16.47 & Quasar\\
BZQ J0850-1213 &  132.54 &  -12.23 & N &  8.91 &   8.58 &   8.70 &  16.74 & Quasar\\
S4 0917+44 &  140.24 &   44.70 & N &  9.00 &   8.85 &   9.24 &  17.64 & Quasar\\
4C +55.17 &  149.41 &   55.38 & N &  9.09 &   9.24 &   9.93 &  18.78 & Quasar\\
MG2 J101241+243 &  153.17 &   24.66 & N &  8.88 &   8.43 &   8.49 &  16.38 & Quasar\\
TXS 1013+054 &  154.01 &    5.22 & N &  8.88 &   8.52 &   8.61 &  16.62 & Quasar\\
FIRSTJ102347.6+003841 &  155.95 &    0.64 & N &  9.06 &   8.55 &   9.24 &  17.91 & LMXB/NS\\
PKS 1124-186 &  171.77 &  -18.96 & N &  8.94 &   8.58 &   8.70 &  16.77 & Quasar\\
TON 599 &  179.88 &   29.25 & N &  8.91 &   8.52 &   8.64 &  16.65 & Quasar\\
1FGLJ1227.9-4852 &  186.98 &  -48.88 & N &  9.57 &   9.72 &  11.31 &  21.42 & Pulsar/PWN\\
PKS 1244-255 &  191.70 &  -25.80 & N &  9.00 &   8.73 &   8.94 &  17.13 & Quasar\\
3C 279 &  194.05 &   -5.79 & N &  8.94 &   8.58 &   8.70 &  16.80 & Quasar\\
PSRB1259-63 &  195.70 &  -63.83 & N &  9.54 &  10.41 &  11.49 &  21.15 & Pulsar/PWN\\
GB 1310+487 &  198.18 &   48.48 & N &  9.03 &   9.03 &   9.57 &  18.18 & Quasar\\
PKS 1329-049 &  203.02 &   -5.16 & N &  8.94 &   8.55 &   8.64 &  16.71 & Quasar\\
CENX-4 &  224.59 &  -31.67            & N & 9.03 &  8.67 &  8.82  &  16.89  &   LMXB/NS\\
PKS 1502+106 &  226.10 &   10.49 & N &  8.97 &   8.61 &   8.76 &  16.86 & Quasar\\
PKS 1510-08 &  228.21 &   -9.10 & N &  9.00 &   8.58 &   8.70 &  16.71 & Quasar\\
B2 1520+31 &  230.54 &   31.74 & N &  9.00 &   8.82 &   9.18 &  17.67 & Quasar\\
SWIFTJ1539.2-6227 &  234.82 &  -62.46 & N &  9.42 &   9.84 &  10.89 &  20.61 & BHC\\
4U1543-624 &  236.98 &  -62.57 & N &  9.90 &  10.32 &  12.45 &  23.52 & LMXB/NS\\
SGR1550-5418 &  237.73 &  -54.31 & N &  9.36 &   9.93 &  11.70 &  21.69 & SGR\\
PKS 1551+130 &  238.39 &   12.95 & N &  9.06 &   8.88 &   9.24 &  17.79 & Quasar\\
USCO &  245.63 &  -17.88 & N & 10.14 &  10.56 &  12.96 &  24.99 & CV\\
PKS 1622-253 &  246.45 &  -25.46 & I &  9.36 &   9.57 &  10.41 &  19.89 & Quasar\\
4C +38.41 &  248.81 &   38.13 & N &  9.09 &   9.12 &   9.75 &  18.66 & Quasar\\
IGRJ16479-4514 &  251.98 &  -45.23 & N & 12.66 &  15.30 &  20.52 &  38.40 & HMXB/SFXT\\
GROJ1655-40 &  253.50 &  -39.85 & I & 10.89 &  12.12 &  15.54 &  29.61 & BHC\\
XTEJ1701-462 &  255.24 &  -46.19 & I & 11.34 &  13.08 &  17.13 &  32.43 & LMXB/NS\\
IGRJ17091-3624 &  257.27 &  -36.40 & N & 11.40 &  12.96 &  16.95 &  32.61 & BHC\\
SWIFTJ1713.4-4219 &  258.36 &  -42.33 & N & 10.83 &  13.11 &  16.17 &  30.84 & BHC\\
4C +51.37 &  265.15 &   52.20 & N &  9.27 &   9.72 &  10.71 &  20.19 & Quasar\\
NGC6440 &  267.22 &  -20.36 & N & 11.40 &  12.90 &  16.83 &  32.07 & Globular Cluster\\
HESSJ1825-137 &  276.13 &  -13.85 & N & 10.38 &  11.13 &  13.89 &  26.49 & Pulsar/PWN\\
4U1822-000 &  276.34 &   -0.01 & N &  9.75 &   9.87 &  11.73 &  22.47 & LMXB/NS\\
LS5039 &  276.56 &  -14.85 & N & 10.35 &  11.10 &  13.86 &  26.43 & Gamma Ray Binary\\
SGR1833-0832 &  278.43 &   -8.52 & N &  9.90 &  10.20 &  12.24 &  23.40 & SGR\\
MAXIJ1836-194 &  278.96 &  -19.39 & I & 10.05 &  10.98 &  13.14 &  25.23 & BHC\\
S4 1849+67 &  282.32 &   67.09 & N &  9.69 &  10.95 &  12.72 &  23.67 & Seyfert 1\\
S4 1851+48 &  283.12 &   48.93 & N &  9.69 &  10.95 &  12.72 &  23.67 & Quasar\\
IGRJ18539+0727 &  283.48 &    7.46 & I &  9.30 &   9.57 &  10.41 &  19.71 & BHC\\
XTEJ1858+034 &  284.65 &    3.35 & I &  9.45 &   9.54 &  10.35 &  14.52 & HMXB/NS\\
TXS 1902+556 &  285.80 &   55.68 & N &  9.72 &  10.44 &  12.57 &  23.94 & BL Lac - type object\\
4U1901+03 &  285.90 &    3.19 & I &  9.81 &  10.02 &  11.55 &  15.48 & HMXB/NS\\
TXS 1920-211 &  290.88 &  -21.08 & N &  9.06 &   8.76 &   9.00 &  17.28 & Quasar\\
IGRJ19294+1816 &  292.48 &   18.64 & N & 9.15 &  9.30 &  9.75  &  18.06 &   HMXB/SFXT\\
PKS 1954-388 &  299.50 &  -38.75 & N &  9.06 &   8.82 &   9.12 &  17.43 & Quasar\\
4U1957+115 &  299.85 &   11.71 & N &  9.24 &   8.88 &   9.84 &  18.93 & LMXB/NS\\
PKS 2052-47 &  314.07 &  -47.25 & N &  9.12 &   9.09 &   9.57 &  18.27 & Quasar\\
V407CYG &  315.54 &   45.78 & N &  9.51 &   9.69 &  11.46 &  21.90 & CV\\
SS CYG &  325.68 &   43.59 & N &  9.51 &   9.60 &  11.43 &  21.72 & CV\\
OX 169 &  325.90 &   17.73 & N &  8.94 &   8.61 &   8.76 &  16.83 & Seyfert 1\\
PKS 2155-304 &  329.72 &  -30.23 & N &  9.06 &   8.67 &   9.27 &  18.06 & BL Lac - type object\\
PKS 2201+171 &  330.86 &   17.43 & N &  8.91 &   8.52 &   8.61 &  16.56 & Quasar\\
CTA 102 &  338.15 &   11.73 & N &  8.94 &   8.49 &   8.76 &  17.13 & Quasar\\
PKS 2255-282 &  344.52 &  -27.97 & N &  8.88 &   8.46 &   8.58 &  16.56 & Seyfert 1\\
B2 2308+34 &  347.77 &   34.42 & N &  8.88 &   8.43 &   8.49 &  16.29 & Quasar\\
CTS 0490 &  351.37 &  -35.97 & N &  8.88 &   8.52 &   8.61 &  16.62 & Quasar\\
PKS 2325+093 &  351.89 &    9.67 & N &  8.82 &   8.28 &   8.22 &  15.90 & Quasar\\
PKS 2326-502 &  352.34 &  -49.93 & N &  9.03 &   9.00 &   9.45 &  18.03 & Quasar\\
PMN J2345-1555 &  356.30 &  -15.92 & N &  8.85 &   8.34 &   8.31 &  16.05 & Quasar\\

\hline
\end{longtable}
\end{center}
\end{tiny}
\end{landscape}

\twocolumn

\subsection{Periodicity Analysis\label{sec:period}}

Using three years of continuous data  from GBM, we have produced folded light curves and performed  searches for periodicities  for all the X-ray binary systems monitored with the Earth occultation technique. For purposes of this catalog, this analysis is intended to increase the confidence of source detections presented in Table~\ref{tbl:main}. 

\begin{deluxetable}{llllll}
\tablewidth{0pt}
\tabletypesize{\scriptsize}
\tablecaption{\scriptsize{Orbital Periods Observed in GBM EOT Light Curves}}  
  
\tablehead{\colhead{Name}  & \colhead{type} & \colhead{Known Porb (days)}       &   \colhead{$\Delta\chi^2$ (dof=10)}   & \colhead{Reference}  }
\startdata
CYGX-3  &  HMXB/BH  &  0.19969  &  1159.42  		    &  \citet{Singh2002} \\
SCOX-1  &  LMXB/NS  &  0.78700  &  1007.48     &  \citet{Cowley1975} \\
LMCX-4  &  HMXB/NS  &  1.40804  &  95.62     &  \citet{Levine2000} \\
HERX-1  &  LMXB/NS  &  1.70017  &  926.97    &  \citet{Wilson1997} \\
CENX-3  &  HMXB/NS  &  2.08700  &  2907.21   &   \citet{Finger2010} \\
4U1700-377  &  HMXB/NS  &  3.41159  &  2490.47   &  \citet{Finger2012a} \\
4U1538-52  &  HMXB/NS  & 3.72836  &  120.00    & \citet{Finger2012b}  \\
SMCX-1  &  HMXB/NS  &  3.89229  &  220.00     & \citet{Wojdowski1998} \\
CYGX-1  &  HMXB/BH  &  5.59985  &  210.58     & \citet{Iorio2008} \\
VELAX-1  &  HMXB/NS  &  8.96436  &  20222.78    & \citet{vanKerkwijk1995}  \\
OAO1657-415  &  HMXB/NS  &  10.44750  &  269.32    &  \citet{Jenke2012} \\
1E1145.1-6141  &  HMXB/NS  &  14.36500  &  145.04   &   \citet{Ray2002} \\
GX301-2  &  HMXB/NS  &  41.50000  &  23760.50     &  \citet{Koh1997} \\
H1417-624  &  HMXB/NS  &  42.12000  &  141.92   &  \citet{Liu2006} \\
EXO2030+375  &  HMXB/NS  &  46.00000  &  819.78    &  \citet{Wilson2008} \\
A0535+262  &  HMXB/NS  &  111.00000  &  88381.29   &  \citet{Bildsten97} \\
GX304-1  &  HMXB/NS  &  133.00000  &  5296.86   &   \citet{Priedhorsky1983}\\
XTEJ1946+274  &  HMXB/NS  &  169.20000  &  193.63  &   \citet{Wilson2003} \\
\enddata
\label{orbperiod}
\end{deluxetable}

\begin{figure}[!h]
\center{\includegraphics[width=3.75cm,height=2.34cm]{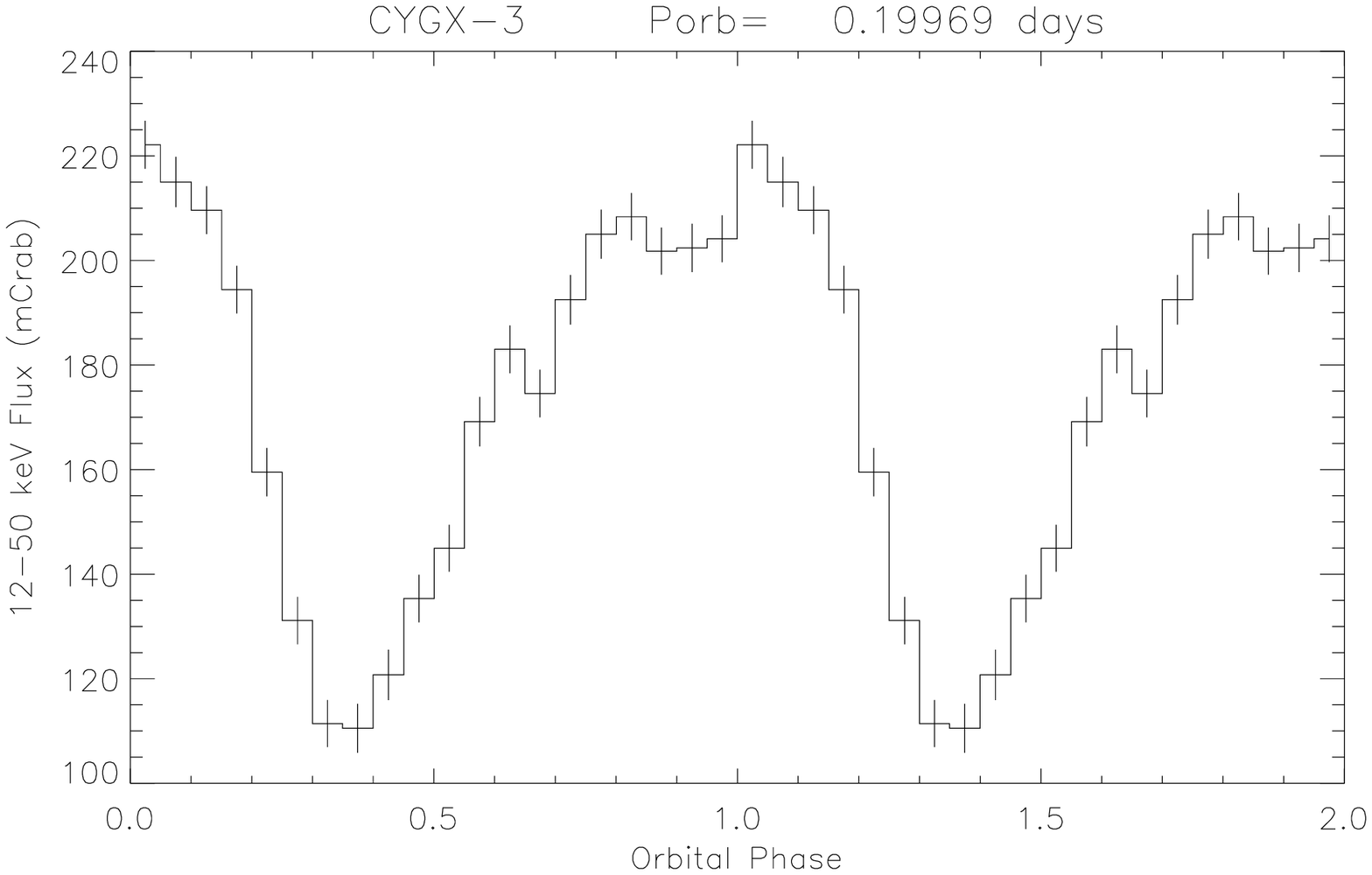} 
\includegraphics[width=3.75cm,height=2.34cm]{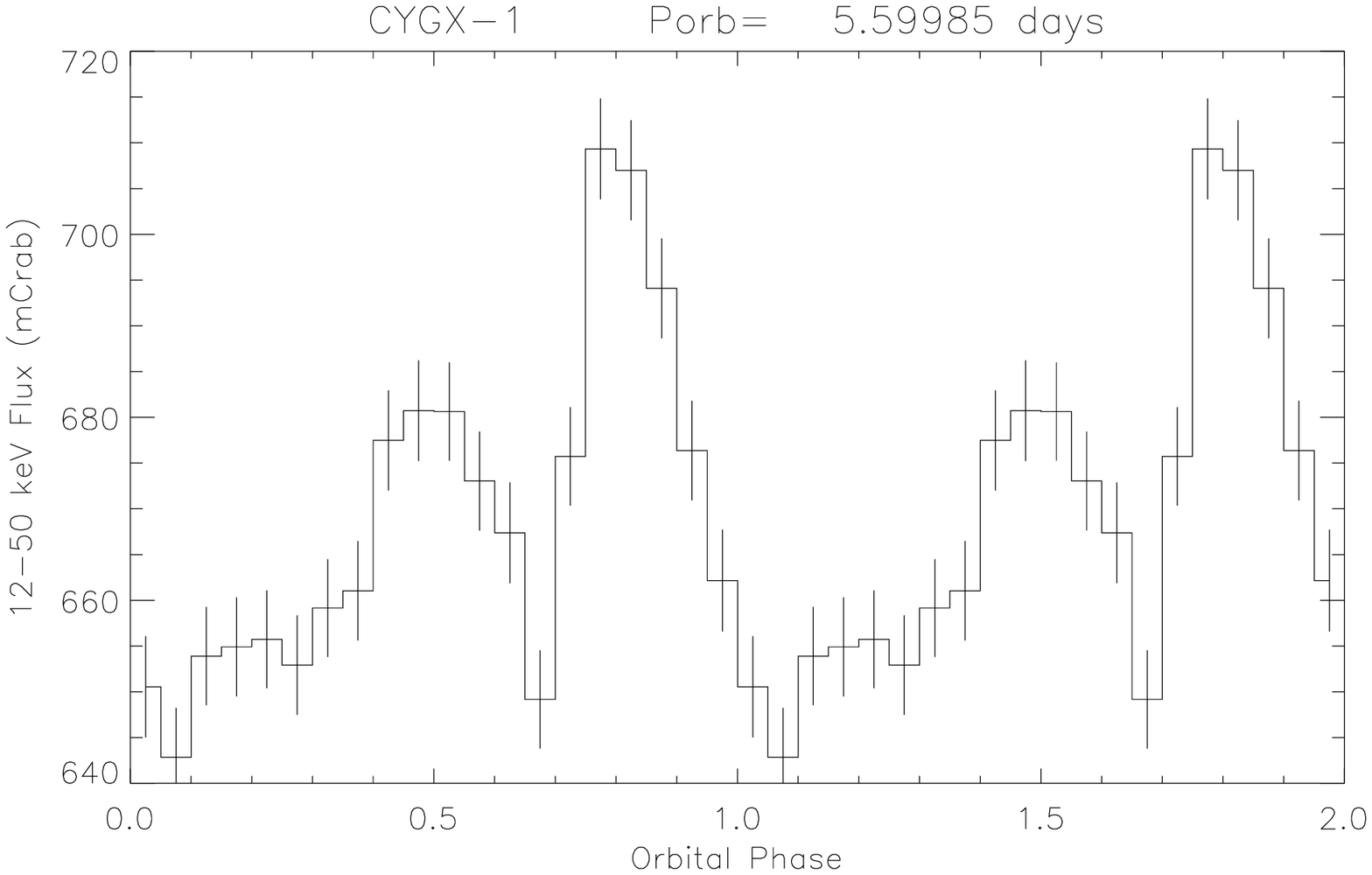}}
\vspace{-0.1in}
\center{\includegraphics[width=3.75cm,height=2.34cm]{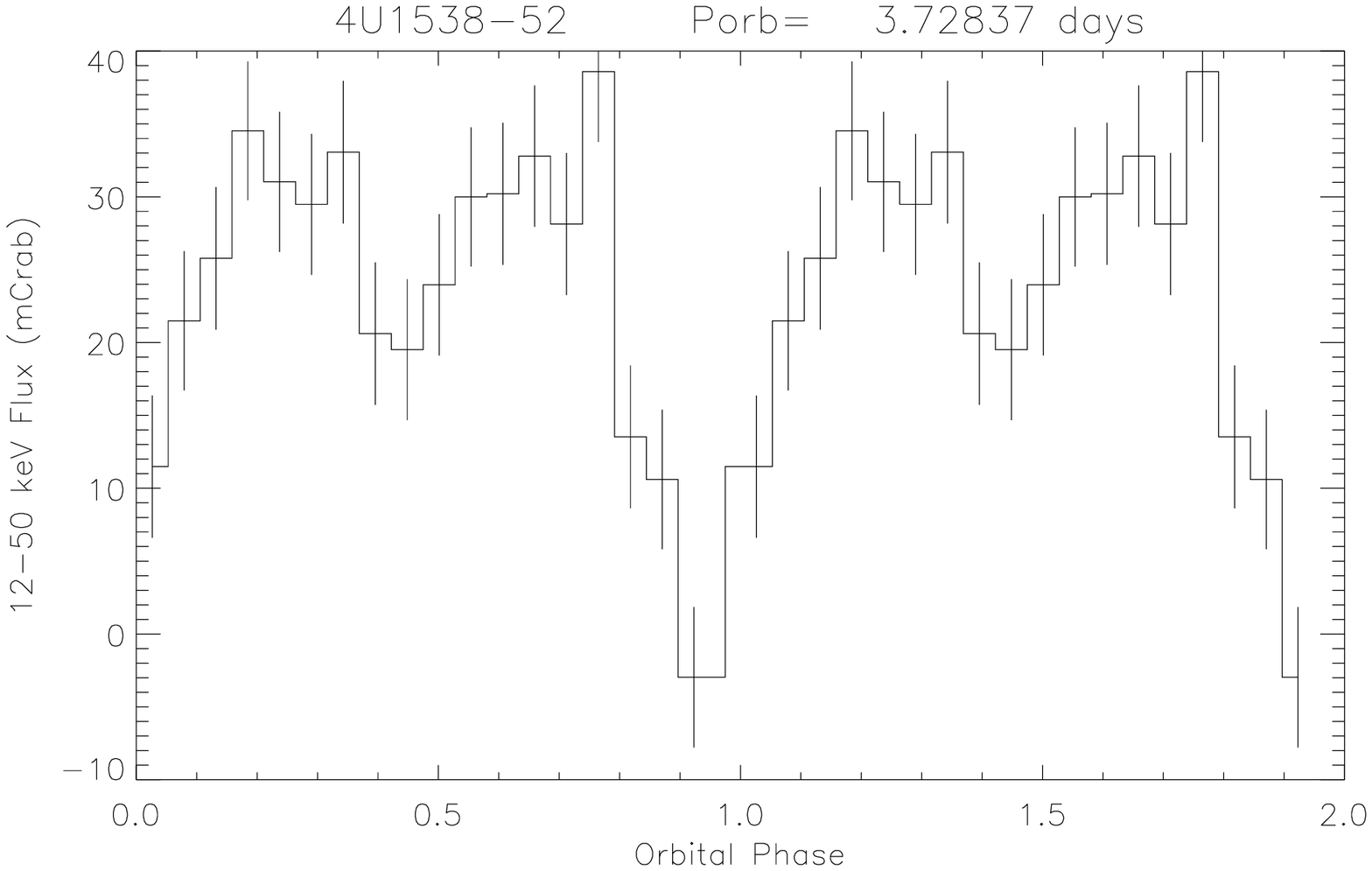} 
\includegraphics[width=3.75cm,height=2.34cm]{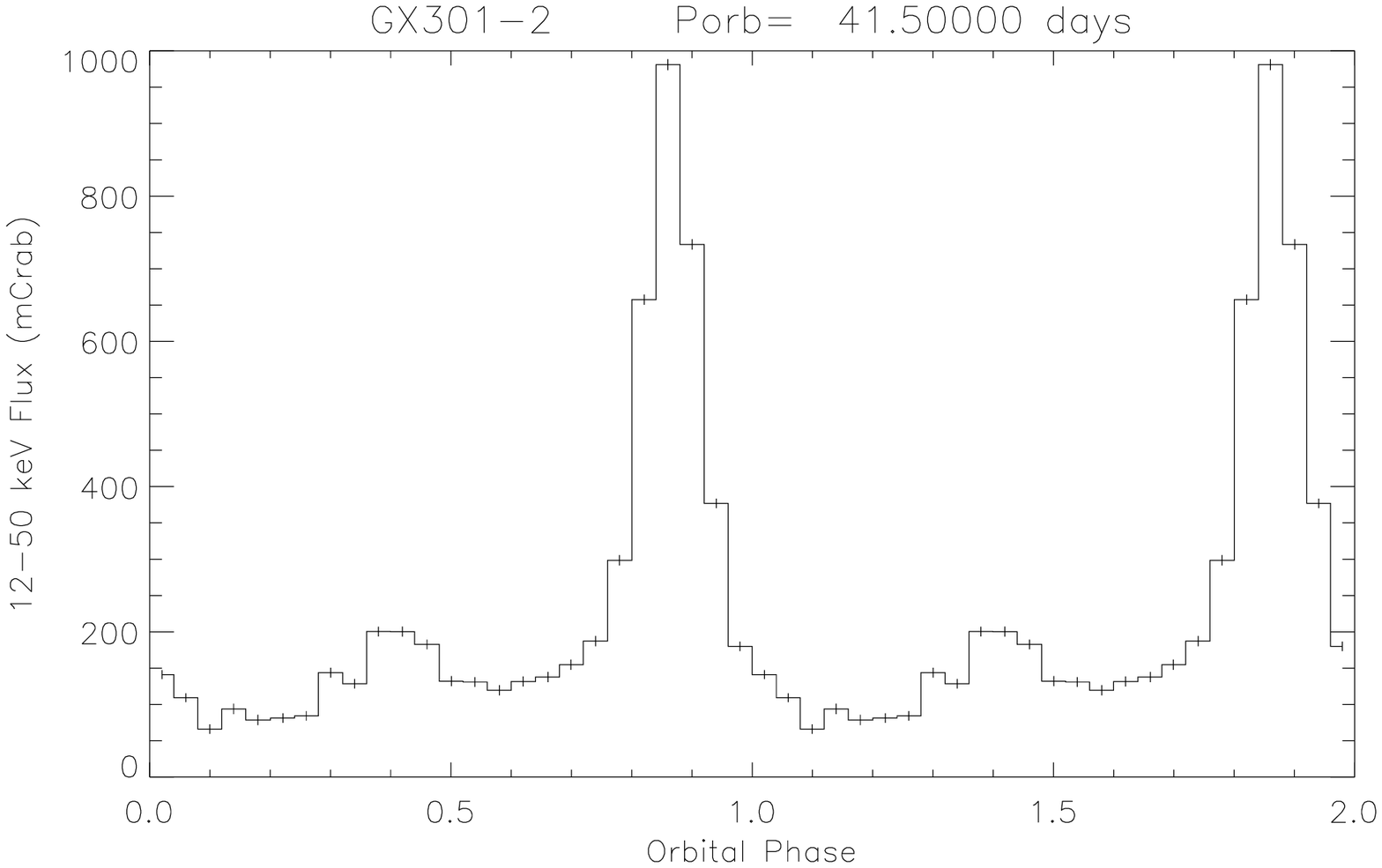}} 
\vspace{-0.1in}
\center{\includegraphics[width=3.75cm,height=2.34cm]{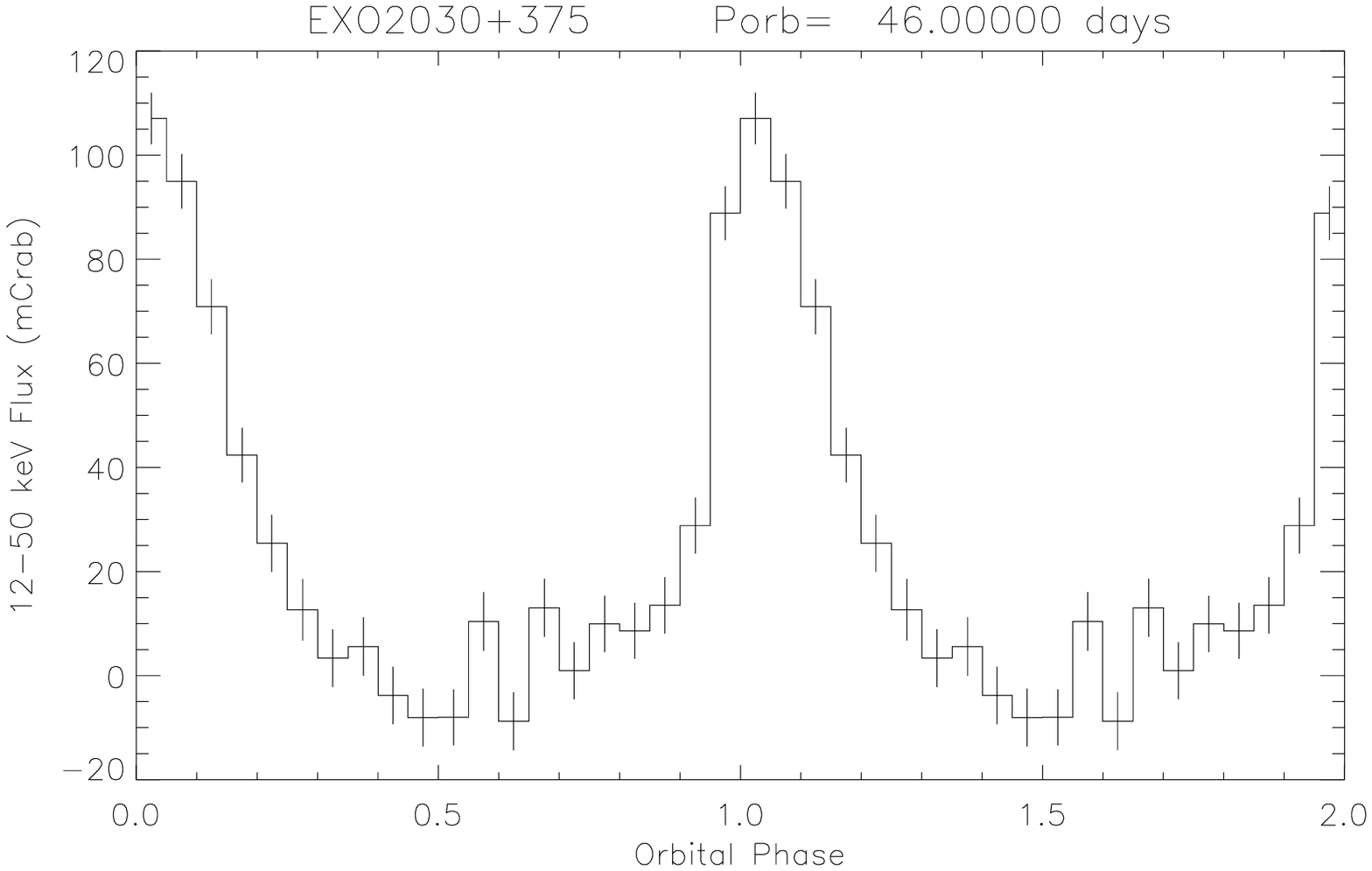}
\includegraphics[width=3.75cm,height=2.34cm]{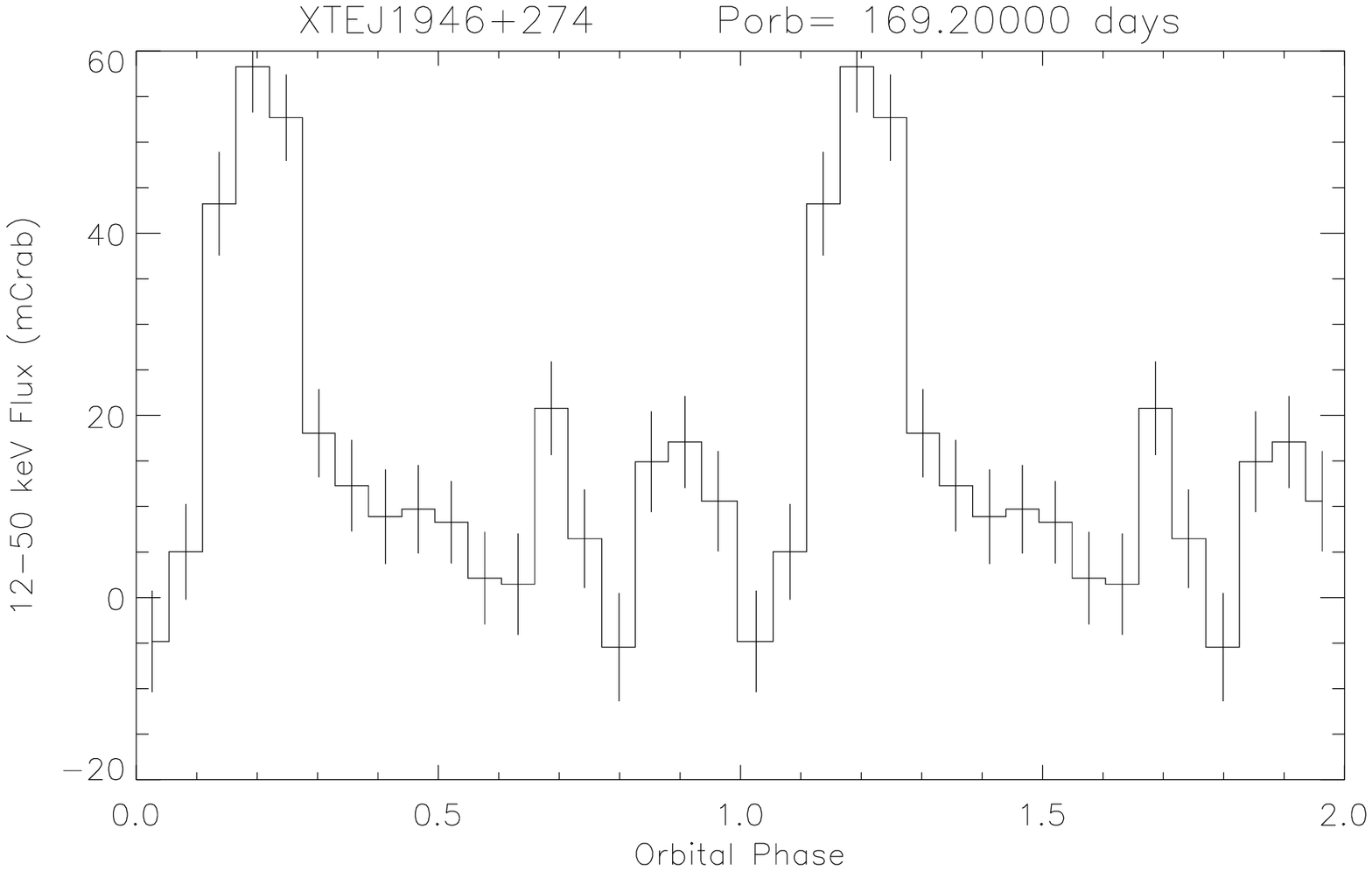}}
\vspace{-0.1in}
\caption{Orbital light curves in the 12--50 keV band for a sample of binaries detected by GBM using EOT. The sample includes the microquasar Cyg\,X--3, the black hole system Cyg\,X--1, the wind-fed HMXRBs 4U\,1538--52 and GX\, 301--2, as well as the Be/X-ray binaries EXO\,2030+375 and XTE\,J1946+274. Arbitrary orbital phases are plotted.\label{orb2}}
\end{figure}

For this simple approach, the lightcurves were folded without barycentering. Values in each phase bin were calculated using weighted averages and statistical (counting) errors, which do not account for intrinsic variability of the source flux. 
Our primary detection criteria uses a $\Delta\chi^2$ fit of the  folded light curve, at the known orbital period, to a constant average flux.  Our initial cut considered an orbital period detected when $\Delta\chi^2 > 40$ for 10 degrees of freedom and the statitistical chance probability is $< 1 \times 10^{-5}$. To estimate systematic effects, we folded the data for 3C 279, an undetected source not expected to show periodicities, at each of the known periods for which we were folding data. The maximum $\Delta\chi^2$ found for 3C 279 was 80 for a period of 18.5 days. Therefore our final detection criteria for orbital periods were $\Delta\chi^2 >80$ for 10 degrees of freedom, corresponding to a statistical chance probablility of $5 \times 10^{-13}$.
 Table~\ref{orbperiod} lists the detected sources, known orbital periods, $\Delta\chi^2$ value, and references for each orbital period. A total of 18 sources were detected, of which 14 are HMXB/NS and two are LMXB/NS systems, and two BH systems.  The detectable orbital periods ranged from minutes to hundreds of days. Figure~\ref{orb2} shows the orbital light curves in the 12--50 keV band for a sample of binaries: the microquasar Cyg\,X--3, the black hole system Cyg\,X--1,the wind-fed HMXBs 4U\,1538--52 and GX 301--2, and the Be/X-ray binaries EXO\,2030+375 and XTE\,J1946+274.
 
\begin{figure}[!h]
\center{\includegraphics[width=3.75cm,height=2.34cm]{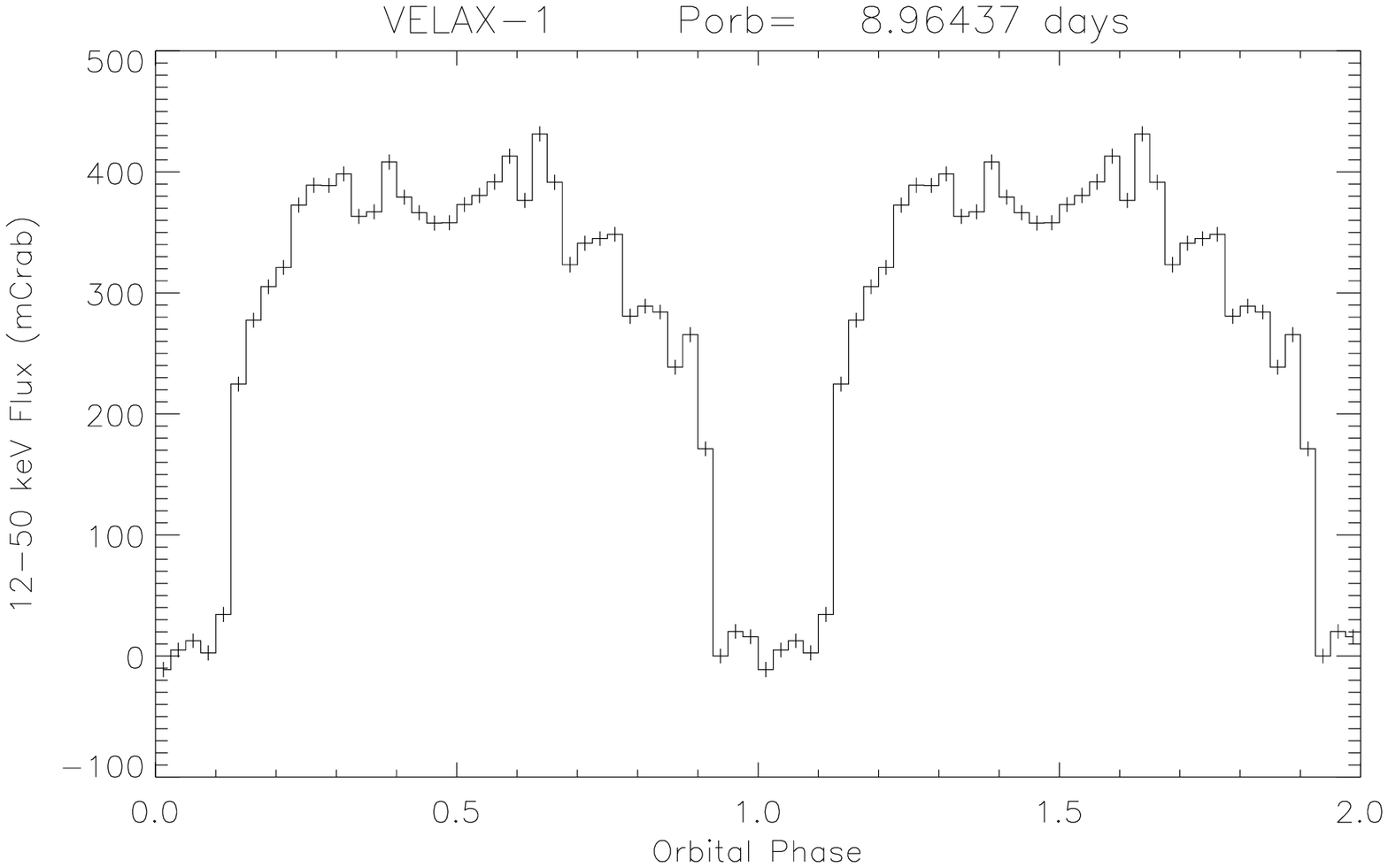}
\includegraphics[width=3.75cm,height=2.34cm]{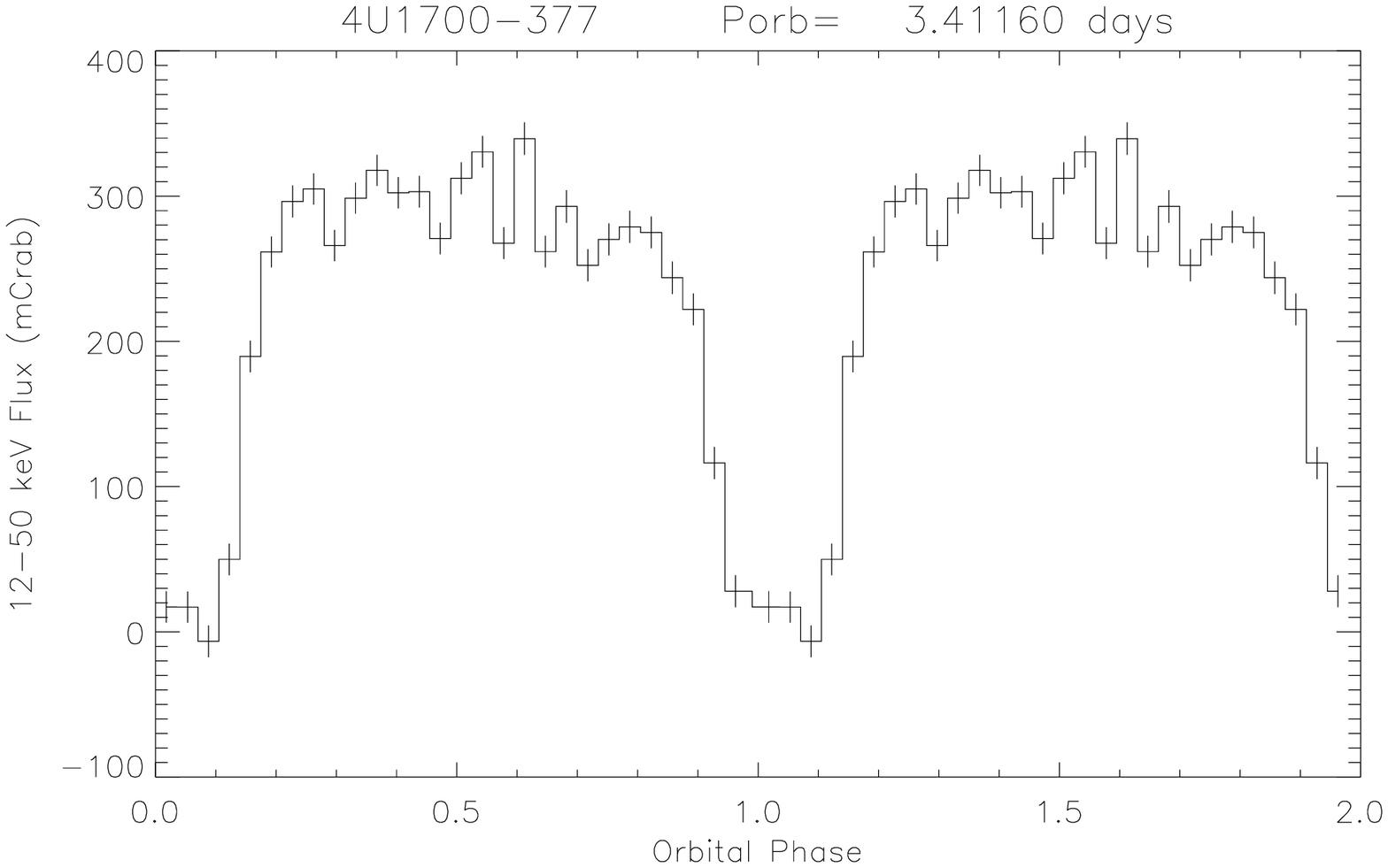}}
\vspace{-0.1in}
\center{\includegraphics[width=3.75cm,height=2.34cm]{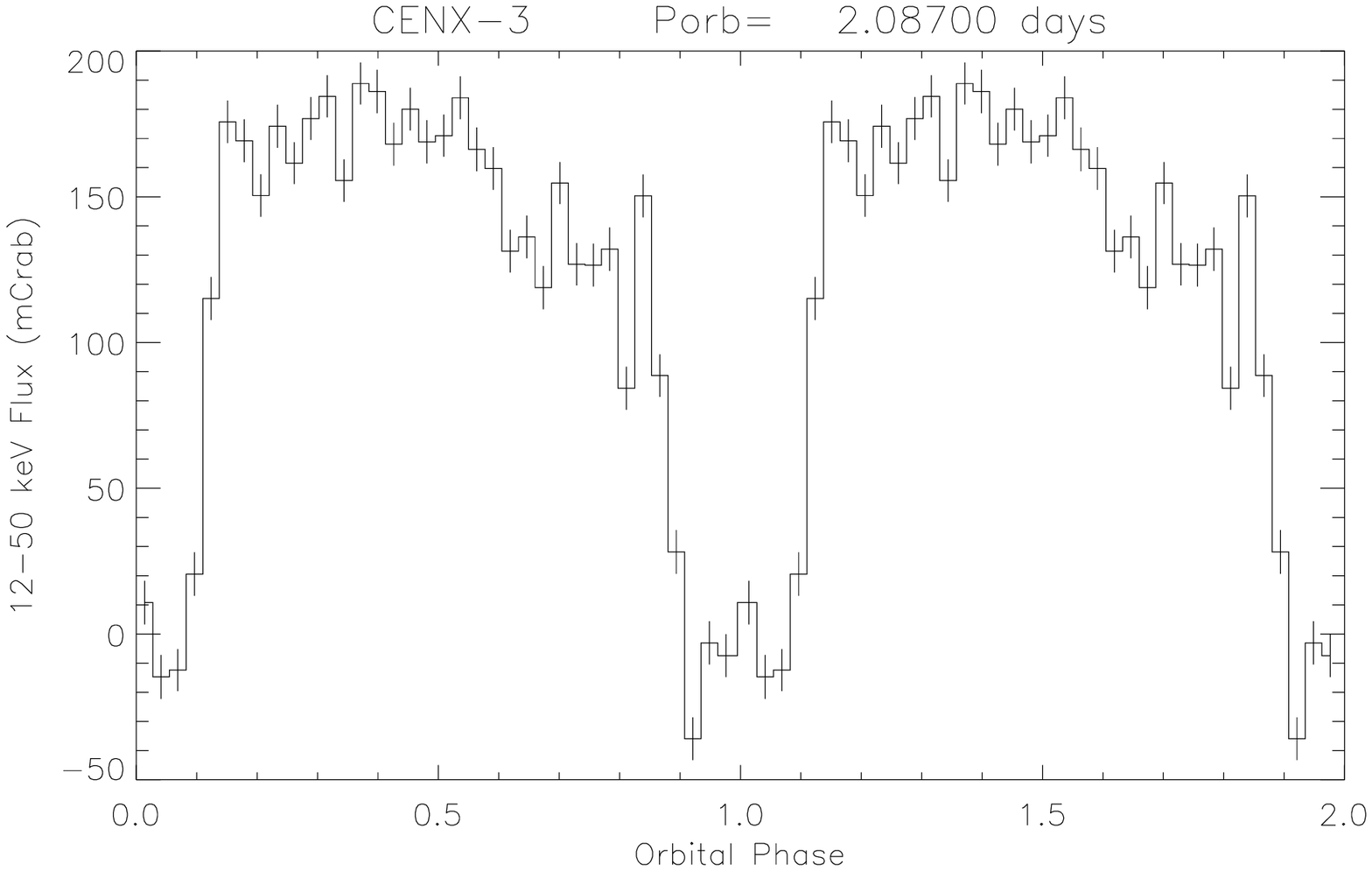}
\includegraphics[width=3.75cm,height=2.34cm]{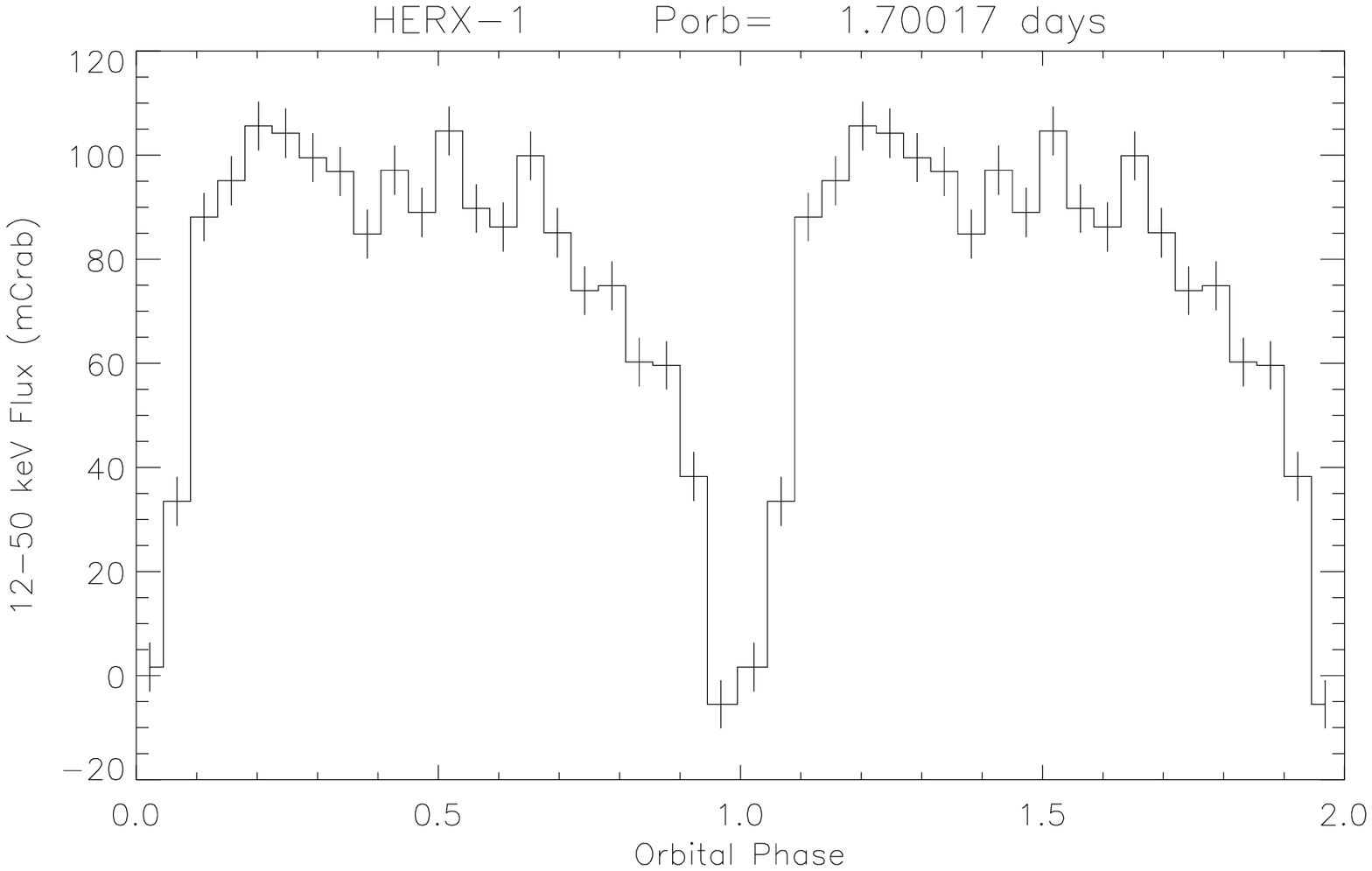}}
\vspace{-0.1in}
\center{\includegraphics[width=3.75cm,height=2.34cm]{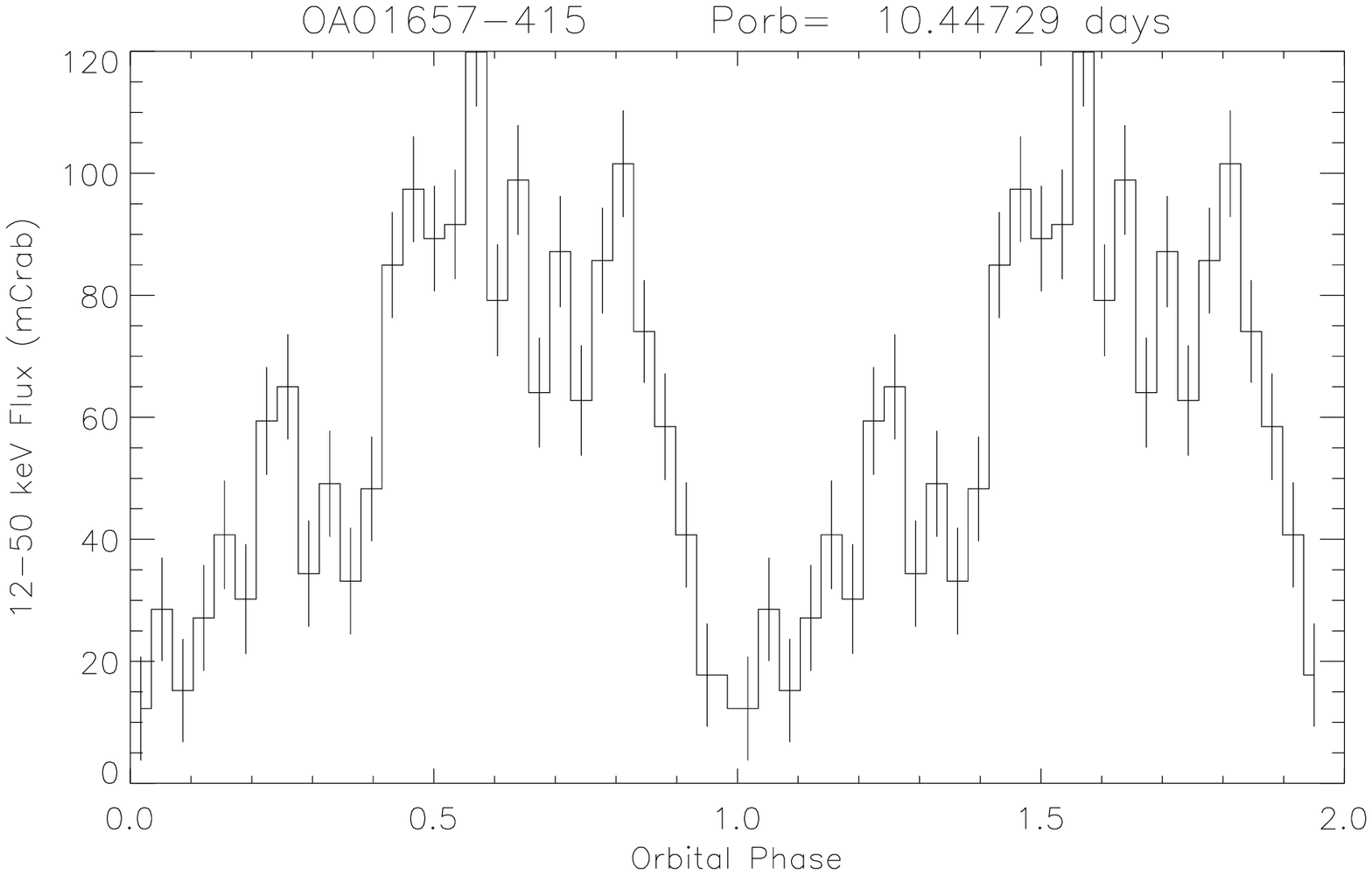}
\includegraphics[width=3.75cm,height=2.34cm]{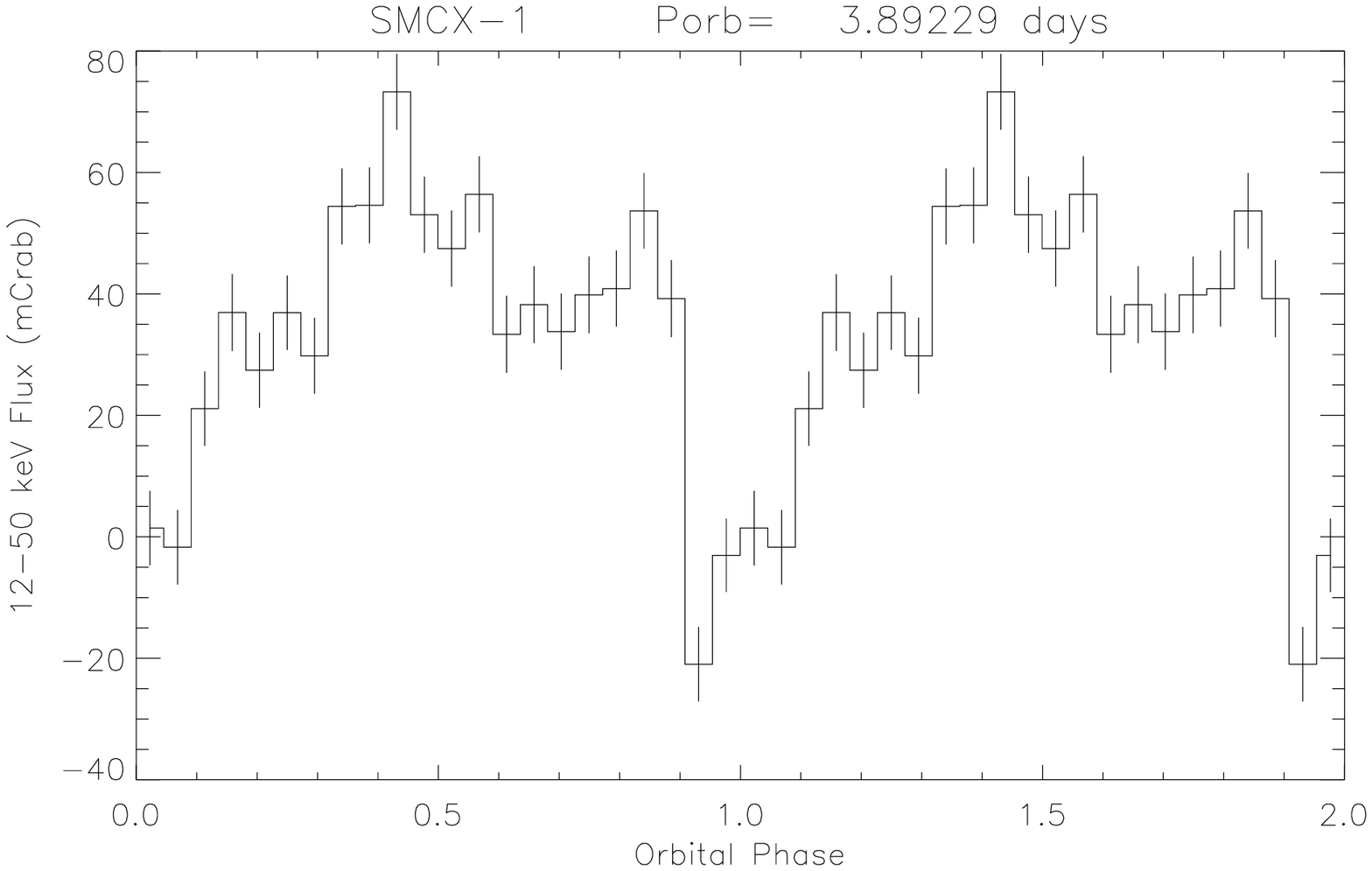}}
\vspace{-0.1in}
\caption{Eclipse profiles in the 12--50 keV band for 6 bright persistent HMXBs: Vela\,X--1, 4U\,1700--377, Cen\,X--3,
Her\,X--1, OAO\,1657--415  and SMC\,X-1. \label{uno}}
\end{figure}


We also attempted a more ambitious periodicity search. Since our data are unevenly sampled, the Lomb-Scargle periodogram \citep{Scargle1982} is better suited to the GBM data than a direct Fourier transform technique. However the non-sinusoidal nature and red-noise in many of the light curves resulted in only eight sources being easily detected using the Lomb-Scargle technique. These bright and frequently active sources included the microquasar Cyg X-3 and 7 HMXB/NS systems:  4U 1700-377, A0535+262, Cen X-3, EXO 2030+375, GX 301-2, GX 304-1, and Her X-1. Because our goal for this initial catalog was confirmation of detections, we did not pursue this analysis further.

Several bright X--ray binaries containing neutron stars undergo complete eclipses by their companions. In  Fig~\ref{uno} we present eclipse profiles in the 12--50 keV band for 6 bright persistent HMXBs: Vela\,X--1, 4U\,1700--377, Cen\,X--3, Her\,X--1, OAO\,1657--415  and SMC\,X-1.  As described in section~\ref{sec:fitting},  during the fitting process eclipsing sources are excluded as interfering source terms when they are in eclipse as defined by the the eclipse database. These sources were strong detections (Table~\ref{orbperiod}) using the folded orbit profiles. Three of these sources were also easily detected using Lomb-Scargle techniques.

\section{Summary and Future Plans}

During the first three years of the \fermi\ mission, from 2008 August 12 to 2011 August 11, we monitored a catalog of 209 sources using the Earth occultation technique. A total of 99 sources were detected at $>5\sigma$ (including systematic errors) in 3-year average fluxes or via transient or periodicity searches. X-ray binaries containing neutron stars dominated the detected sources  with 40 of 52 monitored LMXB/NS systems and  31 of 39 monitored HMXB/NS systems. Black hole systems (BHC) were the next most numerous class detected, with 12 detections out of 19 monitored, including seven detected in the 100-300 keV band and one in the 300-500 keV band. Twelve AGN sources of 71 monitored were detected, with one of these detected in the 100-300 keV 3-year average flux. Five additional AGN sources and two LMXB/NS look promising for detection above 100 keV with more data. All-sky monitoring with the GBM EOT complements other sky monitoring instrument, providing confirming observations below 100 keV, providing coverage when other instruments are limited by solar constraints, and providing important monitoring observations above 100 keV, to reveal e.g. state changes in BHC such as Cyg X-1 and XTE J1752-223. Future work using the BGO detectors with custom CSPEC data above 100 keV will provide additional effective area in the 100-500 keV band and will explore energies above 1 MeV for bright events. Multi-instrument monitoring including GBM was crucial to our discovery of the 7\% decline of the Crab Nebula flux from 2008-2010 \citep{Wilson2011}. 

\acknowledgements 

This work is supported by the NASA \fermi\ Guest Investigator program, NASA/Louisiana Board of Regents Cooperative Agreement \\ NNX07AT62A (LSU), and the Louisiana Board of Regents Graduate Fellowship Program (J. Rodi).
We thank the NOAA Space Prediction Center helpdesk for their support and Sebastian Drave for providing information about SFXT systems with known orbital solutions. Flare database entries and \swift/BAT comparisons relied upon \swift/BAT transient monitor results provided by the \swift/BAT team. This research has made use of the \maxi\ data provided by RIKEN, JAXA and the \maxi\ team and results provided by the \rxte/ASM teams at MIT and GSFC.


\begin{thebibliography}{}
\setlength{\baselineskip}{10pt}
\setlength{\parskip}{-4pt}
\scriptsize
\bibitem[Abdo et al. (2011)]{Abdo11} Abdo, A.A. et al. 2011, arXiv:1108.1435v1
\bibitem[Ackermann et al. (2011a)]{Ackermann11} Ackermann, M. et al. 2011a, arXiv:1108.1420v2
\bibitem[Ackermann et al. (2011b)]{Ackermann2011b} Ackermann, M. et al. 2011b, \apj, accepted, arXiv:1111.7026v1
\bibitem[Baumgartner et al. (2010)]{Baumgartner2010} Baumgartner, W. et al. 2010, \apjs, submitted; and   \protect{\url{http://swift.gsfc.nasa.gov/docs/swift/results/bs58mon/}}
\bibitem[Bildsten et al. (1997)]{Bildsten97} Bildsten, L. et al. 1997, \apjs, 113, 367
\bibitem[Bissaldi et al.(2009)]{Bissaldi2009} Bissaldi, E. et al. 2009, Exp. Astron., 24, 47
\bibitem[Bowyer et al. (1964)]{Bowyer1964} Bowyer, S., Byram, E.T., Chubb, T.A., \& Friedman, H. 1964, Science, 146, 912
\bibitem[Camero-Arranz et al. (2012)]{Camero2012} Camero-Arranz, A. et al. \apj, submitted, arXiv:1109.3924
\bibitem[Case et al. (2011a)]{Case2011a} Case, G.L. et al. 2011a, \apj, 729, 105
\bibitem[Case et al. (2011b)]{Case2011b} Case, G. L., Wilson-Hodge, C. A., Camero-Arranz, A., Chaplin, V., Cherry, M.
L., Finger, M.,\& Jenke, P.  2011b, ATel \# 3636
\bibitem[Case et al. (2011c)]{Case2011c} Case, G. L., Wilson-Hodge, C. A., Camero-Arranz, A., Chaplin, V., Cherry, M.
L., Finger, M.,\& Jenke, P. 2011c, ATel \# 3802
\bibitem[Case et al. (2011d)]{Case2011d} Case, G.L. et al. 2011d, in Proc. 2011 Fermi Symposium, eConf C110509, arXiv:1111.0686
\bibitem[Case et al. (2012)]{Case2012} Case, G.L. et al. 2012, in preparation
\bibitem[Cowley \& Crampton (1975)]{Cowley1975} Cowley, A. P., \& Crampton, D. 1975, \apj, 201, L65
\bibitem[Davison \& Morrison (1977)]{Davidson1977} Davidson, P.J.N \& Morrison, L.V 1977, MNRAS, 178, 53
\bibitem[Finger et al. (2010)]{Finger2010} Finger, M.H., Wilson-Hodge, C.A, Camero-Arranz, A., Jenke, P. 2010, BAAS, 41, 726
\bibitem[Finger (2012a)]{Finger2012a} Finger, M.H. 2012a, in preparation
\bibitem[Finger (2012b)]{Finger2012b} Finger, M.H. 2012b, in preparation
\bibitem[Fukada et al. (1975)]{Fukada1975} Fukada et al. 1975, Nature, 255, 465
\bibitem[Grinberg et al. (2011)]{Grinberg11} Grinberg, V. et al. 2011, ATel \# 3534
\bibitem[Harmon et al.(2002)]{Harmon2002} Harmon, B. A. et al. 2002, \apjs, 138, 149
\bibitem[Harmon et al.(2004)]{Harmon2004} Harmon, B. A. et al. 2004, \apjs, 154, 585
\bibitem[Hoover et al.(2008)]{Hoover2008} Hoover, A. S. et al. 2008, in Gamma-Ray Bursts 2007 (AIP Conf. Proc. 1000), ed. M. Galassi, D. Palmer, \& E. Fenimore (Melville, NY:AIP), 565
\bibitem[Iorio (2008)]{Iorio2008} Iorio, L., 2008, Ap\&SS, 315, 335
\bibitem[Jenke et al. (2012)]{Jenke2012} Jenke, P., Finger, M.H., Wilson-Hodge, C.A., Camero-Arranz, A. 2012, \apj, submitted 
\bibitem[Jourdain \& Roques (2009)]{Jourdain_2009} Jourdain, E. \& Roques, J.P. 2009, \apj, 704, 17
\bibitem[Koh et al. (1997)]{Koh1997} Koh, D. T., Bildsten, L., Chakrabarty, D., Nelson, R. W., Prince, T. A., Vaughan, B. A., Finger, M. H., Wilson, R. B., Rubin, B. C., 1997, \apj, 479, 933
\bibitem[Krivonos et al. (2010)]{Krivonos2010} Krivonos, R. et al. 2010, \aap, 523, A61
\bibitem[Levine et al. (1996)]{Levine1996} Levine, A.M. et al. 1996, 469, L33
\bibitem[Levine, Rappaport, \& Zojcheski (2000)]{Levine2000} Levine, A. M., Rappaport, S. A., \& Zojcheski, G. 2000, \apj, 541, 194
\bibitem[Ling et al.(2000)]{Ling2000} Ling, J. C. et al. 2000, \apjs, 127, 79
\bibitem[Matsuoka et al. (2009)]{Matsuoka2009} Matsuoka, M. et al. 2009, PASJ, 61, 999
\bibitem[Liu, van Paradijs, \& van den Heuvel (2006)]{Liu2006} Liu, Q. Z., van Paradijs, J., \& van den Heuvel, E. P. J. 2006, A\&A, 455, 1165	
\bibitem[Meegan et al.(2009)]{Meegan2009} Meegan, C. et al. 2009, \apj, 702, 791
\bibitem[Negoro et al.(2010)]{Negoro10} Negoro, H. et al. 2010, \# ATel 2711
\bibitem[Negoro et al.(2011)]{Negoro11} Negoro, H. et al. 2011, \# ATel 3534
\bibitem[Nowak et al. (2011)]{Nowak11} Nowak, M. A. et al. 2011, \apj, 728, 13
\bibitem[Priedhorsky \& Terrell (1983)]{Priedhorsky1983} Priedhorsky, W.C. \& Terrell, J., 1983, \apj, 273, 709
\bibitem[Ray \& Chakrabarty (2002)]{Ray2002} Ray, P.S. \& Chakrabarty, D. 2002, \apj, 581, 1293
\bibitem[Rodi et al.(2011)]{Rodi2011} Rodi, J. et al. 2011, in 2011 Fermi Symposium Proceedings, eConf C110509, arXiv:1111.0345
\bibitem[Rushton et al.(2010)]{Rushton10} Rushton, A. et al. 2010, ATel 2714
\bibitem[Quiroz (1961)]{Quiroz61} Quiroz, R.S. 1961, J. Geophys. Res., 66, 2129
\bibitem[Scargle (1982)]{Scargle1982} Scargle, J.D. 1982, \apj, 263, 835
\bibitem[Singh et al.(2002)]{Singh2002} Singh, N.S, Naik, S., Paul, B., Agrawal, P. C., Rao, A. R., Singh, K. Y. 2002, A\&A, 392, 161
\bibitem[Stella et al. (1985)]{Stella1985} Stella, L., White, N. E., Davelaar, J., Parmar, A. N., Blissett, R. J., \& van der Klis, M. 1985, \apj, 288, L45
\bibitem[Toor \& Seward (1974)]{Toor1974} Toor, A., \& Seward, F. D. 1974, \aj, 79, 995
\bibitem[US Standard Atmosphere (1976)]{atm1976}
 US Committee on Extension to the Standard Atmosphere 1976, US Standard Atmosphere, NOAA-S/T 76-1562 (Washington, DC:U.S. Govt Printing Office)
\bibitem[van Kerkwijk et al. (1995)]{vanKerkwijk1995} Van Kerkwijk M.H., van Paradijs J., Zuiderwijk E.J., et al., 1995, A\&A 303, 483
\bibitem[Wilson, Finger, \& Camero-Arranz (2008)]{Wilson2008} Wilson, C.A., Finger, M.H., Camero-Arranz, A., 2008, \apj, 678, 1268
\bibitem[Wilson et al. (2003)]{Wilson2003} Wilson, C. A., Finger, M. H., Coe, M. J. and Negueruela, I., 2003, \apj, 584, 996
\bibitem[Wilson, Scott, \& Finger (1997)]{Wilson1997} Wilson, R.B., Scott, D.M., \& Finger, M.H. 1997 in The Fourth Compton Symposium, AIP conference proceedings 410, 739
\bibitem[Wilson-Hodge \& Case (2010)]{Wilsonhodge10} Wilson-Hodge, C.A. \& Case, G.L. 2010, ATel \# 2721
\bibitem[Wilson-Hodge et al. (2011a)]{Wilson2011} Wilson-Hodge, C.A. et al. 2011a, \apj, 729, 105
\bibitem[Wilson-Hodge et al. (2011b)]{Wilson2011b} Wilson-Hodge, C.A. et al. 2011b, PoS(HTRS 2011)043,\protect{\url{http://pos.sissa.it/}}
\bibitem[Wodjdowski et al. (1998)]{Wojdowski1998} Wodjdowski, P. et al. 1998, \apj, 502, 253
\end{thebibliography}
\end{document}